\documentclass[preprints,article,accept,moreauthors,a4paper,pdftex]{mdpi}

\firstpage{1}
\pubvolume{1}
\issuenum{1}
\articlenumber{0}
\pubyear{2021}
\copyrightyear{2020}
\datereceived{21.05.2021} 
\dateaccepted{30.06.2021} 
\datepublished{} 
\hreflink{https://doi.org/} 
\pdfoutput=1

\usepackage{nicefrac, dsfont, morefloats}
\usepackage[linesnumbered,ruled]{algorithm2e}

\Title{Geometric variational inference}

\TitleCitation{Geometric variational inference}

\Author{Philipp Frank $^{1,2}$*\orcidA{}, Reimar Leike $^{1}$, and Torsten A. En{\ss}lin $^{1,2}$}

\AuthorNames{Philipp Frank, Reimar Leike, Torsten A. En{\ss}lin}

\AuthorCitation{Frank, P.; Leike, R.; En{\ss}lin, T.}

\address{%
	$^{1}$ \quad Max--Planck Institut f{\"u}r Astrophysik, Karl-Schwarzschild-Stra\ss e 1, 85748 Garching, Germany; reimar@mpa-garching.mpg.de (R.L.); ensslin@mpa-garching.mpg.de (T.A.E.)\\
	$^{2}$ \quad Faculty of Physics, Ludwig-Maximilians-Universit{\"a}t M{\"u}nchen, Geschwister-Scholl-Platz 1, 80539 M{\"u}nchen, Germany}

\corres{Correspondence: philipp@mpa-garching.mpg.de}

\abstract{Efficiently accessing the information contained in non-linear and high dimensional probability distributions remains a core challenge in modern statistics. Traditionally, estimators that go beyond point estimates are either categorized as Variational Inference (VI) or Markov-Chain Monte-Carlo (MCMC) techniques. While MCMC methods that utilize the geometric properties of continuous probability distributions to increase their efficiency have been proposed, VI methods rarely use the geometry. This work aims to fill this gap and proposes geometric Variational Inference (geoVI), a method based on Riemannian geometry and the Fisher information metric. It is used to construct a coordinate transformation that relates the Riemannian manifold associated with the metric to Euclidean space. The distribution, expressed in the coordinate system induced by the transformation, takes a particularly simple form that allows for an accurate variational approximation by a normal distribution. Furthermore, the algorithmic structure allows for an efficient implementation of geoVI which is demonstrated on multiple examples, ranging from low-dimensional illustrative ones to non-linear, hierarchical Bayesian inverse problems in thousands of dimensions.}


\begin{document}
\section{Introduction}
In modern statistical inference and machine learning it is of utmost importance to access the information contained in complex and high dimensional probability distributions. In particular in Bayesian inference, it remains one of the key challenges to approximate samples from the posterior distribution, or the distribution itself, in a computationally fast and accurate way. Traditionally, there have been two distinct approaches towards this problem: the direct construction of posterior samples based on Markov Chain Monte-Carlo (MCMC) methods \cite{geyer1992practical,brooks2011handbook,marjoram2003markov}, and the attempt to approximate the probability distribution with a different one, chosen from a family of simpler distributions, known as variational inference (VI) \cite{blei2017variational,hoffman2013stochastic,rezende2015variational,kucukelbir2017automatic} or variational Bayes' (VB) methods \cite{kingma2013auto,fox2012tutorial,vsmidl2006variational}. While MCMC methods are attractive due to their theoretical guarantees to reproduce the true distribution in the limit, they tend to be more expensive compared to variational alternatives. On the other hand, the family of distributions used in VI is typically chosen ad-hoc. While VI aims to provide an appropriate approximation within the chosen family, the entire family may be a poor approximation to the true distribution.

In recent years, MCMC methods have been improved by incorporating geometric information of the posterior, especially by means of Riemannian manifold Hamilton Monte-Carlo (RMHMC) \cite{rmhmc}, a particular hybrid Monte-Carlo (HMC) \cite{duane1987hybrid,betancourt2017conceptual} technique that constructs a Hamiltonian system on a Riemannian manifold with a metric tensor related to the Fisher information metric of the likelihood distribution and the curvature of the prior. For VI methods, however, the geometric structure of the true distribution has rarely been utilized to motivate and enhance the family of distributions used during optimization. One of the few examples being  \cite{doi:10.1080/01621459.2019.1585253} where the Fisher metric has been used to reformulate the task of VI by means of $\alpha$-divergencies in the mean-field setting.

In addition, a powerful variational approximation technique for the family of normal distributions utilizing infinitesimal geometric properties of the posterior is Metric Gaussian Variational Inference (MGVI) \cite{knollmuller2019metric}. In MGVI the family is parameterized in terms of the mean $m$, and the covariance matrix is set to the inverse of the metric tensor evaluated at $m$. This choice ensures that the true distribution and the approximation obtain the same geometric properties infinitesimally, i.E.\ at the location of the mean $m$.
In this work we extend the geometric correspondence used by MGVI to be valid not only at $m$, but also in a local neighborhood of $m$. We achieve this extension by means of an invertible coordinate transformation from the coordinate system used within MGVI, in which the curvature of the prior is the identity, to a new coordinate system in which the metric of the posterior becomes (approximately) the Euclidean metric. We use a normal distribution in these coordinates as the approximation to the true distribution and thereby establish a non-Gaussian posterior in the MGVI coordinate system. The resulting algorithm, called geometric Variational Inference (geoVI) can be computed efficiently and is inherently similar to the implementation of MGVI. This is not by mere coincidence: To linear order, geoVI reproduces MGVI. In this sense, the geoVI algorithm is a non-linear generalization of MGVI that captures the geometric properties encoded in the posterior metric not only infinitesimally, but also in a local neighborhood of this point. We include an implementation of the proposed geoVI algorithm into the software package Numerical Information Field Theory ({\tt NIFTy} \cite{arras2019nifty5}), a versatile library for signal inference algorithms.

\subsection{Mathematical setup}
Throughout this work, we consider the joint distribution $P(d,s)$ of observational data $d \in \Omega$ and the unknown, to be inferred signal $s$. This distribution is factorized into the likelihood of observing the data, given the signal $P(d|s)$, and the prior distribution $P(s)$.
In general, only a subset of the signal, denoted as $s'$, may be directly constrained by the likelihood, such that $P(d|s) = P(d|s')$, and therefore there may be additional hidden variables in $s$, that are unobserved by the data, but part of the prior model.
Thus the prior distribution $P(s)$ may posses a hierarchical structure that summarizes our knowledge about the system prior to the measurement, and $s$ represents everything in the system that is of interest to us, but about which our knowledge is uncertain a priori.
We do not put any constraints on the functional form of $P(s)$, and assume that the signal $s$ solely consists of continuous real valued variables, i.E.\ $s \in X \subset \mathds{R}^M$.
This enables us to regard $s$ as coordinates of the space on which $P(s)$ is defined and to use geometric concepts such as coordinate transformations to represent probability distributions in different coordinate systems.
Probability densities transform in a probability mass preserving fashion. Specifically let $f: \mathds{R}^M \rightarrow X$ be an invertible function, and let $s = f(\xi)$. Then the distributions $P(s)$ and $P(\xi)$ relate via

\begin{equation}
\int P(s) \ \mathrm{d}s = \int P(\xi) \ \mathrm{d}\xi \ .
\end{equation}
This allows us to express $P(s)$ by means of the pushforward of $P(\xi)$ by $f$. We denote the pushforward as

\begin{equation}
P(s) = \left(f \star P(\xi)\right)(s) = \int \delta(s - f(\xi)) \ P(\xi) \ \mathrm{d}\xi = \left.\left(P(\xi) \ \left|\left|\frac{\mathrm{d} f}{\mathrm{d} \xi}\right|\right|^{-1}\right)\right|_{\xi = f^{-1}(s)} \ .
\end{equation}
Under mild regularity conditions on the prior distribution, there always exists an $f$ that relates the complex hierarchical form of $P(s)$ to a simple distribution $P(\xi)$ \cite{bogachev2005triangular}. We choose $f$ such that $P(\xi)$ takes the form of a normal distribution with zero mean and unit covariance and call such a distribution a \emph{standard distribution}:

\begin{equation}
P(\xi) = \mathcal{N}(\xi; 0, \mathds{1}) \ ,
\end{equation}
where $\mathcal{N}(\xi; m, D)$ denotes a multivariate normal distribution in the random variables $\xi$ with mean $m$ and covariance $D$.

We may express the likelihood in terms of $\xi$ as

\begin{equation}\label{eq:likelihood}
P(d|\xi) \equiv P(d|s'=f'(\xi)) \ ,
\end{equation}
where $f'$ is the part of $f$ that maps onto the observed quantities $s'$. In general, $f'$ is a non-invertible function and is commonly referred to as \emph{generative model} or \emph{generative process}, as it encodes all the information necessary to transform a standard distribution into the observed quantities, subject to our prior beliefs. Using equation \eqref{eq:likelihood} we get by means of Bayes' theorem, that the posterior takes the form

\begin{equation}\label{eq:posterior}
P(\xi|d) = \frac{P(\xi,d)}{P(d)} = \frac{P(d|\xi) \ \mathcal{N}(\xi; 0,\mathds{1})}{P(d)} \ .
\end{equation}
Using the push-forward of the posterior, we can recover the posterior statistics of $s$ via

\begin{equation}\label{eq:pushposterior}
P(s|d) = \left(f \star P(\xi|d)\right)(s) \ ,
\end{equation}
which means that we can fully recover the posterior properties of $s$, which typically has a physical interpretation as opposed to $\xi$. In particular equation \eqref{eq:pushposterior} implies that if we are able to draw samples from $P(\xi|d)$ we can simply generate posterior samples for $s$ since $s = f(\xi)$.

\section{Geometric properties of posterior distributions}\label{sec:geopost}
In order to access the information contained in the posterior distribution $P(\xi|d)$, in this work, we wish to exploit the geometric properties of the posterior, in particular with the help of Riemannian geometry. Specifically, we define a Riemannian manifold using a metric tensor, related to the Fisher Information metric of the likelihood and a metric for the prior, and establish a (local) isometry of this manifold to Euclidean space. The associated coordinate transformation gives rise to a coordinate system in which, hopefully, the posterior takes a simplified form despite the fact that probabilities do not transform in the same way as metric spaces do. As we will see, in cases where the isometry is global, and in addition the transformation is (almost) volume-preserving, the complexity of the posterior distribution can be absorbed (almost) entirely into this transformation.

To begin our discussion, we need to define an appropriate metric for posterior distributions. To this end, consider the negative logarithm of the posterior, sometimes also referred to as information Hamiltonian, which takes the form

\begin{equation}
\mathcal{H}(\xi|d) \equiv - \log\left(P(\xi|d)\right) = \mathcal{H}(d|\xi) + \mathcal{H}(\xi) - \mathcal{H}(d) \ .
\end{equation}
A common choice to extract geometric information from this Hamiltonian is the Hessian $\mathcal{C}$ of $\mathcal{H}$. Specifically

\begin{equation}\label{eq:postcurve}
\mathcal{C}(\xi) \equiv \frac{\partial^2 \mathcal{H}(\xi|d)}{\partial \xi \partial \xi'} = \frac{\partial^2 \mathcal{H}(d|\xi)}{\partial \xi \partial \xi'} + \mathds{1} \equiv \mathcal{C}_{d|\xi}(\xi) + \mathds{1} \ ,
\end{equation}
where the identity matrix arises from the curvature of the prior (information Hamiltonian). While $\mathcal{C}$ provides information about the local geometry, it turns out to be unsuited for our approach to construct a coordinate transformation, as it is not guaranteed to be positive definite for all $\xi$. An alternative, positive definite, measure for the curvature can be obtained by replacing the Hessian of the likelihood with its Fisher information metric \cite{fisher1925theory}, defined as

\begin{equation}
\mathcal{M}_{d|\xi}(\xi) = \left<\frac{\partial \mathcal{H}}{\partial \xi} \frac{\partial \mathcal{H}}{\partial \xi'}\right>_{P(d|\xi)} = \left<\frac{\partial^2 \mathcal{H}(d|\xi)}{\partial \xi \partial \xi'}\right>_{P(d|\xi)} = \left<\mathcal{C}_{d|\xi}(\xi)\right>_{P(d|\xi)} \ .
\end{equation}
The Fisher information metric can be understood as a Riemannian metric defined over the statistical manifold associated with the likelihood \cite{rao1992information}, and is a core element in the field of information geometry \cite{amari2000methods} as it provides a distance measure between probability distributions \cite{cencov2000statistical}.
Replacing $\mathcal{C}_{d|\xi}$ with $\mathcal{M}_{d|\xi}$ in equation \eqref{eq:postcurve} we find

\begin{equation}\label{eq:postmetric}
\mathcal{M}(\xi) \equiv \mathcal{M}_{d|\xi}(\xi) + \mathds{1} = \left<\mathcal{C}_{d|\xi}(\xi)\right>_{P(d|\xi)} + \mathds{1} = \left<\mathcal{C}(\xi)\right>_{P(d|\xi)} \ ,
\end{equation}
which, from now on, we refer to as the metric $\mathcal{M}$. As the Fisher metric of the likelihood is a symmetric, positive-semidefinite matrix, we get that $\mathcal{M}$ is a symmetric, positive-definite matrix for all $\xi$.
It is noteworthy that upon insertion, we find that the metric $\mathcal{M}$ is defined as the expectation value of the Hessian of the posterior Hamiltonian $\mathcal{C}$ w.r.t.\ the likelihood $P(d|\xi)$. Therefore, in some way, we may regard $\mathcal{M}$ as the measure for the curvature in case the observed data $d$ is unknown, and the only information given is the structure of the model itself, as encoded in $P(d|\xi)$. This connection is only of qualitative nature, but it highlights a key limitation of $\mathcal{M}$ when used as the defining property of the posterior geometry. From a Bayesian perspective, only the data $d$ that is actually observed is of relevance as the posterior is conditioned on $d$. Therefore a curvature measure that arises from marginalization over the data must be sub-optimal compared to a measure conditional to the data, as it ignores the local information that we gain from observing $d$. Nevertheless, we find that in many practical applications $\mathcal{M}$ encodes enough relevant information about the posterior geometry that it provides us with a valuable metric to construct a coordinate transformation. It is noteworthy that attempts have been provided to resolve this issue via a more direct approach to recover a positive definite matrix from the Hessian of the posterior while retaining the local information of the data. E.g.\ in \cite{betancourt2013general}, the SoftAbs non-linearity is applied to the Hessian and the resulting positive definite matrix is used as a curvature measure. In our practical applications, however, we are particularly interested in solving very high dimensional problems, and applying a matrix non-linearity is currently too expensive to give rise to a scalable algorithm for our purposes. Therefore we rely on the metric $\mathcal{M}$ as a measure for the curvature of the posterior, and leave possible extensions to future research.

\begin{figure*}[ht]
	\centering
	\includegraphics[scale=1., angle=0]{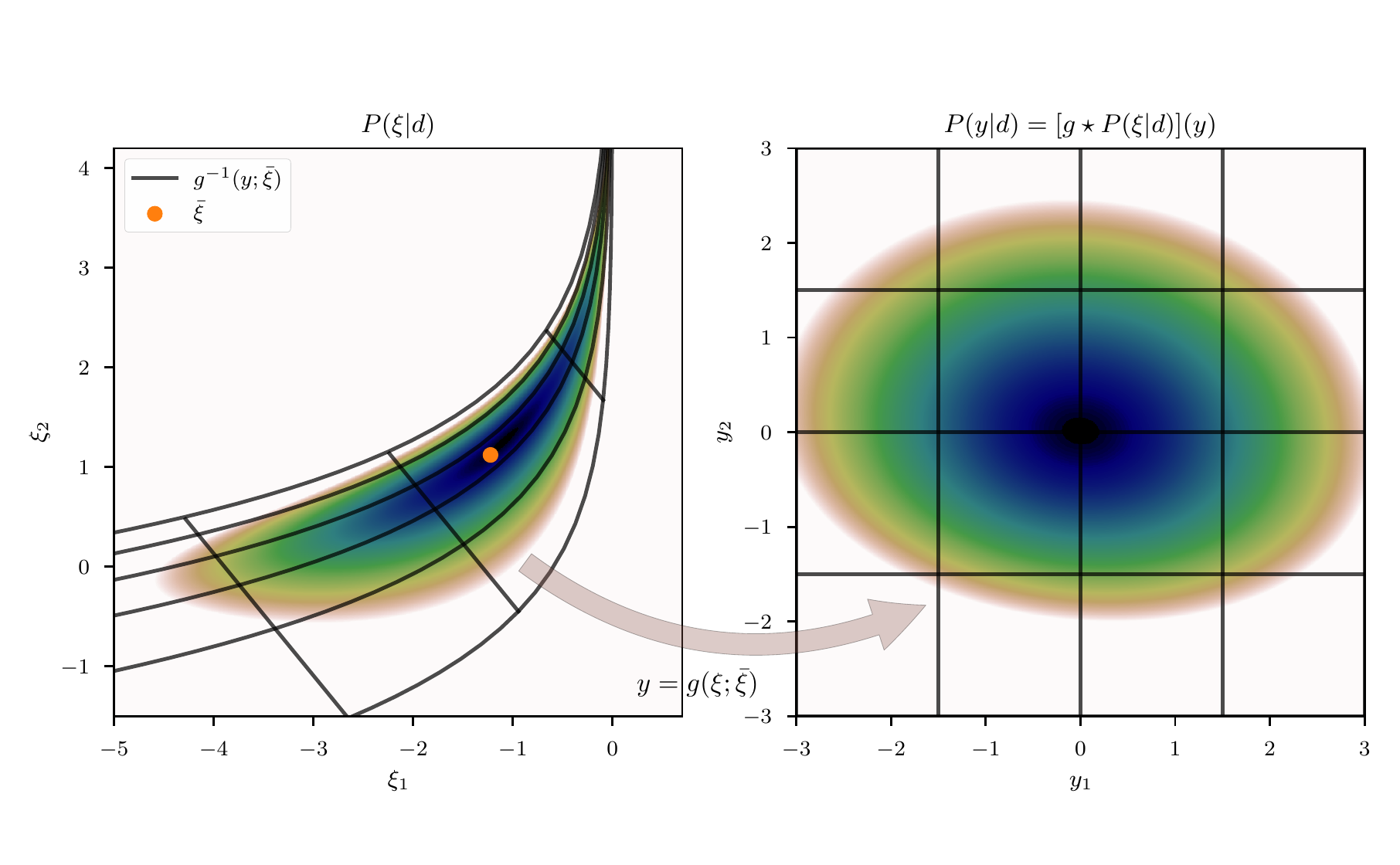}
	\centering
	\caption{Non-linear posterior distribution $P(\xi|d)$ in the standard coordinate system of the prior distribution $\xi$ (left) and the transformed distribution $P(y|d)$ (right) in the coordinate system $y$ where the posterior metric becomes (approximately) the identity matrix. $P(y|d)$ is obtained from $P(\xi|d)$ via the push-forward through the transformation $g$ which relates the two coordinate systems. The functional form of $g$ is derived in section \ref{sec:coord_trafo} and depends on an expansion point $\bar{\xi}$ (orange dot in the left image), and $g$ is set up such that $\bar{\xi}$ coincides with the origin in $y$. To visualize the transformation, the coordinate lines of $y$ (black mesh grid on the right) are transformed back into $\xi$-coordinates using the inverse coordinate transformation $g^{-1}$ and are displayed as a black mesh in the original space on the left. In addition, note that while the transformed posterior $P(y|d)$ arguably takes a simpler form compared to $P(\xi|d)$, it does not become trivial (e.g.\ identical to a standard distribution) as there remain small asymmetries in the posterior density. There are multiple reasons for these deviations which are discussed in more detail in section \ref{sec:properties_error} once we established how the transformation $g$ is constructed.}\label{fig:vis_trafo}
\end{figure*}

\subsection{Coordinate transformation}\label{sec:coord_trafo}
Our goal is to construct a coordinate system $y$ and an associated transformation $g$, that maps from $\xi$ to $y$, in which the posterior metric $\mathcal{M}$ takes the form of the identity matrix $\mathds{1}$. The motivation is that if $\mathcal{M}$ captures the geometric properties of the posterior, a coordinate system in which this metric becomes trivial should also be a coordinate system in which the posterior distribution takes a particularly simple form. For an illustrative example see figure \ref{fig:vis_trafo}.
To do so, we require the Fisher metric of the likelihood $\mathcal{M}_{d|\xi}$ to be the pullback of the Euclidean metric. Specifically we propose a function $x(\xi)$ such that

\begin{equation}\label{eq:lhmetdecomp}
\mathcal{M}_{d|\xi} \overset{!}{=} \left(\frac{\partial x}{\partial \xi}\right)^T \frac{\partial x}{\partial \xi} \ ,
\end{equation}
where $T$ denotes the adjoint of a matrix.
As outlined in Appendix \ref{ap:isolhs}, for many practically relevant likelihoods such a decomposition is possible by means of an inexpensive to evaluate function $x$.\footnote{Here with ``inexpensive'' we mean that applying the function $x(\xi)$ has a similar computational cost compared to applying the likelihood function $P(d|\xi)$ to a specific $\xi$.}
Given $x$, we can rewrite the posterior metric $\mathcal{M}$ as

\begin{equation}\label{eq:metric}
\mathcal{M} = \left(\frac{\partial x}{\partial \xi}\right)^T \frac{\partial x}{\partial \xi} + \mathds{1} \ .
\end{equation}
In order to relate this metric to Euclidean space, we aim to find the isometry $g$ that relates the Riemannian manifold associated with the metric $\mathcal{M}$ to Euclidean space. 
Specifically we seek to find an invertible function $g$ satisfying

\begin{equation}\label{eq:metriciso}
\mathcal{M}(\xi) = \left(\frac{\partial x}{\partial \xi}\right)^T \frac{\partial x}{\partial \xi} + \mathds{1} \overset{!}{=} \left(\frac{\partial g}{\partial \xi}\right)^T \frac{\partial g}{\partial \xi} \ .
\end{equation}
In general, i.E. for a general function $x(\xi)$, however, this decomposition does not exist globally. Nevertheless, there exists a transformation $g(\xi;\bar{\xi})$ based on an approximative Taylor series around an expansion point $\bar{\xi}$, that results in a metric $\tilde{\mathcal{M}}(\xi)$ such that

\begin{equation}\label{eq:approxmet}
\mathcal{M}(\xi) \approx \tilde{\mathcal{M}}(\xi) \equiv \left(\frac{\partial g(\xi; \bar{\xi})}{\partial \xi}\right)^T \frac{\partial g(\xi; \bar{\xi})}{\partial \xi} \ ,
\end{equation}
in the vicinity of $\bar{\xi}$. This transformation $g$ can be obtained up to an integration constant by Taylor expanding equation \eqref{eq:approxmet} around $\bar{\xi}$ and solving for the Taylor coefficients of $g$ in increasing order.
We express $g$ in terms of its Taylor series using the Einstein sum convention as

\begin{equation}
g(\xi; \bar{\xi})^i = \bar{g}^i + \bar{g}^i_{\ ,j} \left(\xi - \bar{\xi}\right)^j + \bar{g}^i_{\ ,jk} \left(\xi - \bar{\xi}\right)^j \left(\xi - \bar{\xi}\right)^k + ... \ ,
\end{equation}
where repeated indices get summed over, $a_{,i}$ denotes the partial derivative of $a$ w.r.t.\ the $i$th component of $\xi$, and $\bar{s}$ denotes a (tensor) field $s(\xi)$, evaluated at the expansion point $\bar{\xi}$.
We begin to expand equation \eqref{eq:approxmet} around $\bar{\xi}$ and obtain for the zeroth order

\begin{equation}\label{eq:zeroorder}
\bar{\mathcal{M}}_{ij} \equiv \bar{x}^\alpha_{\ ,i} \bar{x}^\alpha_{\ ,j} + \delta_{ij} \overset{!}{=} \bar{g}^\alpha_{\ ,i} \ \bar{g}^\alpha_{\ ,j} \ .
\end{equation}
Expanding equation \eqref{eq:approxmet} to first order yields

\begin{equation}\label{eq:firstorderexpand}
\bar{x}^\alpha_{\ ,ik} \bar{x}^\alpha_{\ ,j} + \bar{x}^\alpha_{\ ,i} \bar{x}^\alpha_{\ ,jk} \overset{!}{=} \bar{g}^\alpha_{\ ,ik} \bar{g}^\alpha_{\ ,j} + \bar{g}^\alpha_{\ ,i} \bar{g}^\alpha_{\ ,jk} \ ,
\end{equation}
and therefore

\begin{equation}\label{eq:firstorder}
\bar{g}^i_{\ ,jk} = \bar{\mathcal{M}}^{i \gamma} \bar{g}^\beta_{\ ,\gamma} \bar{x}^\alpha_{\ ,\beta} \bar{x}^\alpha_{\ ,jk} \ ,
\end{equation}
where $\bar{\mathcal{M}}^{ij} = \left(\bar{\mathcal{M}}^{-1}\right)_{ij}$ denotes the components of the inverse of $\bar{\mathcal{M}}$.

Thus, to first order in the metric (meaning to second order in the transformation) the expansion remains solvable for a general $x$. Proceeding with the second order, however, we get that

\begin{equation}\label{eq:taylorsecond}
\bar{x}^\alpha_{\ ,ikl} \bar{x}^\alpha_{\ ,j} + \bar{x}^\alpha_{\ ,ik} \bar{x}^\alpha_{\ ,jl} + \bar{x}^\alpha_{\ ,il} \bar{x}^\alpha_{\ ,jk} + \bar{x}^\alpha_{\ ,i} \bar{x}^\alpha_{\ ,jkl} \overset{!}{=} \bar{g}^\alpha_{\ ,ikl} \bar{g}^\alpha_{\ ,j} + \bar{g}^\alpha_{\ ,ik} \bar{g}^\alpha_{\ ,jl} + \bar{g}^\alpha_{\ ,il} \bar{g}^\alpha_{\ ,jk} + \bar{g}^\alpha_{\ ,i} \bar{g}^\alpha_{\ ,jkl} \ ,
\end{equation}
which does not exhibit a general solution for $\bar{g}^i_{\ ,jkl}$ in higher dimensions due to the fact that the third derivative has to be invariant under arbitrary permutation of the latter three indices $jkl$. However, in analogy to equation \eqref{eq:firstorder}, we may set

\begin{equation}\label{eq:secondorder}
\bar{g}^i_{\ ,jkl} = \bar{\mathcal{M}}^{i \gamma} \bar{g}^\beta_{\ ,\gamma} \bar{x}^\alpha_{\ ,\beta} \bar{x}^\alpha_{\ ,jkl} \ ,
\end{equation}
which cancels the first and the last term of equation \eqref{eq:taylorsecond}, and study the remaining error which takes the form

\begin{align}\label{eq:error_full}
	\epsilon_{ijkl} &= \bar{x}^\alpha_{\ ,ik} \bar{x}^\alpha_{\ ,jl} + \bar{x}^\alpha_{\ ,il} \bar{x}^\alpha_{\ ,jk} - \bar{g}^\alpha_{\ ,ik} \bar{g}^\alpha_{\ ,jl} - \bar{g}^\alpha_{\ ,il} \bar{g}^\alpha_{\ ,jk} \notag\\ &= \bar{x}^\alpha_{\ ,ik} \bar{x}^\alpha_{\ ,jl} + \bar{x}^\alpha_{\ ,il} \bar{x}^\alpha_{\ ,jk} - \bar{x}^\alpha_{\ ,ik} \bar{x}^\gamma_{\ ,\alpha} \bar{\mathcal{M}}^{\gamma \delta} \bar{x}^\delta_{\ ,\beta} \bar{x}^\beta_{\ ,jl} - \bar{x}^\alpha_{\ ,il} \bar{x}^\gamma_{\ ,\alpha} \bar{\mathcal{M}}^{\gamma \delta} \bar{x}^\delta_{\ ,\beta} \bar{x}^\beta_{\ ,jk} \notag\\ 
	&= \bar{x}^\alpha_{\ ,ik} \left(\delta_{\alpha \beta} - \bar{x}^\gamma_{\ ,\alpha} \bar{\mathcal{M}}^{\gamma \delta} \bar{x}^\delta_{\ ,\beta}\right) \bar{x}^\beta_{\ ,jl} + \bar{x}^\alpha_{\ ,il} \left(\delta_{\alpha \beta} - \bar{x}^\gamma_{\ ,\alpha} \bar{\mathcal{M}}^{\gamma \delta} \bar{x}^\delta_{\ ,\beta}\right) \bar{x}^\beta_{\ ,jk}\ .
\end{align}
Let $X^i_{\ j}\equiv \bar{x}^i_{\ ,j}$, the expression in the parentheses takes the form

\begin{equation}\label{eq:defining_M}
\mathds{1} - X \bar{\mathcal{M}}^{-1} X^T = \mathds{1} - X \left(\mathds{1} + X^T X\right)^{-1} X^T = \left(\mathds{1} + X X^T\right)^{-1} \equiv M\ ,
\end{equation}
and thus equation \eqref{eq:error_full} reduces to

\begin{equation}\label{eq:error}
\epsilon_{ijkl} = \bar{x}^\alpha_{\ ,ik} M_{\alpha \beta} \bar{x}^\beta_{\ ,jl} + \bar{x}^\alpha_{\ ,il} M_{\alpha \beta} \bar{x}^\beta_{\ ,jk} \ .
\end{equation}
The impact of this error contribution can be qualitatively studied using the spectrum $\lambda(M)$ of the matrix $M$. This spectrum may exhibit two extreme cases, a so-called likelihood dominated regime, where the spectrum $\lambda(X X^T) \gg 1$, and a prior dominated regime where $\lambda(X X^T) \ll 1$. In the likelihood dominated regime, we get that $\lambda(M) \ll 1$ and thus the contribution of the error is small, whereas in the prior dominated regime $\lambda(M) \approx 1$ which yields an $\mathcal{O}(1)$ error. However, in the prior dominated regime, the entire metric $\mathcal{M}$ is close to the identity as we are in the standard coordinate system of the prior $\xi$ and therefore higher order derivatives of $x$ are small. As a consequence, the error is of the order $\mathcal{O}(1)$ only in regimes where the third (and higher) order of the expansion is negligible compared to the first and second order. An exception occurs when the expansion point is close to a saddle point of $x$, i.E. in cases where the first derivative of $x$ becomes small (and therefore the metric is close to the identity), but higher order derivatives of $x$ may be large. For the moment, we proceed under the assumption that the change of $x$, as a function of $\xi$, is sufficiently monotonic throughout the expansion regime. We discuss the implications of violating this assumption in section \ref{sec:pathological}.

If we proceed to higher order expansions of equation \eqref{eq:approxmet}, we notice that a repetitive picture emerges: The leading order derivative tensor $\bar{g}^i_{\ ,j ...}$ that appears in the expansion may be set in direct analogy to equations \eqref{eq:firstorder} and \eqref{eq:secondorder} as

\begin{equation}
\bar{g}^i_{\ ,j ...} = \bar{\mathcal{M}}^{i \gamma} \bar{g}^\beta_{\ ,\gamma} \bar{x}^\alpha_{\ ,\beta} \bar{x}^\alpha_{\ ,j ...} \ ,
\end{equation}
where $...$ denotes the higher order derivatives. The remaining error contributions at each order take a similar form as in equation \eqref{eq:error}, where the matrix $M$ reappears in between all possible combinations of the remaining derivatives of $x$ that appear using the Leibniz rule. Note that for increasing order, the number of terms that contribute to the error also increases. Specifically for the $n$th order expansion of equation \eqref{eq:approxmet} we get $m = \sum_{k=1}^{n-1} \binom{n}{k}$ contributions to the error. Therefore, even if each individual contribution by means of $M$ is small, the expansion error eventually becomes large once high order expansions become relevant. Therefore the proposed approximation only remains locally valid around $\bar{\xi}$.

Nevertheless, we may proceed to combine the derivative tensors of $g$ determined above in order to get the Jacobian of the transformation $g$ as

\begin{align}
g^i_{\ ,j}(\xi) &\equiv \bar{g}^i_{\ ,j} + \bar{g}^i_{\ ,jk} \left(\xi - \bar{\xi}\right)^k + \frac{1}{2} \bar{g}^i_{\ ,jkl} \left(\xi - \bar{\xi}\right)^k \left(\xi - \bar{\xi}\right)^l + ... \notag\\ &= \bar{g}^i_{\ ,j} + \bar{\mathcal{M}}^{i \alpha} \bar{g}^\beta_{\ ,\alpha} \bar{x}^\gamma_{\ ,\beta} \left(\bar{x}^\gamma_{\ ,j k} (\xi-\bar{\xi})^k + \frac{1}{2} \bar{x}^\gamma_{\ ,j k l} (\xi-\bar{\xi})^k (\xi-\bar{\xi})^l + ...\right) \ ,
\end{align}
or equivalently

\begin{equation}\label{eq:jacexpand}
\bar{g}^\alpha_{\ ,i} g^\alpha_{\ ,j}(\xi) = \delta_{ij} + \bar{x}^\alpha_{\ ,i} \left(\bar{x}^\alpha_{\ ,j} + \bar{x}^\alpha_{\ ,j k} (\xi-\bar{\xi})^k + \frac{1}{2} \bar{x}^\alpha_{\ ,j k l} (\xi-\bar{\xi})^k (\xi-\bar{\xi})^l + ...\right) \ .
\end{equation}
From the zeroth order, equation \eqref{eq:zeroorder}, we get that $\bar{g}^i_{\ ,j} = \left(\sqrt{\bar{\mathcal{M}}}\right)^i_j$ up to a unitary transformation, and we can sum up the Taylor series in $x$ of equation \eqref{eq:jacexpand} to arrive at an index free representation of the Jacobian as

\begin{equation}
\frac{\partial g}{\partial \xi} = \sqrt{\bar{\mathcal{M}}}^{-1} \left(\mathds{1} + \left(\left.\frac{\partial x}{\partial \xi}\right|_{\bar{\xi}}\right)^T \frac{\partial x}{\partial \xi}\right) \ .
\end{equation}
Upon integration, this yields a transformation

\begin{equation}\label{eq:trafodist}
g(\xi) - g(\bar{\xi}) = \sqrt{\bar{\mathcal{M}}}^{-1} \left(\xi - \bar{\xi} + \left(\left.\frac{\partial x}{\partial \xi}\right|_{\bar{\xi}}\right)^T \left(x(\xi) - x(\bar{\xi})\right)\right) \ .
\end{equation}
The resulting transformation takes an intuitive form: The approximation to the distance between a point $g(\xi)$ and the transformed expansion point $g(\bar{\xi})$ consists of the distance w.r.t.\ the prior measure $\left(\xi - \bar{\xi}\right)$ and the distance w.r.t.\ the likelihood measure $\left(x(\xi) - x(\bar{\xi})\right)$, back-projected into the prior domain using the local transformation at $\bar{\xi}$. Finally, the metric at $\bar{\xi}$ is used as a measure for the local curvature. Equation \eqref{eq:trafodist} is only defined up to an integration constant, and therefore, without loss of generality, we may set $g(\bar{\xi}) = 0$ to obtain the final approximative coordinate transformation as

\begin{equation}\label{eq:trafo}
g(\xi; \bar{\xi}) = \sqrt{\bar{\mathcal{M}}}^{-1} \left(\xi - \bar{\xi} + \left(\left.\frac{\partial x}{\partial \xi}\right|_{\bar{\xi}}\right)^T \left(x(\xi) - x(\bar{\xi})\right)\right) \equiv \sqrt{\bar{\mathcal{M}}}^{-1} \tilde{g}(\xi;\bar{\xi}) \ .
\end{equation}

\subsection{Basic properties}\label{sec:properties_error}
In order to study a few basic properties of this transformation, for simplicity, we first consider a posterior distribution with a metric that allows for an exact isometry $g_{\mathrm{iso}}$ to Euclidean space. Specifically let $g_{\mathrm{iso}}$ be a coordinate transformation satisfying equation \eqref{eq:metriciso}. The posterior distribution in coordinates $y = g_{\mathrm{iso}}(\xi)$ is given via the push-forward of $P(\xi|d)$ through $g_{\mathrm{iso}}$ as

\begin{equation}
P(y|d) \propto \left(g_{\mathrm{iso}} \star P(\xi|d)\right)(y) = \left.\left(P(\xi|d) \ \left|\left|\frac{\partial g_{\mathrm{iso}}}{\partial \xi}\right|\right|^{-1} \right)\right|_{\xi = g_{\mathrm{iso}}^{-1}(y)} = \left.\frac{P(\xi|d)}{\sqrt{\left|\mathcal{M}(\xi)\right|}}\right|_{\xi = g_{\mathrm{iso}}^{-1}(y)} \ ,
\end{equation}
and the information Hamiltonian takes the form

\begin{equation}
\mathcal{H}(y|d) = \left.\left(\mathcal{H}(\xi|d) + \frac{1}{2} \log\left(\left|\mathcal{M}(\xi)\right|\right)\right)\right|_{\xi = g_{iso}^{-1}(y)} + \mathcal{H}_0 \equiv \tilde{\mathcal{H}}(\xi = g_{iso}^{-1}(y)) + \mathcal{H}_0 \ ,
\end{equation}
where $\mathcal{H}_0$ denotes $y$ independent contributions. We may study the curvature of the posterior in coordinates $y$ given as:

\begin{align}
\mathcal{C}(y) = \frac{\partial \xi}{\partial y} \left(\frac{\partial^2 \tilde{\mathcal{H}}(\xi)}{\partial \xi \partial \xi'} \right) \left(\frac{\partial \xi'}{\partial y'}\right)^T &+ \frac{\partial \tilde{\mathcal{H}}(\xi)}{\partial \xi} \frac{\partial^2 \xi}{\partial y \partial y'} \quad \text{with} \\ \quad \xi = g_{\mathrm{iso}}^{-1}\left(y\right) \quad &\text{and} \quad \xi' = g_{\mathrm{iso}}^{-1}\left(y'\right) \ , \notag
\end{align}
which we can use to construct a metric $\mathcal{M}(y)$ in analogy to equation \eqref{eq:postmetric} by taking the expectation value of the curvature w.r.t.\ the likelihood. This yields

\begin{align}\label{eq:metrictrafo}
	\mathcal{M}(y) &= \frac{\partial \xi}{\partial y} \mathcal{M}(\xi) \left(\frac{\partial \xi'}{\partial y'}\right)^T + \frac{1}{2} \frac{\partial \xi}{\partial y} \left(\frac{\partial^2 \log\left(\left|\mathcal{M}(\xi)\right|\right)}{\partial \xi \partial \xi'}\right) \left(\frac{\partial \xi'}{\partial y'}\right)^T + \left<\frac{\partial \tilde{\mathcal{H}}(\xi)}{\partial \xi}\right>_{P(d|\xi)} \frac{\partial^2 \xi}{\partial y \partial y'} \notag\\ &\equiv \frac{\partial \xi}{\partial y} \mathcal{M}(\xi) \left(\frac{\partial \xi'}{\partial y'}\right)^T + \mathcal{R}(y) = \mathds{1} + \mathcal{R}(y) \ .
\end{align}
The first terms yields the identity, as it is the defining property of $g_{\mathrm{iso}}$. Furthermore, in case we are able to say that $\mathcal{R}(y)$ is small compared to the identity, we notice that the quantity $\mathcal{M}(\xi)$ (equation \eqref{eq:postmetric}), that we referred to as the posterior metric, approximately transforms like a proper metric under $g_{\mathrm{iso}}$. In this case we find that the isometry $g_{\mathrm{iso}}$ between the Riemannian manifold associated with $\mathcal{M}(\xi)$ and the Euclidean space is also a transformation that removes the complex geometry of the posterior. To further study $\mathcal{R}(y)$, we consider its two contributions separately, where for the first part, the log-determinant (or logarithmic volume), we get that it becomes small compared to the identity if

\begin{equation}\label{eq:consistency_one}
\mathcal{M}(\xi) \overset{!}{\gg} \frac{1}{2} \left(\frac{\partial^2 \log\left(\left|\mathcal{M}(\xi)\right|\right)}{\partial \xi \partial \xi'}\right) \ .
\end{equation}
Therefore, the curvature of the log-determinant of the metric has to be much smaller then the metric itself. To study the second term of $\mathcal{R}(y)$, we may again split the discussion into a prior and a likelihood dominated regime, depending on the $\xi$ at which we evaluate the expression. In a prior dominated regime we get that

\begin{equation}
\frac{\partial^2 \xi}{\partial y \partial y'} \approx 0 \ ,
\end{equation}
as the metric is close to the identity in this regime (and therefore $\xi \approx y$). In a likelihood dominated regime we get that $\tilde{\mathcal{H}} \approx \mathcal{H}(d|\xi)$ and therefore

\begin{equation}
\left<\frac{\partial \tilde{\mathcal{H}}(\xi)}{\partial \xi}\right>_{P(d|\xi)} \approx \left<\frac{\partial \mathcal{H}(d|\xi)}{\partial \xi}\right>_{P(d|\xi)} = - \left<\frac{1}{P(d|\xi)} \frac{\partial P(d|\xi)}{\partial \xi}\right>_{P(d|\xi)} = 0 \ .
\end{equation}
So at least in a prior dominated regime, as well as a likelihood dominated regime, the posterior Hamiltonian transforms in an analogous way as the manifold, under the transformation $g_{\mathrm{iso}}$, if equation \eqref{eq:consistency_one} also holds true.

For a practical application, however, in all but the simplest cases the isometry $g_{\mathrm{iso}}$ is not accessible, or might not even exist. Therefore, in general, we have to use the approximation $g(\xi; \bar{\xi})$, as defined in equation \eqref{eq:trafo}, instead. We may express the transformation of the metric using $g(\xi; \bar{\xi})$, and find that

\begin{align}
\mathcal{M}(y) &= \sqrt{\bar{\mathcal{M}}} \left(\mathds{1} +  \left(\frac{\partial x}{\partial \xi}\right)^T \left.\frac{\partial x}{\partial \xi}\right|_{\bar{\xi}}\right)^{-1} \left(\mathds{1} + \left(\frac{\partial x}{\partial \xi}\right)^T \frac{\partial x'}{\partial \xi'} \right) \left(\mathds{1} + \left(\left.\frac{\partial x}{\partial \xi}\right|_{\bar{\xi}}\right)^T \frac{\partial x'}{\partial \xi'}\right)^{-1} \sqrt{\bar{\mathcal{M}}} \notag\\&+ \tilde{\mathcal{R}}(y) \ ,
\end{align}
now with $\xi = g^{-1}\left(y; \bar{\xi}\right)$ and analogous for $\xi'$. $\tilde{\mathcal{R}}$ is defined by replacing $g_{\mathrm{iso}}$ with $g$ for the entire expression of $\mathcal{R}$. We notice that this transformation does not yield the identity, except when evaluated at the expansion point $\xi = \bar{\xi}$. Therefore, in addition to the error $\tilde{\mathcal{R}}$ there is a deviation from the identity related to the expansion error as one moves away from $\bar{\xi}$.

At this point we would like to emphasize that the posterior Hamiltonian $\mathcal{H}$ and the Riemannian manifold constructed from $\mathcal{M}$ are only loosely connected due to the errors described by $\tilde{\mathcal{R}}$ and the additional expansion error. They are arguably small in many cases and in the vicinity of $\bar{\xi}$, but we do not want to claim that this correspondence is valid in general (see section \ref{sec:pathological}). Nevertheless, we find that in many cases this correspondence works well in practice. Some illustrative examples are given in section \ref{sec:simple_demos}.

\section{Posterior approximation}
Utilizing the derived coordinate transformation for posterior approximation is mainly based on the idea that in the transformed coordinate system, the posterior takes a simpler form. In particular we aim to remove parts (if not most) of the complex geometry of the posterior, such that a simple probability distribution, e.g.\ a Gaussian distribution, yields a good approximation.

\subsection{Direct approximation}\label{sec:direct approx}
Assuming that all the errors discussed in the previous section are small enough, we may attempt to directly approximate the posterior distribution via a unit Gaussian in the coordinates $y$ as in this case the transformed metric $\mathcal{M}(y)$ is close to the identity. As the coordinate transformation $g$, defined via equation \eqref{eq:trafo}, is only known up to an integration constant by construction, the posterior approximation is achieved by a shifted unit Gaussian in $y$. This shift needs to be determined, which we can do by maximizing the transformed posterior distribution

\begin{equation}
P(y|d) \propto \left.\left(P(\xi, d) \left|\left|\frac{\partial g}{\partial \xi}\right|\right|^{-1}\right)\right|_{\xi = g^{-1}(y; \bar{\xi})} \ ,
\end{equation}
w.r.t.\ $y$. Here $g^{-1}(y; \bar{\xi})$ denotes the inverse of $g(\xi; \bar{\xi})$ w.r.t.\ its first argument. Equivalently we can minimize the information Hamiltonian $\mathcal{H}(y|d)$, defined as

\begin{equation}\label{eq:hamiltony}
\mathcal{H}(y|d) \equiv - \log\left(P(y|d)\right) = \left.\left(\mathcal{H}(\xi, d) + \frac{1}{2} \log\left(\left|\tilde{\mathcal{M}}\right|\right)\right)\right|_{\xi = g^{-1}(y; \bar{\xi})} \equiv \tilde{\mathcal{H}}(\xi = g^{-1}(y; \bar{\xi}))\ .
\end{equation}
Minimizing $\mathcal{H}(y|d)$ yields the maximum a posterior solution $y^*$ which, in case the posterior is close to a unit Gaussian in the coordinates $y$, can be identified with the shift in $y$.
As $g$ is an invertible function, we may instead minimize $\tilde{\mathcal{H}}$ w.r.t.\ $\xi$ and apply $g$ to the result in order to obtain $y^*$. Specifically

\begin{equation}\label{eq:opity}
y^* \equiv \underset{y}{\mathrm{argmin}} \left(\mathcal{H}(y|d)\right) =  g\left(\underset{\xi}{\mathrm{argmin}} \left(\tilde{\mathcal{H}}(\xi)\right)\right) \ .
\end{equation}
Therefore we can circumvent the inversion of $g$ at any point during optimization.
Now suppose that we use any gradient based optimization scheme to minimize for $\xi$, starting from some initial position $\xi^0$. If we set the expansion point $\bar{\xi}$, used to construct $g$, to be equal to $\xi^0$, we notice that 

\begin{align}\label{eq:metric_consistency}
	\tilde{\mathcal{M}}(\bar{\xi}) &= \mathcal{M}(\bar{\xi}) \\ 
	\left.\frac{\partial \tilde{\mathcal{M}}}{\partial \xi}\right|_{\xi = \bar{\xi}} &= \left.\frac{\partial \mathcal{M}}{\partial \xi}\right|_{\xi = \bar{\xi}} \ ,
\end{align}
as the expansion of the metric is valid to first order by construction. Therefore if we set the expansion point $\bar{\xi}$ to the current estimate of $\xi$ after every step, we can replace the approximated metric $\tilde{\mathcal{M}}$ with the true metric $\mathcal{M}$ and arrive at an optimization objective of the form

\begin{equation}\label{eq:optexpand_direct}
\bar{\xi} = \underset{\xi}{\mathrm{argmin}} \left(\mathcal{H}(\xi, d) + \frac{1}{2} \log\left(\left|\mathcal{M}(\xi)\right|\right)\right) \ .
\end{equation}
Note that $g(\bar{\xi}; \bar{\xi}) = 0$ by construction, and therefore $y^* = 0$, as there is a degeneracy between a shift in $y$ and a change of the expansion point $\bar{\xi}$.
Once the optimal expansion point $\bar{\xi}$ is found, we directly retrieve a generative process to sample from our approximation to the posterior distribution. Specifically

\begin{align}
	P(y|d) &\approx \mathcal{N}(y; 0, \mathds{1}) \\
	\xi &= g^{-1}(y; \bar{\xi}) \ ,
\end{align}
where $g^{-1}$ is only implicitly defined using equation \eqref{eq:trafo} and therefore its inverse application has to be approximated numerically in general.

\subsubsection{Numerical approximation to sampling}\label{sec:sampling}
Recall that

\begin{equation}\label{eq:relationy}
y = g(\xi; \bar{\xi}) = \sqrt{\bar{\mathcal{M}}}^{-1} \tilde{g}(\xi; \bar{\xi}) \ .
\end{equation}
To generate a posterior sample for $\xi$ we have to draw a random realization for $y$ from a unit Gaussian, and then solve equation \eqref{eq:relationy} for $\xi$. To avoid the matrix square root of $\bar{\mathcal{M}}$, we may instead define

\begin{equation}\label{eq:relationz}
z \equiv \sqrt{\bar{\mathcal{M}}} \ y = \tilde{g}(\xi; \bar{\xi}) \quad \text{with} \quad P(z) = \mathcal{N}(z; 0, \bar{\mathcal{M}}) \ .
\end{equation}
Sampling from $P(z)$ is much more convenient then constructing the matrix square root, since

\begin{equation}
\bar{\mathcal{M}} = \mathds{1} + \left.\left(\left(\frac{\partial x}{\partial \xi}\right)^T \frac{\partial x}{\partial \xi}\right)\right|_{\bar{\xi}} \ ,
\end{equation}
and therefore a random realization may be generated using

\begin{equation}\label{eq:zsam}
z = \eta_1 + \left(\left.\frac{\partial x}{\partial \xi}\right|_{\bar{\xi}}\right)^T \eta_2 \quad \text{with} \quad \eta_i \sim \mathcal{N}(\eta_i; 0, \mathds{1}) \ , \ i \in \left\lbrace 1,2 \right\rbrace \ .
\end{equation}
Finally, a posterior sample $\xi$ is retrieved by inversion of equation \eqref{eq:relationz}. We numerically approximate the inversion by minimizing the squared difference between $z$ and $\tilde{g}(\xi)$. Specifically,

\begin{equation}\label{eq:miniinv}
\xi = \underset{\xi}{\mathrm{argmin}} \left(\frac{1}{2} \left(z - \tilde{g}(\xi)\right)^T \left(z - \tilde{g}(\xi)\right)\right) \ .
\end{equation}
Note that if $g$ is invertible then also $\tilde{g}$ is invertible as $\bar{\mathcal{M}}$ is a symmetric positive definite matrix. Therefore the quadratic form of equation \eqref{eq:miniinv} has a unique global optimum at zero which corresponds to the inverse of equation \eqref{eq:relationz}.

In practice, this optimum is typically only reached approximately. For an efficient numerical approximation, throughout this work, we employ a second order quasi-Newton method, named NewtonCG \cite{NoceWrig06}, as implemented in the {\tt NIFTy} framework. Within the NewtonCG algorithm, we utilize the metric $\tilde{\mathcal{M}}(\xi)$ as a positive-definite approximation to the curvature of the quadratic form in equation \eqref{eq:miniinv}. Furthermore, its inverse application, required for the second order optimization step of NewtonCG, is approximated with the conjugate gradient (CG) \cite{hestenes1952methods} method, which requires the metric to be only implicitly available via matrix-vector products.
In addition, in practice we find that the initial position $\xi^0$ of the minimization procedure can be set to be equal to the prior realization $\eta_1$ used to construct $z$ (equation \eqref{eq:zsam}) in order to improve convergence as $\xi = \eta_1$ is the solution of equation \eqref{eq:relationz} for all degrees of freedom unconstrained by the likelihood. Alternatively, for weakly non-linear problems, initializing $\xi^0$ as the solution of the linearized problem

\begin{equation}\label{eq:linsolve}
	z = \left.\tilde{g}(\xi; \bar{\xi})\right|_{\xi = \bar{\xi}} + \left.\frac{ \partial \tilde{g}}{\partial \xi}\right|_{\xi = \bar{\xi}} \left(\xi - \bar{\xi}\right) = \left.\left(\mathds{1} + \left(\frac{\partial x}{\partial \xi}\right)^T \frac{\partial x}{\partial \xi}\right)\right|_{\xi=\bar{\xi}} \left(\xi - \bar{\xi}\right) \ ,
\end{equation}
can significantly improve the convergence. The full realization of the sampling procedure is summarized in algorithm \ref{alg:sampling}.

\begin{algorithm}
	\caption{Approximate posterior samples using inverse transformation}\label{alg:sampling}
	\SetKwFunction{sample}{drawSample}
	\SetKwFunction{en}{Energy}
	\SetKwProg{func}{Function}{:}{}
	\SetKwFunction{newton}{NewtonCG}
	\SetKwFunction{sol}{Solve}
	\func{\sample{Location $\bar{\xi}$, Transformation $x(\xi)$, Jacobian $\frac{\partial x}{\partial \xi}$}}{
		$A \leftarrow \left.\frac{\partial x}{\partial \xi}\right|_{\xi = \bar{\xi}}$ \\
		$\eta_1 \sim \mathcal{N}(\eta_1; 0, \mathds{1})$ \\
		$\eta_2 \sim \mathcal{N}(\eta_2; 0, \mathds{1})$ \\
		$z \leftarrow \eta_1 + A^T \eta_2$ \\
		$\xi^0 \leftarrow \eta_1$ or $\xi^0 \leftarrow$ \sol{$z = \left(\mathds{1}+A^T A\right) \left(\xi^0 - \bar{\xi}\right)$} for $\xi^0$ $\quad$ (see Eq.\ \eqref{eq:linsolve})\\
		\func{\en{$\xi$}}{
			$\tilde{g} \leftarrow \xi - \bar{\xi} + A^T\left(x(\xi) - x(\bar{\xi})\right)$ \\
			\KwRet{$\frac{1}{2} \left(z - \tilde{g}\right)^T \left(z - \tilde{g}\right)$}
		}
		$\xi^* \leftarrow$ \newton{\en, $\xi^0$} \\
		\KwRet{$\xi^*$}
	}
\end{algorithm}

\subsubsection{Properties}\label{sec:simple_demos}
We may qualitatively study some basic properties of the coordinate transformation and the associated approximation using illustrative one and two dimensional examples. To this end, consider a one dimensional log-normal prior model with zero mean and standard deviation $\sigma_p$ of the form

\begin{equation}
	s(\xi) = e^{\sigma_p \xi} \quad \text{with} \quad P(\xi) = \mathcal{N}(\xi; 0, 1) \ ,
\end{equation}
from which we obtain a measurement $d$ subject to independent, additive Gaussian noise with standard deviation $\sigma_n$ such that the likelihood takes the form

\begin{equation}
	P(d|\xi) = \mathcal{N}(d; s(\xi), \sigma_n^2) \ .
\end{equation}
The posterior distribution is given as

\begin{equation}\label{eq:post1d}
	P(\xi|d) \propto P(d|\xi) \ P(\xi) = \mathcal{N}(d; s(\xi), \sigma_n^2) \ \mathcal{N}(\xi; 0, 1) \ ,
\end{equation}
and its metric takes the form

\begin{equation}\label{eq:met1d}
	\mathcal{M}(\xi) = \left(\frac{1}{\sigma_n} \frac{\partial s(\xi)}{\partial \xi}\right)^2 + 1 = \left(\frac{\sigma_p}{\sigma_n}\right)^2 e^{2 \sigma_p \xi} + 1 \ .
\end{equation}

\begin{figure*}[htp]
	\centering
	\includegraphics[scale=1., angle=0]{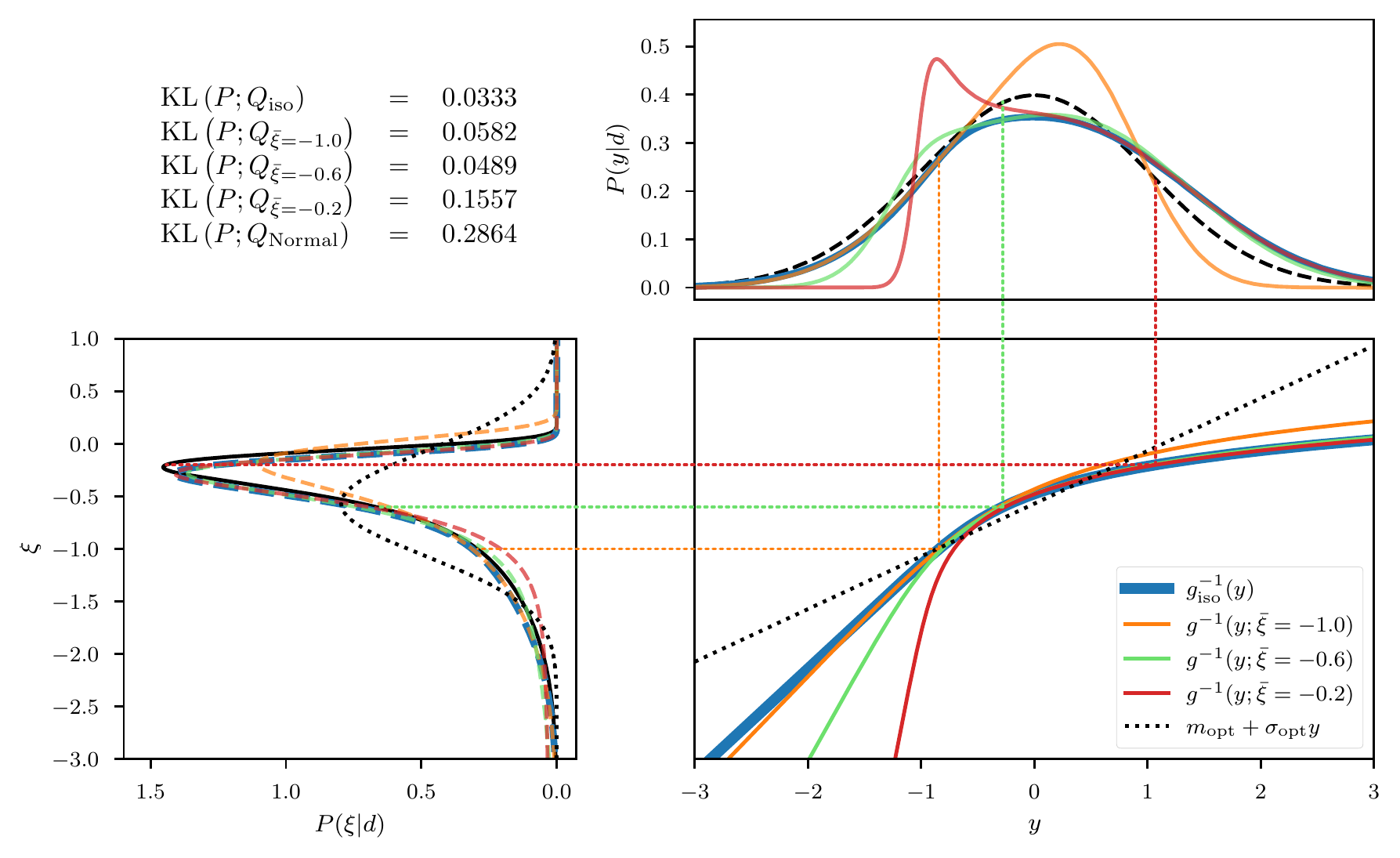}
	\centering
	\caption{Illustration of the coordinate transformation for the one-dimensional log-normal model (equation \eqref{eq:post1d}). The true posterior $P(\xi|d)$, displayed as the black solid line in the left panel, is transformed into the coordinate system $y$ using the optimal transformation $g_{\mathrm{iso}}$ (blue), as well as three approximations $g$ thereof with expansion points $\bar{\xi} \in \left\lbrace-1,-0.6,-0.2 \right\rbrace$ (orange, green, red). The resulting distributions $P(y|d)$ are displayed in the top panel of the figure as solid lines, color coded according the used transformation $g$ (or $g_{\mathrm{iso}}$ in case of blue). The black, dashed line in the top panel displays a standard distribution in $y$. The location of the expansion point $\bar{\xi}$, and its associated point in $y$, is highlighted via the color coded, dotted lines. Finally, the direct approximations to the posterior associated with the transformations, meaning the push-forwards of the standard distribution in $y$ using the inverse of the various transformations $g^{-1}$, are displayed in the left panel as dashed lines, color coded according to their used transformation. As a comparison, the ``optimal linear approximation'' (black dotted line in the central panel), which corresponds to the optimal approximation of the posterior with a normal distribution in $\xi$ (black dotted line in left panel), is displayed as a comparison. To numerically quantify the information distance between the true distribution $P$ and its approximations $Q_{\bullet}$, the Kullback-Leibler (KL) divergences between $P$ and $Q_{\bullet}$ are displayed in the top left of the image. The numerical values of the KL are given in nats (meaning the KL is evaluated in the basis of the natural logarithm).}\label{fig:lognormal_1d}
\end{figure*}

In this one dimensional example we can construct the exact transformation $g_{\mathrm{iso}}$ that maps from $\xi$ to the transformed coordinates $y$, by integrating the square root of equation \eqref{eq:met1d} over $\xi$. The resulting transformation can be seen in the central panel of figure \ref{fig:lognormal_1d}, for an example with $\sigma_p = 3$ and $\sigma_n = 0.3$ and measured data $d = 0.5$. In addition, we depict the approximated transformation $g(\xi;\bar{\xi})$ for multiple expansion points $\bar{\xi} \in \left\lbrace -1, -0.6, -0.2 \right\rbrace$. We see that the function approximation quality depends on the choice of the expansion point $\bar{\xi}$ as the approximation error is smallest in the vicinity of $\bar{\xi}$. In order to transform the posterior distribution $P$ (Eq.\ \eqref{eq:post1d}) into the new coordinated system, not all parts of the transformation are equally relevant and therefore different expansion points result in more/less complex transformed distributions (see top panel of figure \ref{fig:lognormal_1d}). Finally, if we use a standard distribution in the transformed coordinates $y$ and transform it back using the inverse transformations $g^{-1}(y;\bar{\xi})$, we find that the approximation quality of the resulting distributions $Q_{\bar{\xi}}$ depends on $\bar{\xi}$. The distributions are illustrated in the left panel of figure \ref{fig:lognormal_1d} together with the Kullback-Leibler divergence $\mathrm{KL}$ between the true posterior distribution $P$ and the approximations $Q_{\bar{\xi}}$. We also illustrate the ``geometrically optimal'' approximation using a standard distribution in $y$ and the optimal transformation $g_{\mathrm{iso}}$ and find that while the approximation error becomes minimal in this case, it remains non-zero. Considering the discussion in section \ref{sec:properties_error}, this result is to be expected due to the error contribution from the change in volume associated with the transformation $g$. As a comparison we also depict the optimal linear approximation of $P$, that is a normal distribution in the coordinates $\xi$ with optimally chosen mean and standard deviation. We see that even the worst expansion point $\bar{\xi} = - 0.2$, that is far away from the optimum, still yields a better approximation of the posterior.

\begin{figure*}[htp]
	\centering
	\includegraphics[scale=1., angle=0]{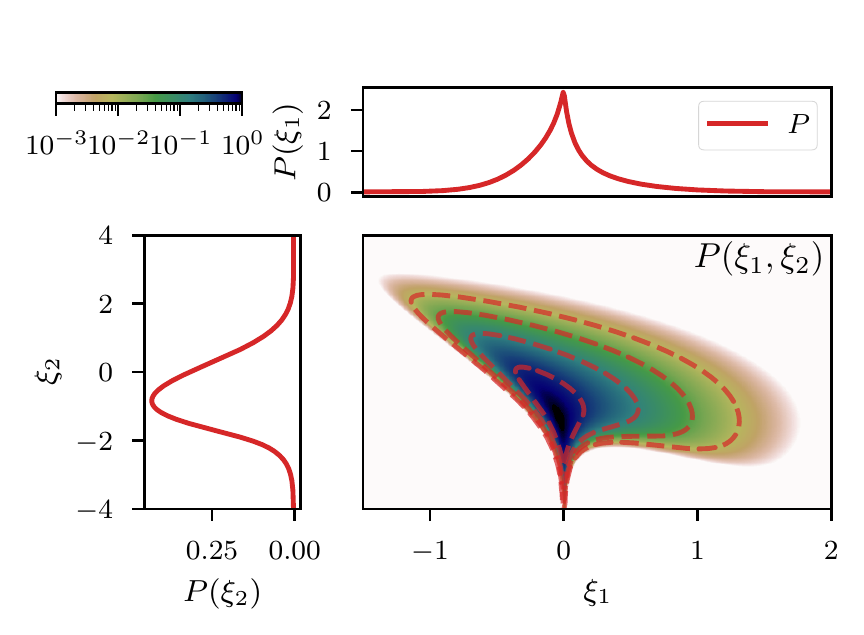}
	\includegraphics[scale=1., angle=0]{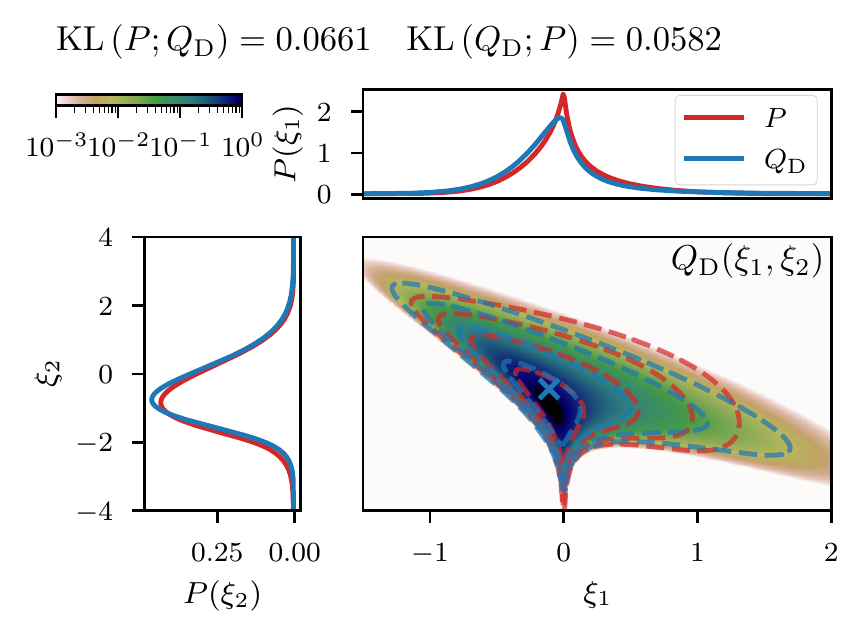}
	\centering
	\caption{Left: posterior distribution $P$ in the standard coordinates $\xi_{\nicefrac{1}{2}}$ for the inference of the mean and variance of a normal distribution (equations \eqref{eq:2dlh} and \eqref{eq:2dpr}). The central panel shows the two dimensional density and the red dashed lines are logarithmically spaced contours. The top and left sub-panels display the marginal posterior distributions for $\xi_1$ and $\xi_2$, respectively.
	Right: Approximation $Q_{\mathrm{D}}$ to the posterior distribution using the direct method (section \ref{sec:direct approx}). As a comparison, the contours (red dashed) and the marginal distributions (red solid) of the true posterior distribution $P$ are displayed in addition to the approximation. The blue cross in the central panel denotes the location of the expansion point used to construct $Q_{\mathrm{D}}$.
	Above the panel we display the optimal ($\mathrm{KL}(P;Q_\mathrm{D})$) and variational ($\mathrm{KL}(Q_\mathrm{D};P)$) Kullback-Leibler divergences between $P$ and $Q_{\mathrm{D}}$.}\label{fig:vis_2d_direct}
\end{figure*}

As a second example we consider the task of inferring the mean $m$ and variance $v$ of a single, real valued Gaussian random variable $d$. In terms of $s=\left(m, v\right)$, the likelihood takes the form

\begin{equation}\label{eq:2dlh}
	P(d|s) = \mathcal{N}(d; m, v) \ .
\end{equation}
Furthermore we assume a prior model for $s$ by means of a generative model of the form

\begin{equation}\label{eq:2dpr}
	m = \xi_1 \quad \text{and} \quad v = \exp\left[3 (\xi_2 + 2 \xi_1)\right] \ ,
\end{equation}
where $\xi_1$ and $\xi_2$ follow standard distributions a priori. This artificial model results in a linear prior correlation between the mean and the log-variance and thus introduces a non-linear coupling between $m$ and $v$. The resulting two dimensional posterior distribution $P(\xi_1,\xi_2)$ can be seen in the left panel of figure \ref{fig:vis_2d_direct}, together with the two marginals $P(\xi_1)$ and $P(\xi_2)$ for a given measurement $d=0$. We approximate this posterior distribution following the direct approach described in section \ref{sec:direct approx}, where the expansion point $\bar{\xi}$ is obtained from minimizing the sum of the posterior Hamiltonian and the log-determinant of the metric (see Eq.\ \eqref{eq:optexpand_direct}). The resulting approximative distribution $Q_{\mathrm{D}}$ is shown in the right panel of figure \ref{fig:vis_2d_direct}, where the location of $\bar{\xi}$ is indicated as a blue cross. In comparison to the true distribution, we see that both, the joint distribution as well as the marginals are in a good agreement qualitatively, which is also supported quantitatively by a small difference of the KL between $P$ and $Q_{\mathrm{D}}$ (see figure \ref{fig:vis_2d_direct}). The difference between $P$ and $Q_{\mathrm{D}}$ appears to increase in regions further away from the expansion point, which is to be expected due to the local nature of the approximation. However, non-linear features such as the sharp peak at the ``bottom'' of $P$ (figure \ref{fig:vis_2d_direct}), are also present in $Q_{\mathrm{D}}$, although slightly less prominent. This demonstrates that relevant non-linear structure can, to some degree, be captured by the coordinate transformation $g$ derived from the metric $\mathcal{M}$ of the posterior.

Although these low-dimensional, illustrative examples appear promising, there remains one central issue left to be addressed before the approach can be applied to high-dimensional problems. In particular, the direct approach possesses a substantial additional computational burden compared to e.g.\ a maximum a posteriori (MAP) estimate in $\xi$ which is obtained by minimizing the posterior Hamiltonian $\mathcal{H}$. For the direct approach, the optimization objective Eq.\ \eqref{eq:optexpand_direct} consists not only of $\mathcal{H}$, but also of the log-determinant of the metric $\mathcal{M}$. In all but the simplest examples this term cannot be computed directly but has to be approximated numerically as in high dimensions an explicit representation of the matrix becomes infeasible and $\mathcal{M}$ is only implicitly accessible through matrix vector products (MVPs). There are a variety of stochastic log-determinant (more specifically trace-log) estimators based on combining Hutchinsons' trace-estimation \cite{hutchinson1989stochastic} with approximations to the matrix logarithm using e.g.\ Chebychev polynomials \cite{han2015large}, Krylov subspace methods \cite{ubaru2017fast}, or moment constrained estimation based on Maximum Entropy \cite{fitzsimons2017entropic}. While all these methods provide a significant improvement in performance compared to directly computing the determinant, they nevertheless typically require many MVPs in order to yield an accurate estimate. For large and complex problems, evaluating an MVP of $\mathcal{M}$ is dominated by applying the Jacobian of $x$, more precisely of the generative process $s'(\xi)$, and its adjoint to a vector. Similarly, evaluating the gradient of $\mathcal{H}$ is also dominated by an MVP that invokes applying the adjoint Jacobian of $s'(\xi)$. Therefore the computational overhead compared to a MAP estimate in $\xi$ is, roughly, multiplicative in the number of MVPs. For large, non-linear problems, this quickly becomes infeasible as nonlinear optimization typically requires many steps to reach a sensible approximation to the optimum. 

Nevertheless there remain some important exceptions, in which a fast and scalable algorithm emerges. In particular recall that

\begin{equation}
	\log\left(\left|\mathcal{M}\right|\right) = \mathrm{tr}\left(\log\left(\mathds{1} + \left(\frac{\partial x}{\partial \xi}\right)^T \frac{\partial x}{\partial \xi}\right)\right) = \mathrm{tr}\left(\log\left(\mathds{1} + \frac{\partial x}{\partial \xi} \left(\frac{\partial x}{\partial \xi}\right)^T\right)\right) \ ,
\end{equation}
where the last equality arises from applying the matrix determinant lemma. Therefore in cases where the dimensionality of the so-called data-space (i.E.\ the target space of $x$) is much smaller then the dimensionality of the signal space (the domain of $\xi$), the latter representation of the metric is of much smaller dimension. Thus in cases where either the signal- or the data-space is small, or in weakly non-linear cases (i.E.\ if $\mathcal{M}$ is close to the identity), the log-determinant may be approximated efficiently enough to give rise to a fast and scalable algorithm. For the (arguably most interesting) class of problems where neither of these assumptions is valid, however, the direct approach to obtain the optimal expansion point becomes too expensive for practical purposes as none of the log-determinant estimators scale linearly with the size of the problem in general.

\subsection{Geometric Variational inference (geoVI)}
As we shall see, it is possible to circumvent the need to compute the log-determinant of the metric at any point, if we employ a specific variant of a variational approximation to obtain the optimal expansion point. To this end, we start with a variational approximation to the posterior $P$, assuming that the approximative distribution $\tilde{Q}$ is given as the unit Gaussian in $y$ transformed via $g$. To this end let

\begin{equation}
\tilde{Q}(\xi|\bar{\xi}) = \mathcal{N}\left(g(\xi;\bar{\xi}); 0, \mathds{1}\right) \left|\left|\frac{\partial g(\xi ; \bar{\xi})}{\partial \xi}\right|\right| \ ,
\end{equation}
denote the approximation to the posterior conditional to the expansion point $\bar{\xi}$. The variationally optimal $\bar{\xi}$ can be found by optimization of the forward Kullback-Leibler divergence between $\tilde{Q}$ and $P$, as given via

\begin{align}
	\mathrm{KL}\left(\tilde{Q}|P\right) &\equiv \int \log\left(\frac{\tilde{Q}(\xi| \bar{\xi})}{P(\xi|d)}\right) \tilde{Q}(\xi| \bar{\xi}) \mathrm{d}\xi \notag\\
	&= \left<\mathcal{H}(\xi|d)\right>_{\tilde{Q}(\xi|\bar{\xi})} - \left< \mathcal{H}_{\tilde{Q}}(\xi|\bar{\xi}) \right>_{\tilde{Q}(\xi|\bar{\xi})} \notag\\ &= \left<\mathcal{H}(\xi| d)\right>_{\tilde{Q}(\xi|\bar{\xi})} + \frac{1}{2} \left<\log\left(\left|\tilde{\mathcal{M}}(\xi)\right|\right)\right>_{\tilde{Q}(\xi|\bar{\xi})} + \mathrm{KL}_0 \ ,
\end{align}
where $\mathrm{KL}_0$ denotes contributions independent of $\bar{\xi}$, and $\mathcal{H}(\xi|d)$ and $\mathcal{H}_{\tilde{Q}}$ denote the Hamiltonians of the posterior and the approximation, respectively. We notice that in this form, a minimization of the KL w.r.t.\ $\bar{\xi}$ does not circumvent a computation of the log-determinant of the metric. Within the KL, this term arises from the entropy of the approximation $\tilde{Q}$, and can be understood as a measure of the volume associated with the distribution. In order to avoid this term, our idea is to propose an alternative family of distributions $Q_m(\xi|\bar{\xi})$, defined as a shifted version of $\tilde{Q}$. Specifically we let $\xi \rightarrow m + \xi - \bar{\xi}$ such that the distribution may be written as
\begin{equation}
	Q_m(\xi|\bar{\xi}) = \left.\tilde{Q}(\xi|\bar{\xi})\right|_{\xi = \xi + \bar{\xi} - m} \equiv \left.Q(r|\bar{\xi})\right|_{r = \xi - m} \quad \text{with} \quad r = \xi - \bar{\xi} \ ,
\end{equation}
where we also introduced the residual $r$, which measures the deviations from $\bar{\xi}$, and the associated distribution $Q(r|\bar{\xi})$. In words, $Q_m(\xi|\bar{\xi})$ is the distribution using the residual statistics $r$, around an expansion point $\bar{\xi}$, but shifted to $m$. One can easily verify that the entropy related to $Q_m$ becomes independent of $m$, as shifts are volume-preserving transformations. Therefore we may use some fixed expansion point $\bar{\xi}$, and find the optimal shift $m$ using the KL which now may be written as
\begin{align}
\mathrm{KL}\left(Q_m|P\right) &= \left<\mathcal{H}(\xi = m + r , d)\right>_{Q(r|\bar{\xi})} + \frac{1}{2} \left<\left.\log\left(\left|\tilde{\mathcal{M}}(\xi)\right|\right)\right|_{\xi = \bar{\xi} + r}\right>_{Q(r|\bar{\xi})} + \mathrm{KL}_0 \notag\\ 
\widehat{\mathrm{KL}} &= \left<\mathcal{H}(\xi = m + r , d)\right>_{Q(r|\bar{\xi})}\ ,
\end{align}
where $\widehat{\mathrm{KL}}$ denotes the KL up to $m$ independent contributions. After optimization for $m$, we can update to a new expansion point, and use it to define a new family of distributions $Q_m$ which are a more appropriate class of approximations.
In general, the expectation value in $\widehat{\mathrm{KL}}$ cannot be computed analytically, but it can be approximated using a set of $N$ samples $\left\lbrace r^*_i \right\rbrace_{i \in \lbrace 1, ... , N \rbrace}$, drawn from $Q(r|\bar{\xi})$, which yields

\begin{equation}\label{eq:kl_estimate}
\widehat{\mathrm{KL}}\left(Q_m|P\right) \approx \frac{1}{N} \sum_{i=1}^N \mathcal{H}(\xi = m + r^*_i , d) \quad \text{with} \quad r^*_i \sim Q(r|\bar{\xi}) \ .
\end{equation}
Sampling from $Q(r|\bar{\xi})$ is defined as in section \ref{sec:sampling}, where the sampling procedure for $\tilde{Q}(\xi|\bar{\xi})$ is described, with the addition that a sample $r^*$ is is obtained from a sample for $\xi^*$ as $r^* = \xi^* - \bar{\xi}$.

\begin{figure*}[ht]
	\centering
	\includegraphics[scale=1., angle=0]{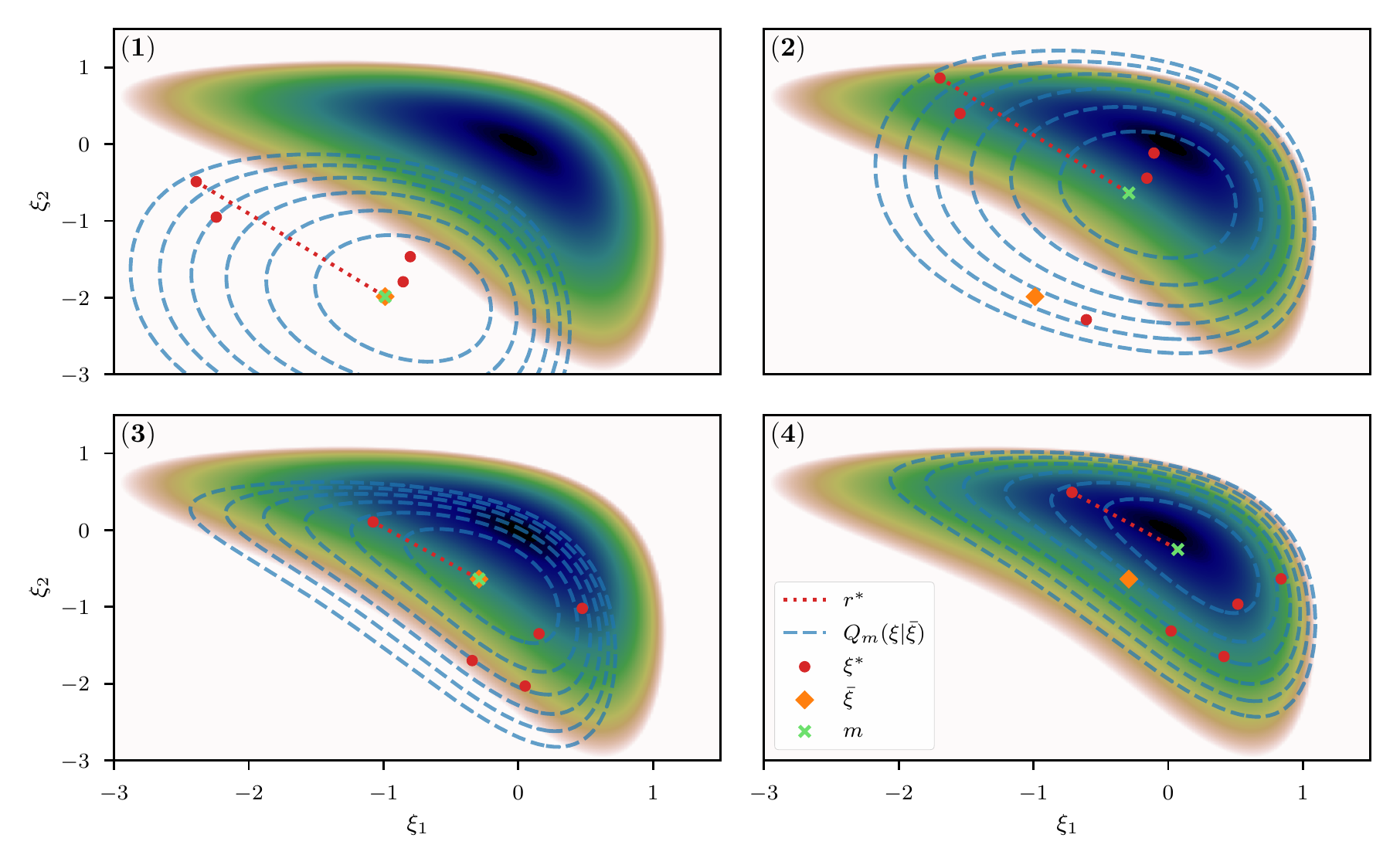}
	\centering
	\caption{$(1) - (4)$: Visualization of the geoVI steps. $(1)$: A randomly initialized shift $m$ (green cross) is used to set the initial expansion point $\bar{\xi}$ (orange dot) which in turn defines the initial approximation $Q_m(\xi|\bar{\xi}$ (blue dashed contours) used to generate a set of samples $\xi^*$ (red dots). $(2)$: The KL (equation \eqref{eq:kl_estimate}), estimated from the samples, is used to optimize for $m$, which results in a shift of $Q_m(\xi|\bar{\xi}$ away from the expansion point $\bar{\xi}$. The residual statistics $r^*$ derived from the geometry around $\bar{\xi}$, however, remains unchanged during this shift and therefore, at the new location $m$, becomes a bad representation of the local geometry. Thus, in $(3)$, the expansion point is set to the current estimate of $m$, which yields an update to the approximation $Q_m(\xi|\bar{\xi}$. Finally, we generate samples from this update and use them to optimize the re-estimated KL for $m$ which again results in a shift as seen in $(4)$. Within the full geoVI algorithm this procedure is iterated until convergence.}\label{fig:vis_steps}
\end{figure*}

Optimizing $\widehat{\mathrm{KL}}$ w.r.t.\ $m$ yields the variational optimum for the distribution $Q_m(\xi| \bar{\xi})$, given a fixed, predetermined expansion point $\bar{\xi}$. In order to move the expansion point $\bar{\xi}$ towards the optimal point, its location is updated subsequently and the KL is re-estimated using novel samples from $Q(r|\bar{\xi})$ with an updated $\bar{\xi}$. Specifically, we initialize the optimization algorithm at some position $m^0$, set $\bar{\xi} = m^0$ to obtain a set of samples $\left\lbrace r^*_i \right\rbrace^{(0)}_{i \in \lbrace 1, ... , N \rbrace}$, and use this set to approximate the KL. This approximation is then used to obtain an optimal shift $m^1$. Given this optimal shift, a new expansion point $\bar{\xi} = m^1$ is defined and used to obtain a novel set of samples $\left\lbrace r^*_i \right\rbrace^{(1)}_{i \in \lbrace 1, ... , N \rbrace}$ which defines a new estimate for the KL. This estimate is furthermore used to obtain a novel optimal $m$, and so on. An illustrative view of this procedure is given in figure \ref{fig:vis_steps}.
Finally, the entire procedure of optimizing the KL for $m$ and re-estimation of the KL via a novel expansion point is repeated until the algorithm converges to an optimal point $m^* = \bar{\xi}^*$. To optimize $\widehat{\mathrm{KL}}$ for $m$, we again employ the NewtonCG algorithm, and use the average of the metric $\mathcal{M}$ as a proxy for the curvature of $\widehat{\mathrm{KL}}$ to perform the optimization step. Specifically we use

\begin{equation}
	\widehat{\mathcal{M}}(m) = \frac{1}{N} \sum_{i=1}^N \mathcal{M}(\xi = m + r^*_i) \ ,
\end{equation}
as the metric of $\widehat{\mathrm{KL}}$.
We call this algorithm the \textit{geometric Variational Inference} (geoVI) method. A pseudo-code summary of geoVI is given in algorithm \ref{alg:geovi}.

\begin{algorithm}[ht]
	\caption{Geometric Variational Inference (geoVI)}\label{alg:geovi}	
	\SetKwData{Mm}{m}
	\SetKwData{Mn}{m$^*$}
	
	\SetKwData{postSams}{posteriorSamples}
	\SetKwData{Sams}{samples}
	\SetKwData{Kl}{kl}
	\SetKwData{Sample}{$r^*$}
	\SetKwFunction{en}{Energy}
	\SetKwFunction{kl}{geoKL}
	\SetKwFunction{sample}{drawSample}
	\SetKwFunction{newton}{NewtonCG}
	\SetKwProg{func}{Function}{:}{}
	\KwIn{Likelihood $\mathcal{H}(d|\xi)$, Transformation $x(\xi)$, Jacobian $\frac{\partial x}{\partial \xi}$}
	\func{\en{$\xi$}}{
		\KwRet{$\mathcal{H}(d|\xi) + \frac{1}{2} \xi^T\xi$}
	}
	\Mm $ \sim \mathcal{N}(m, 0, \mathds{1})$ \\
	\While{\Mm not converged}{
		$\bar{\xi} \leftarrow$ \Mm \\
		\Sams $\leftarrow$ \emph{empty list} \\
		\For{$i = 1$ to $N$}{
			$\xi^* \leftarrow$ \sample{$\bar{\xi}$, $x$, $\frac{\partial x}{\partial \xi}$} \quad (see Algorithm \ref{alg:sampling})\\
			\Sample $\leftarrow \xi^* - \bar{\xi}$ \\
			Insert \Sample into \Sams
		}
		\func{\kl{$\xi$}}{
			\Kl $\leftarrow 0$ \\
			\For{\Sample in \Sams}{
				\Kl $\leftarrow$ \Kl $+$ \en{$\xi$ + \Sample} \\
			}
			\KwRet{$\frac{1}{N}$ \Kl}
		}
		\Mn $\leftarrow$ \newton{\kl, \Mm} \\
		\Mm $\leftarrow$ \Mn
	}
	\postSams $\leftarrow$ \emph{empty list} \\
	\For{\Sample in \Sams}{
		$\xi^* \leftarrow$ \Mm + \Sample \\
		Insert $\xi^*$ into \postSams \\
	}
	\KwOut{\postSams}
\end{algorithm}

\subsubsection{Numerical sampling within geoVI}
It is noteworthy that, as described in section \ref{sec:sampling}, an implementation of the proposed sampling procedure for the residual $r$, and as a result also of the geoVI method itself, inevitably relies on numerical approximations to realize a sample for $r$. To better understand the impact of such approximations, we have to consider its impact on the distribution $Q(r|\bar{\xi})$.
To this end, we denote with $f$ the function that, given the expansion point $\bar{\xi}$, turns two standard distributed random vectors $\eta_1$ and $\eta_2$ into a random realization of $r$. Specifically

\begin{equation}
	r = f(\eta_1, \eta_2; \bar{\xi}) \quad \text{with} \quad \eta_{\nicefrac{1}{2}} \sim \mathcal{N}\left(\eta_{\nicefrac{1}{2}}; 0, \mathds{1}\right) \ ,
\end{equation}
where the functional form of $f$ is defined by combination of equation \eqref{eq:zsam} and \eqref{eq:miniinv}. using $f$ we may write the geoVI distribution $Q$ as

\begin{equation}
	Q(r|\bar{\xi}) = \int \int \delta\left(r - f(\eta_1, \eta_2; \bar{\xi})\right) \ \mathcal{N}\left(\eta_{1}; 0, \mathds{1}\right) \ \mathcal{N}\left(\eta_{2}; 0, \mathds{1}\right) \ \mathrm{d}\eta_1 \mathrm{d}\eta_2 \ .
\end{equation}
Any numerical algorithm used to approximate the sampling, irrespective of its exact form, may be described by replacing the function $f$, leading to exact sampling from $Q$, with some approximation $\widehat{f}$ which leads to an approximation of the distribution for $r$, which we denote as $\widehat{Q}(r|\bar{\xi})$. Therefore, in a way, the geoVI result using a numerical approximation for sampling can be understood as the variational optimum chosen from the family of distributions $\widehat{Q}$, rather than $Q$. Therefore, even for a non-zero approximation error in $\widehat{f}$, the result remains a valid optimum of a variational approximation, it is simply the family of distributions used for approximation that has changed. This finding is of great relevance in practice, as there is typically a trade off between numerical accuracy of the generated samples and computational efforts. Thus we may achieve faster convergence at a cost of accuracy in the approximation, but without completely detaching from the theoretical optimum, so long as $\widehat{f}$ remains sufficiently close to $f$. Nevertheless, as motivated in the introduction, it is important for the chosen family to contain distributions close to the true posterior, and therefore it remains important that the family $\widehat{Q}$ remains close to the family of $Q$ as only for $Q$ the geometric correspondence to the posterior has been established. A detailed study to further quantify this result, is left to future work.

\begin{figure*}[htp]
	\centering
	\includegraphics[scale=1., angle=0]{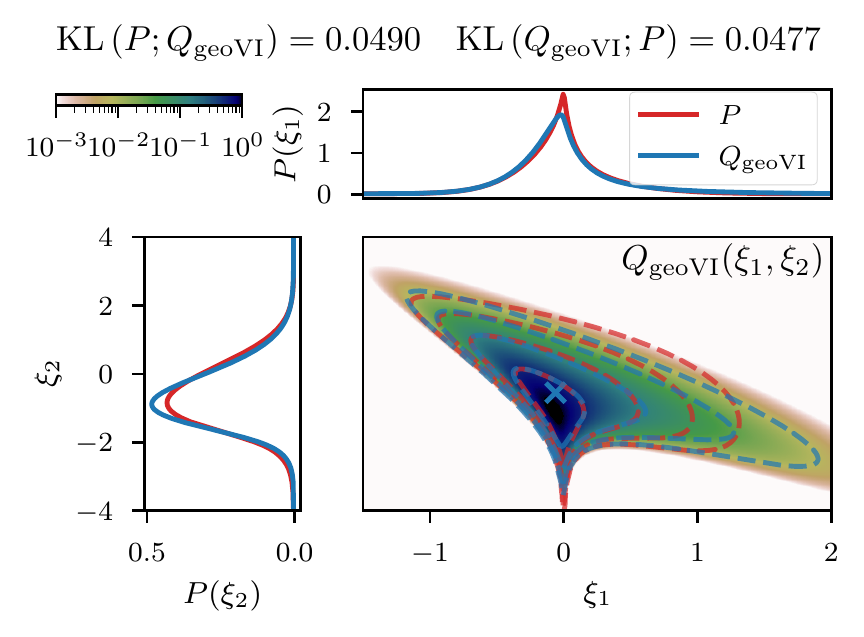}
	\includegraphics[scale=1., angle=0]{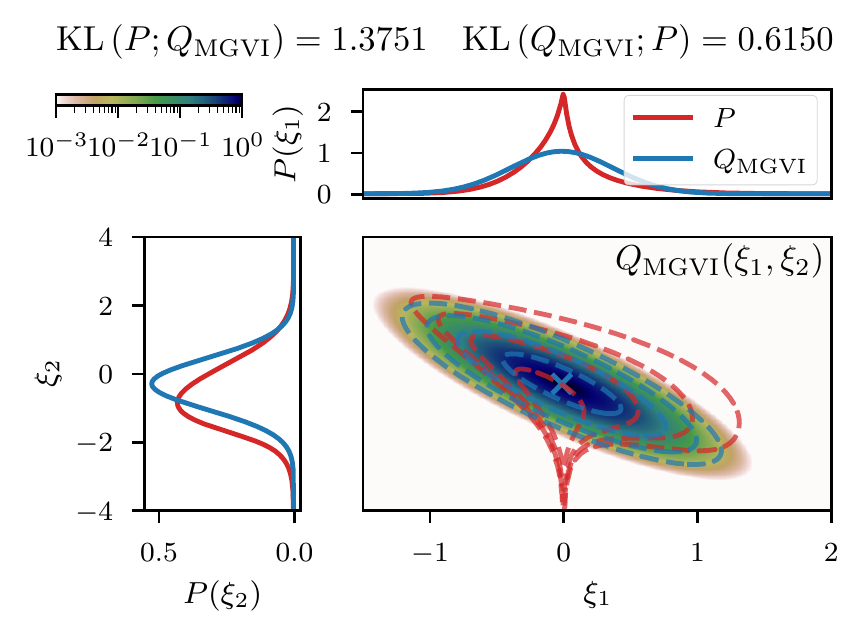}
	\centering
	\caption{The geoVI and MGVI approximations of the two-dimensional example described in section \ref{sec:simple_demos}. We display the same quantities as for the direct approximation shown in figure \ref{fig:vis_2d_direct}.}\label{fig:vis_2d_geo_mgvi}
\end{figure*}

\subsubsection{MGVI as a first order approximation}
We can compare the geoVI algorithm to the aforementioned variational approximation technique called Metric Gaussian variational inference (MGVI), and notice some key similarities. In particular the optimization heuristics with repeated alternation between sampling of $r^*$ and optimization for $m$ is entirely equivalent. The difference occurs in the distribution $Q(r|\bar{\xi})$ used for approximation. In MGVI, $Q$ is assumed to be a Gaussian distribution in $r$, as opposed to the Gaussian distribution in the transformed space $y$ used in geoVI. Specifically

\begin{equation}
Q_{\mathrm{MGVI}}(r | \bar{\xi}) \equiv \mathcal{N}(r; 0, \bar{\mathcal{M}}^{-1}) \ ,
\end{equation}
where the inverse of the posterior metric $\mathcal{M}$, evaluated at the expansion point $\bar{\xi}$, is used as the covariance. As it turns out, the distribution $Q_{\mathrm{MGVI}}$ arises naturally as a first order approximation to the coordinate transformation used in the geoVI approach. Specifically if we consider the geoVI distribution of $r$ given in terms of a generative process

\begin{equation}
r = g^{-1}(y ; \bar{\xi}) - \bar{\xi} \quad \text{with} \quad y \sim \mathcal{N}(y; 0, \mathds{1}) \ ,
\end{equation}
and expand it around $y=0$ to first order, we get that

\begin{align}\label{eq:mgviapprox}
	r &= g^{-1}(0, \bar{\xi}) - \left(\left.\frac{\partial g(\xi, \bar{\xi})}{\partial \xi}\right|_{\xi = \bar{\xi}}\right)^{-1} y  + \mathcal{O}\left(y^2\right) - \bar{\xi} \notag\\ &= \bar{\xi}- \bar{\xi} - \sqrt{\bar{\mathcal{M}}} \left(\mathds{1} + \left.\left(\left(\frac{\partial x}{\partial \xi}\right)^T \frac{\partial x}{\partial \xi}\right)\right|_{\xi = \bar{\xi}}\right)^{-1} y + \mathcal{O}\left(y^2\right) \notag\\ &= - \left(\sqrt{\bar{\mathcal{M}}}\right)^{-1} y + \mathcal{O}\left(y^2\right) \ .
\end{align}
Therefore, to first order in $y$, we get that

\begin{equation}
Q(r|\bar{\xi}) = \int \delta\left(r + \left(\sqrt{\bar{\mathcal{M}}}\right)^{-1} y \right) \ \mathcal{N}(y; 0,\mathds{1}) \ \mathrm{d}y = \mathcal{N}(r; 0, \bar{\mathcal{M}}^{-1}) = Q_{\mathrm{MGVI}}(r | \bar{\xi}) \ .
\end{equation}
This correspondence shows that geoVI is a generalization of MGVI in non-linear cases. This is a welcome result, as numerous practical applications \cite{hutschenreuter2021galactic,welling2021reconstructing,arras2021comparison} have shown that already MGVI provides a sensible approximation to the posterior distribution. On the other hand, it provides further insight in which cases the MGVI approximation remains valid, and when it reaches its limitations. In particular if

\begin{equation}\label{eq:consistencymgvi}
\mathcal{M}(m + r)  \approx \bar{\mathcal{M}} \ , \quad \forall r = g^{-1}(y,\bar{\xi}) - \bar{\xi} \quad \text{with} \quad y \sim \mathcal{N}(y; 0, \mathds{1}) \ ,
\end{equation}
we get that the first order approximation of equation \eqref{eq:mgviapprox} yields a close approximation of the inverse and geoVI reduces to the MGVI algorithm. In contrast, geoVI with its non-linear inversion requires the log determinant of the metric $\mathcal{M}$ to be approximately constant throughout the sampling regime. This is a much less restrictive requirement then equation \eqref{eq:consistencymgvi}, as the variation of eigenvalues of $\mathcal{M}$ is considered on a logarithmic scale whereas it is considered on linear scale in equation \eqref{eq:consistencymgvi}. Furthermore, the log-determinant is invariant under unitary transformations which means that local rotations of the metric, and therefore changes in orientation as we move along the manifold, can be captured by the non-linear approach, whereas equation \eqref{eq:consistencymgvi} does not hold any more if the orientation varies as a function of $r$. Therefore we expect the proposed approach to be applicable in a more general context, while still retaining the MGVI properties, as it reproduces MGVI in the linear limit.

\subsection{Examples}\label{sec:examples}
We can visually compare the geoVI and the MGVI algorithm using the two-dimensional example previously mentioned in section \ref{sec:simple_demos}. In analogy to figure \ref{fig:vis_2d_direct} we depict the approximation to the posterior density together with its two marginals in figure \ref{fig:vis_2d_geo_mgvi}. We see that geoVI yields a similar result compared to the direct approach here, while it provides a significant improvement compared to the approximation capacity of MGVI.

\begin{figure*}[htp]
	\centering
	\includegraphics[scale=1., angle=0]{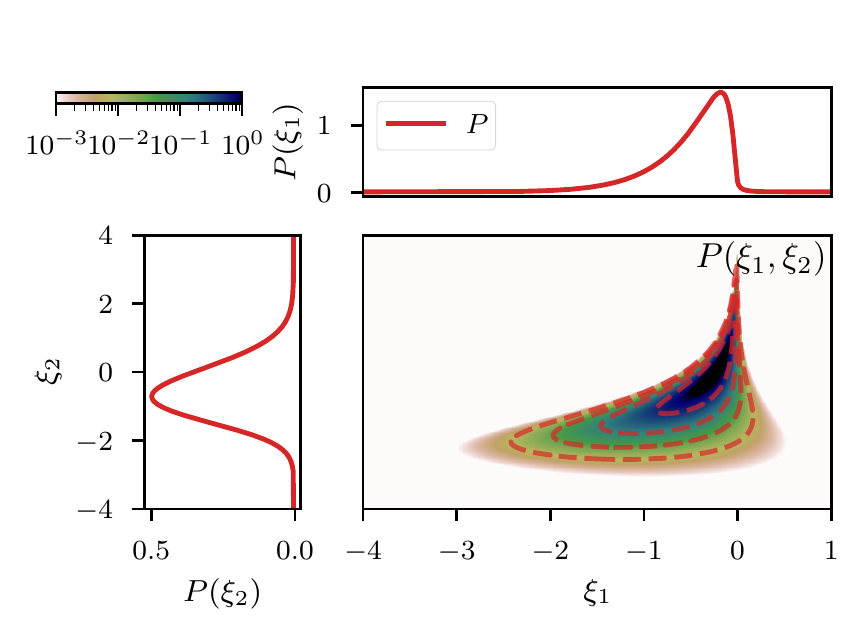}
	\includegraphics[scale=1., angle=0]{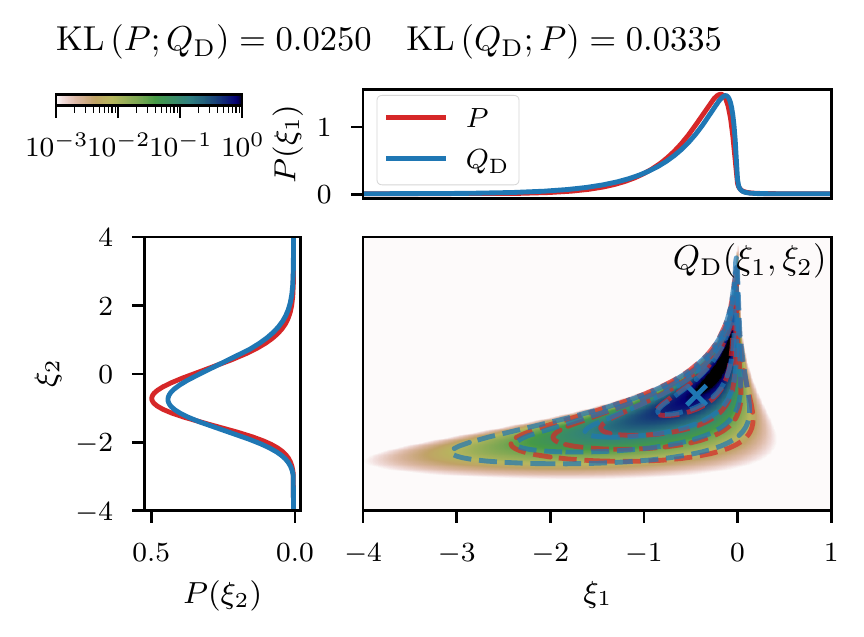}
	\includegraphics[scale=1., angle=0]{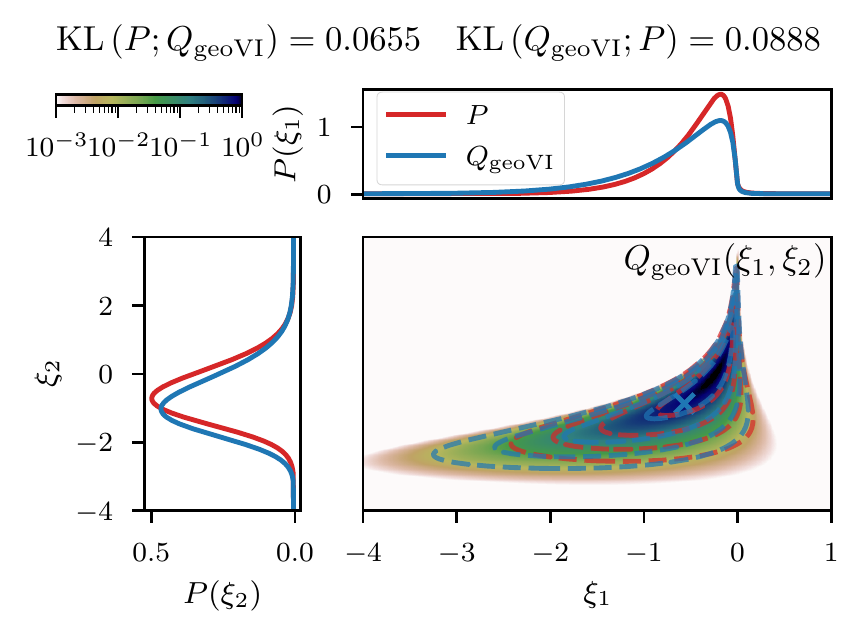}
	\includegraphics[scale=1., angle=0]{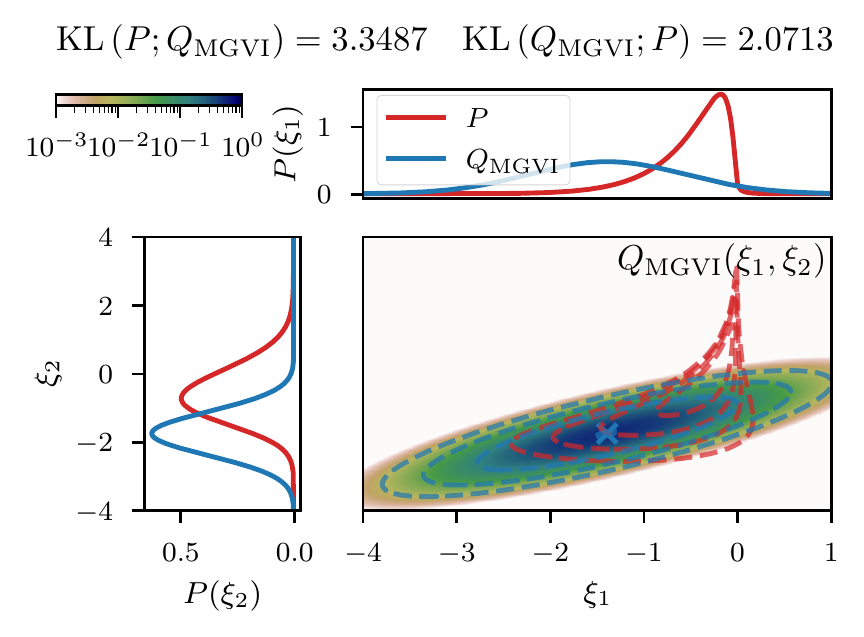}
	\includegraphics[scale=1., angle=0]{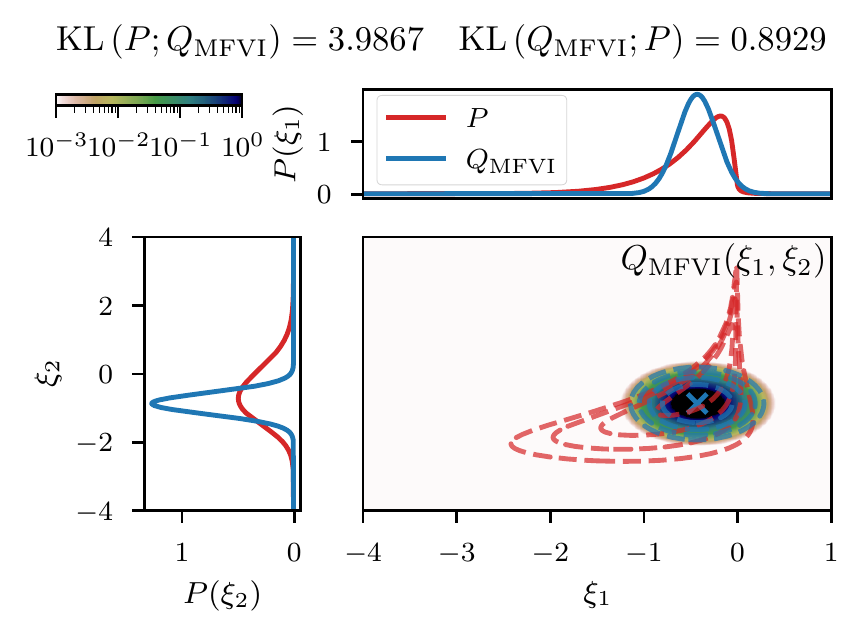}
	\includegraphics[scale=1., angle=0]{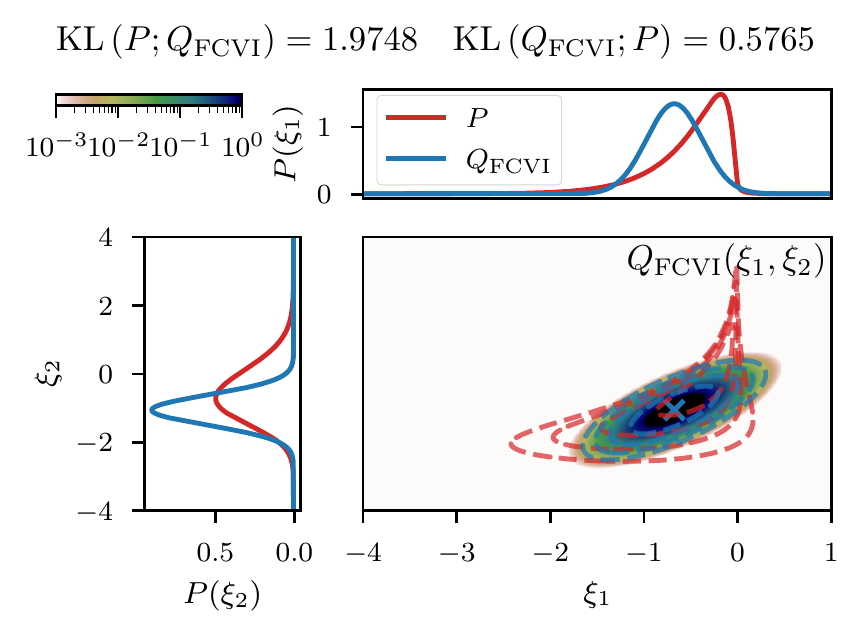}
	\centering
	\caption{Same setup as in figures \ref{fig:vis_2d_direct} and \ref{fig:vis_2d_geo_mgvi} but for a Gaussian measurement of the product of a normal distributed quantity $\xi_1$ and a log-normal distributed one $\xi_2$ as described in the second example of section \ref{sec:examples}. From top to bottom and from left to right: ground truth $P$, direct approximation $Q_{\mathrm{D}}$, geoVI approximation $Q_{\mathrm{geoVI}}$, MGVI approximation $Q_{\mathrm{MGVI}}$, mean-field approximation $Q_{\mathrm{MFVI}}$, and the normal approximation with a full-rank covariance $Q_{\mathrm{FCVI}}$.}\label{fig:lognorm_2d}
\end{figure*}

To conclude the illustrative examples, we consider a single observation of the product of a normal and a log-normal distributed quantity subject to independent, additive Gaussian noise. The full model consists of a likelihood and a prior of the form

\begin{equation}
	P(d|\xi_1, \xi_2) = \mathcal{N}(d; \xi_1 e^{\xi_2}, \sigma_n^2) \quad \text{with} \quad \xi_{\nicefrac{1}{2}} \sim \mathcal{N}(\xi_{\nicefrac{1}{2}}; 0, 1) \ .
\end{equation}
This example should serve as an illustration of the challenges that arise when attempting a separation of non-linearly coupled quantities from a single observation. Such separation problems reappear in section \ref{sec:applications} in much more intertwined and high dimensional examples, but much of the structural challenges can already be seen in this simple two-dimensional problem. Figure \ref{fig:lognorm_2d} displays the results of the direct approach as well as the geoVI and MGVI methods for a measurement setting of $d=-0.3$ and $\sigma_n = 0.1$. As a comparison, we also depict the results from performing a variational approximation using a normal distribution with a diagonal covariance, also known as a mean-field approximation (MFVI), as well as an approximation with a normal distribution using a full-rank matrix as its covariance (FCVI). Both, the diagonal as well as the full-rank covariance are considered parameters of the distribution, and have to be optimized for in addition to the mean of the normal distribution. An efficient implementation thereof is described in  \cite{kucukelbir2017automatic}.
We notice that both, the direct and the geoVI approach manage to approximate the true posterior distribution well, although the KL values indicate that the approximation by geoVI is worse by $\approx 0.016$ nats compared to the direct approach. Here the passive update of the expansion point used in this approach reaches its limitations as in cases where the posterior distribution becomes increasingly narrow towards the optimal expansion point, the static sample statistics of $r$ can get stuck during optimization and increasingly repeated re-sampling becomes necessary as one moves closer to the optimum. Nevertheless, the geoVI approximation remains a good approximation to the true distribution, especially when compared to the approaches using a normal distribution such as MGVI, MFVI, and FCVI.

\section{Applications}\label{sec:applications}
To investigate the performance of the geoVI algorithm in high dimensional imaging problems, we apply it to two mock data examples and compare it to the results using MGVI. In the first example, which serves as an illustration, the geoVI results are additionally compared to the results obtained from applying a Hamiltonian Monte-Carlo (HMC) sampler \cite{duane1987hybrid} to the mock example (see section \ref{sec:hmc} for further information on HMC). The second example is an illustration of a typical problem encountered in astrophysical imaging.
Both examples consist of hierarchical Bayesian models with multiple layers which are represented as a generative process. 
One particularly important process for the class of problems at hand are statistically homogeneous and isotropic Gaussian processes with unknown power spectral density, for which a flexible generative model has been presented in \cite{arras2020variable}. This process is at the core of a variety of astrophysical imaging applications \cite{reimardust,arras2020variable,arras2019unified,hutschfaraday}, and therefore an accurate posterior approximation of problems involving this model is crucial. To better understand the inference challenges that arise in problems using this particular model, we briefly summarize some of its key properties.

\subsection{Gaussian processes with unknown power spectra}\label{sec:gauprocess}
Consider a zero mean, square integrable random process $s_x$ defined on a $L$-dimensional space subject to periodic boundary conditions which, for simplicity, we assume to have size one. Specifically let $x \in \Lambda = [0,1]^{L}$ and thus $s \in \mathcal{L}^2\left(\Lambda\right)$. A Gaussian process 

\begin{equation}
	P(s) = \mathcal{N}(s; 0, S) \ ,
\end{equation}
with mean zero and covariance function $S_{xy}$ is said to be statistically homogeneous and isotropic, if $S$ is a function of the Euclidean distance between two points i.E.\

\begin{equation}
	S_{xy} = S(|x-y|) \ .
\end{equation}
Furthermore, as implied by the Wiener Wiener-Khinchin theorem \cite{wiener_extrapolation_1950}, the linear operator associated with $S$ becomes diagonal in the Fourier space associated with $\Lambda$, and therefore $s$ may be represented in terms of a Fourier series with coefficients $\tilde{s}_k$, where $k$ labels the Fourier coefficients. These coefficients are independent, zero mean Gaussian random variables with variance

\begin{equation}
	\left< |\tilde{s}_k|^2\right>_{P(s)} \equiv P_s(|k|) \ ,
\end{equation}
which is also known as the power spectrum $P_s$ of $s$. As $P_s$ encodes the correlation structure of $s$, its functional form is crucial to determine the prior statistical properties of $s$. In \cite{arras2020variable} a flexible, non-parametric prior process for the power spectrum has been proposed by means of a Gauss-Markov process on log-log-scale. This process models the spectrum as a straight line on log-log-scale (resulting in a power law in $|k|$ on linear scale) with possible continuous deviations thereof. These deviations are itself defined as a Gauss-Markov process (specifically an integrated Wiener process) and their respective variance is, among others, an additional scalar parameter steering the properties of this prior process that are also considered to be random variables that have to be inferred. These parameters are summarized in Table \ref{table:cf_params}. A more formal derivation of this model in terms of a generative process relating standard distributed random variables $\xi_p$ to a random realization $P_s(\xi_p)$ of this prior model, is given in appendix \ref{ap:cf_model}.

\begin{table}[ht]
\begin{tabular}{l p{7cm} r}
	Name & Description & Prior distribution \\
	offset std. & Prior standard deviation of the overall offset of $s$ from zero & Log-normal \\
	fluctuations & Prior amplitude of the variation of $s$ around its offset & Log-normal \\
	slope & Exponent of the power law related to $P_s$ & Normal \\
	flexibility & Amplitude of deviations from the power-law on log-log-scale & Log-normal \\
	asperity & Smoothness of the deviations as a function of $\log(|k|)$ & Log-normal
\end{tabular}
\caption{Table of additional parameters}\label{table:cf_params}
\end{table}

In order to use this prior within a larger inference model, the underlying space has to be discretized such that the solution of the resulting discrete problem remains consistent with the continuum. We achieve this discretization by means of a truncated Fourier series for $s$ such that $s$ may be written as
\begin{equation}
	s = \mathcal{F}^{\dagger} \left(\sqrt{P_k(\xi_p)} \xi \right) \quad \text{with} \quad P(\xi) = \mathcal{N}(\xi; 0, \mathds{1}) \ ,
\end{equation}
where $\mathcal{F}$ denotes a discrete Fourier transformation (DFT) and $\mathcal{F}^{\dagger}$ its back-transformation. If we additionally evaluate $s$ on a regular grid on $\Lambda$, we can replace the DFT with a fast Fourier transformation (FFT) which is numerically more efficient. For a detailed description on how the spatial discretization is constructed please refer to \cite{1930-8337_2009_1_87,fdi_adp}. In this work, however, we are primarily interested in evaluating the approximation quality of the proposed algorithm geoVI, and therefore, from now on, we regard all inference problems involving this random process to be high, but finite, dimensional Bayesian inference problems and ignore the fact that it was constructed from a corresponding continuous, infinite dimensional, inference problem.

\begin{figure*}[ht]
	\centering
	\includegraphics[scale=1., angle=0]{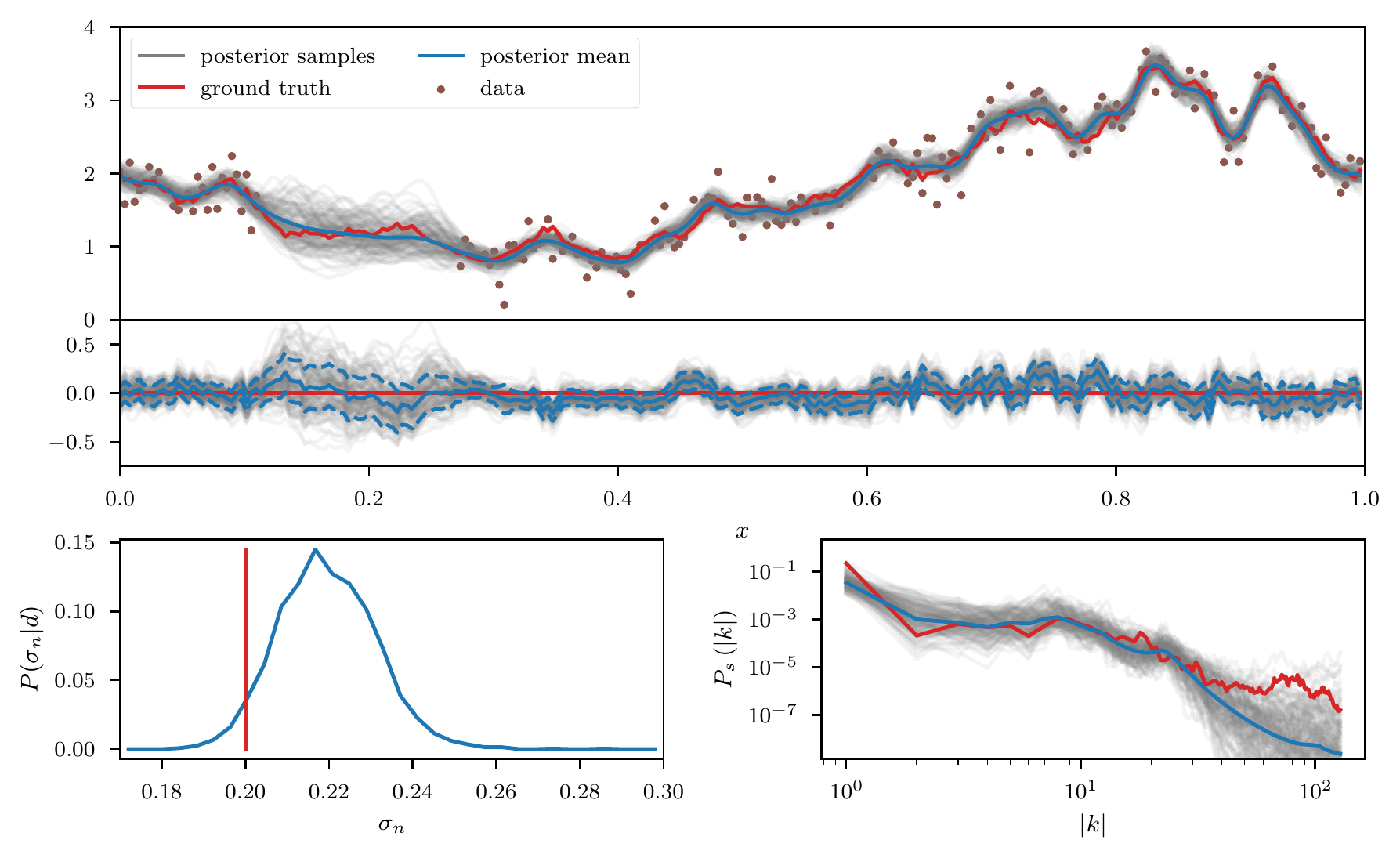}
	\centering
	\caption{Posterior approximation using the geoVI algorithm for the log-normal process described in section \ref{sec:applications_logn}. Top: The ground truth realization of the log-normal process $e^s$ (red line) and the corresponding data (brown dots) used for reconstruction. The blue line is the posterior mean, and the gray lines are a subset of the posterior samples obtained from the geoVI approximation. Below we depict the residual between the ground truth and reconstruction, including the residuals for the posterior samples. The blue dashed line corresponds to the one-sigma uncertainty of the reconstruction. Bottom left: Approximation to the marginal posterior distribution (blue) of the noise standard deviation $\sigma_n$. The red vertical line indicates the true value of $\sigma_n=0.2$ used to construct the data. Bottom right: Power spectrum $P_s$ of the logarithmic quantity $s$. Red displays the ground truth, blue the posterior mean, and the gray lines are posterior samples of the power spectrum.}\label{fig:ne_signal_geo}
\end{figure*}

\begin{figure*}[htp]
	\centering
	\includegraphics[scale=1., angle=0]{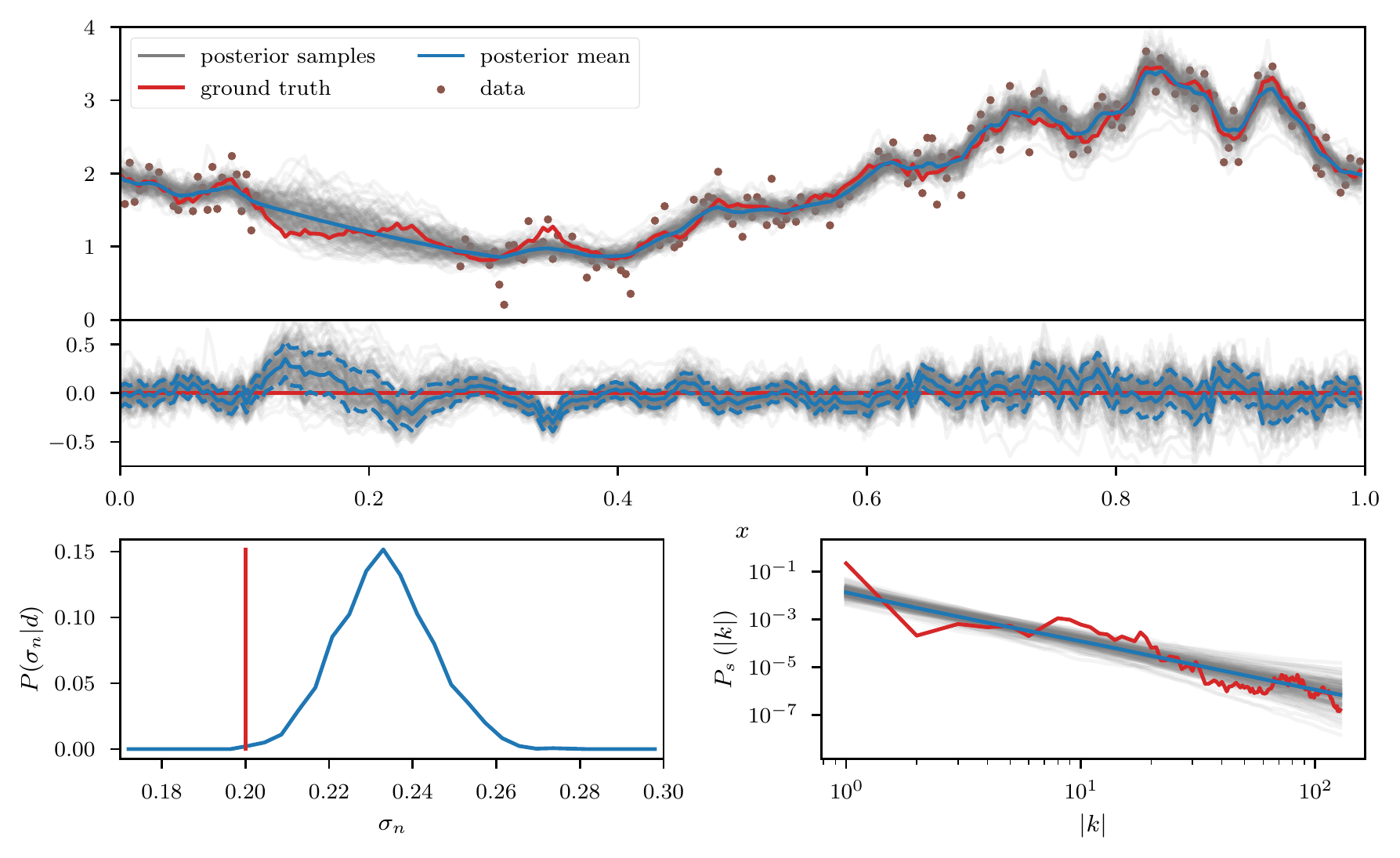}
	\centering
	\caption{Same setup as in figure \ref{fig:ne_signal_geo}, but for the approximation using the MGVI algorithm.}\label{fig:ne_signal_mgvi}
\end{figure*}

\begin{figure*}[htp]
	\centering
	\includegraphics[scale=1., angle=0]{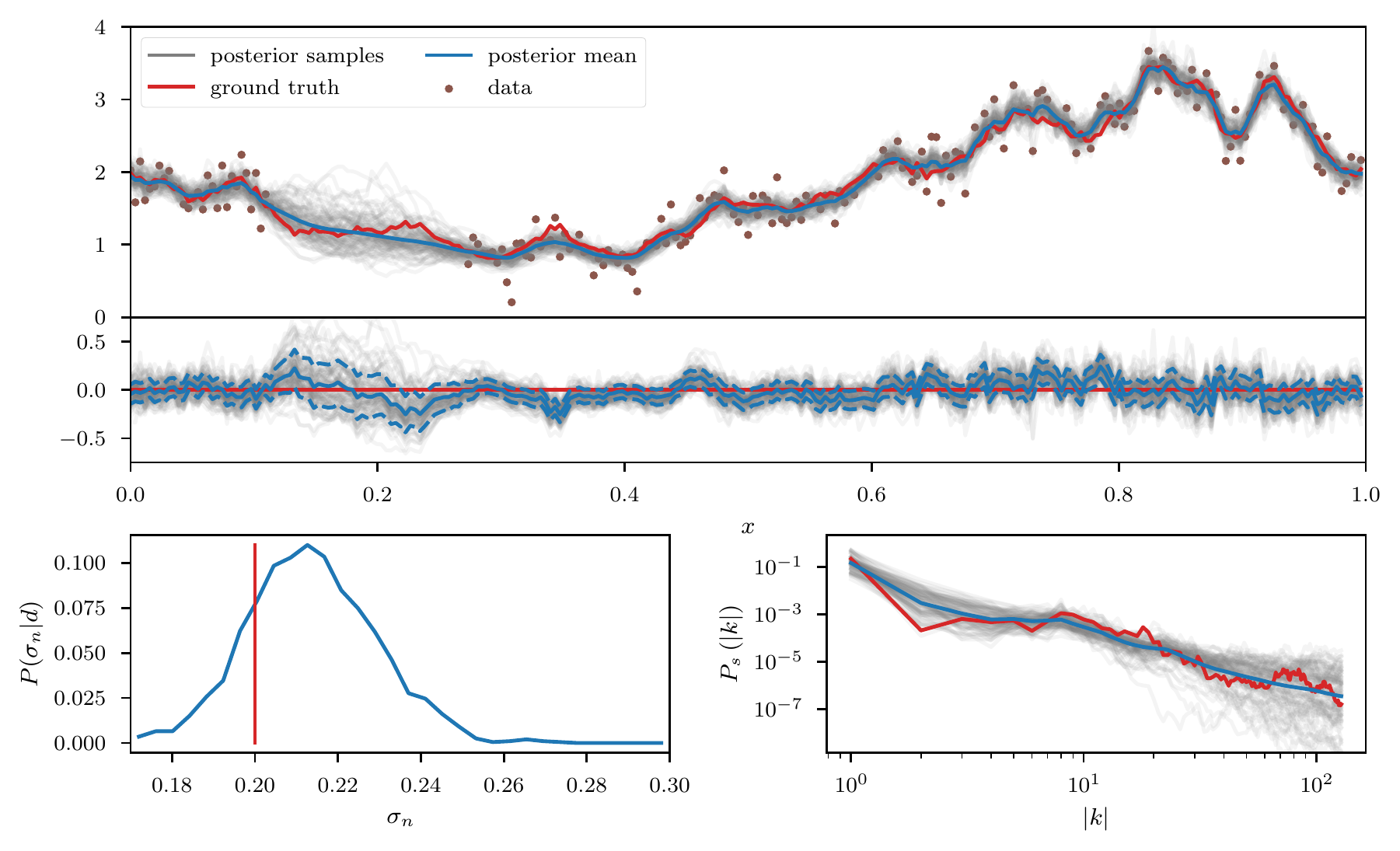}
	\centering
	\caption{Same setup as in figure \ref{fig:ne_signal_geo}, but for the approximation using the HMC sampling.}\label{fig:ne_signal_hmc}
\end{figure*}

\subsection{Log-normal process with noise estimation}\label{sec:applications_logn}
As a first example we consider a log-normal process $e^s$, defined over a one-dimensional space, with $s$ being a priori distributed according to the aforementioned Gaussian process prior with unknown power spectrum. The observed data $d$ (see top panel of figure \ref{fig:ne_signal_geo}) consists of a partially masked realization of this process subject to additive Gaussian noise with standard deviation $\sigma_n$. In addition to $s$ and its power spectrum $P_s$, we also assume $\sigma_n$ to be unknown prior to the observation and place a log normal prior on it. Therefore the corresponding likelihood takes the form

\begin{equation}
	P(d|s, \sigma_n) = \mathcal{N}(d; R e^s, \sigma_n^2) \ .
\end{equation}
We apply the geoVI algorithm (figure \ref{fig:ne_signal_geo}),the MGVI algorithm (figure \ref{fig:ne_signal_mgvi}), and an HMC sampler (figure \ref{fig:ne_signal_hmc}) to this problem and construct a set of 3000 approximate posterior samples for all methods. The HMC results serve as the true reference here, as the true posterior distribution is too high dimensional to be accessible directly and HMC is known to reproduce the true posterior in the limit of infinite samples.
Considering solely the reconstruction of $e^s$, we see that both methods, geoVI and MGVI, agree with the true signal largely within their corresponding uncertainties. Overall we find that the geoVI solution is slightly closer to the ground truth compared to MGVI and the posterior uncertainty is smaller for geoVI in most regions, with the exception of the unobserved region, where it is larger compared to MGVI (see residual plot of figures \ref{fig:ne_signal_geo} and \ref{fig:ne_signal_mgvi}). In this region MGVI appears to slightly underestimate the posterior uncertainty. In addition, in the bottom panels of figures \ref{fig:ne_signal_geo} and \ref{fig:ne_signal_mgvi}, we depict the posterior distribution of the noise standard deviation $\sigma_n$ as well as the posterior mean of the power spectrum $P_s$, together with corresponding posterior samples. Here the difference between geoVI and MGVI becomes evident more visibly, which, to some degree, is to be expected due to the more non-linear coupling of $\sigma_n$ and $P_s$ to the data compared to $e^s$. Indeed we find that the posterior distribution of $\sigma_n$ recovered using MGVI is overestimating the noise level of the reconstruction. The geoVI algorithm is also slightly overestimating $\sigma_n$, however we find that for geoVI the posterior yields $\sigma_n^{\mathrm{geoVI}} = 0.220 \pm 0.026$ which places the true value of $\sigma_n = 0.2$ approximately $0.8$-sigma away from the posterior mean. In contrast, for MGVI, we get that $\sigma_n^{\mathrm{MGVI}} = 0.233 \pm 0.011$ for with the ground truth is almost a $3$-sigma event. Considering the HMC results (bottom panel of figure \ref{fig:ne_signal_hmc}), the geoVI results appear to be closer to the HMC distribution compared to MGVI, although the HMC distribution for $\sigma_n$ is broader and even closer to the true value then geoVI. 
In addition we find that the overall reconstruction quality of the power spectrum $P_s$ is significantly increased when moving from MGVI to geoVI. While MGVI manages to recover the overall slope of the power-law, it fails to reconstruct the devations from this power-law as well as the overall statistical properties of $P_s$ as encoded in the parameters of table \ref{table:cf_params}. In contrast, the geoVI algorithm is able to pick up some of these features and recovers posterior statistical properties of the power spectrum similar to the ground truth. In addition the posterior uncertainty appears to be on a reasonable scale, as opposed to the MGVI reconstruction which significantly underestimates the posterior uncertainty. The structures on the smallest scales (largest values for $|k|$), however, appear to be underestimated by the geoVI mean, although the posterior uncertainty increases significantly in these regimes. In comparison to HMC we find that the results are in agreement for the large scales, although the geoVI uncertainties appear to be slightly larger. On small scales, the methods deviate stronger, and the under-estimation of the spectrum seen by geoVI is absent in the HMC results.

\begin{figure*}[htp]
	\centering
	\includegraphics[scale=1., angle=0]{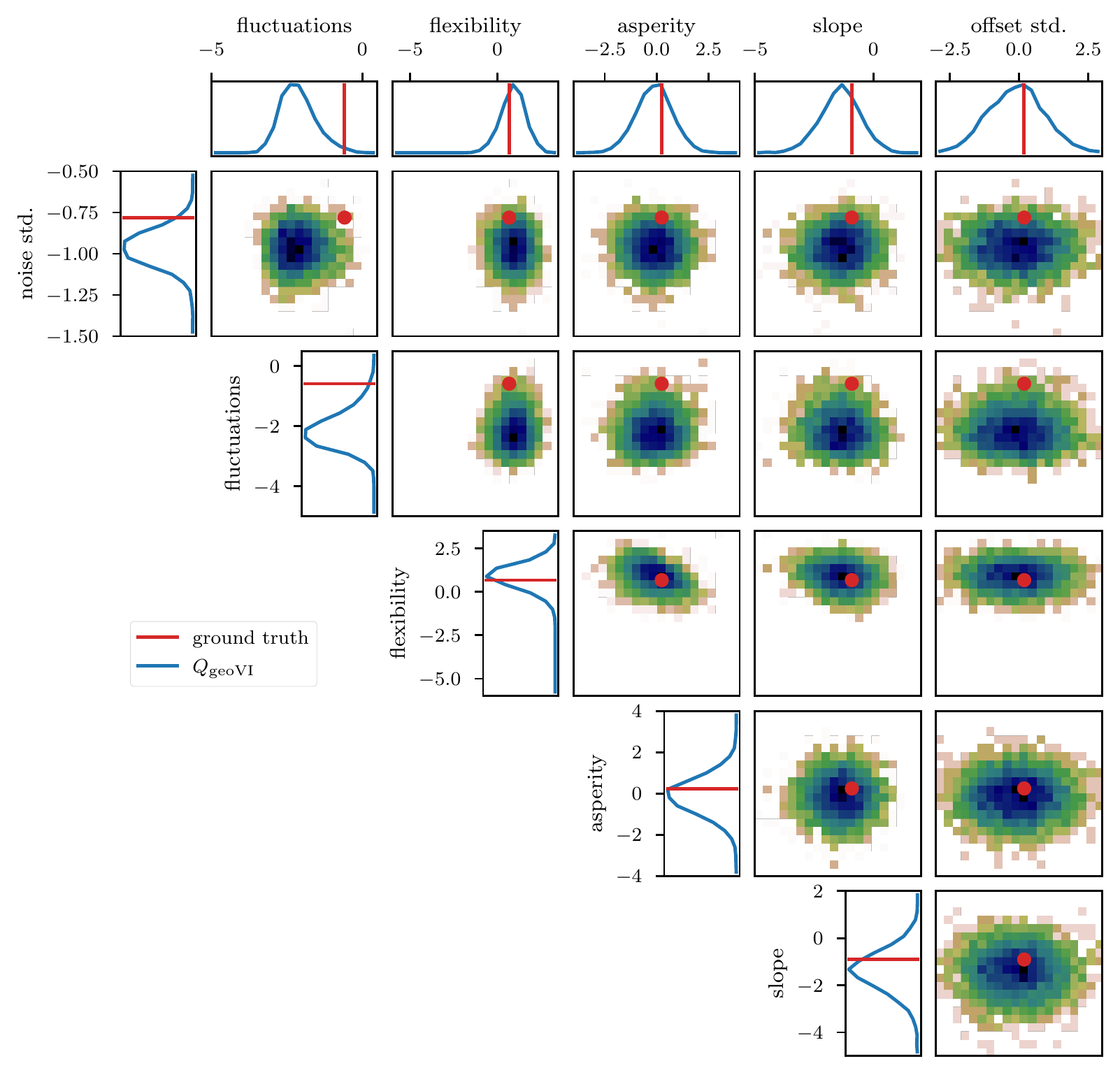}
	\centering
	\caption{Posterior distributions of the scalar parameters that enter the forward model of the power spectrum (Table \ref{table:cf_params}), and the noise standard deviation. All parameters, including the noise parameter, are given in their corresponding prior standard coordinate system, i.E. have a normal distribution with zero mean and variance one as a prior distribution. Each square panel corresponds to the joint posterior of the parameter in the respective row and column. In addition, for each row and each column the one-dimensional marginal posteriors of the corresponding parameter are displayed as blue lines. The red lines in the 1-D, and the red dots in the 2-D plots denote the values of the ground truth used to realize the ground truth values of the spectrum $P_s$, the signal $e^s$, and finally the observed data $d$.}\label{fig:ne_panels_geo}
\end{figure*}

\begin{figure*}[htp]
	\centering
	\includegraphics[scale=1., angle=0]{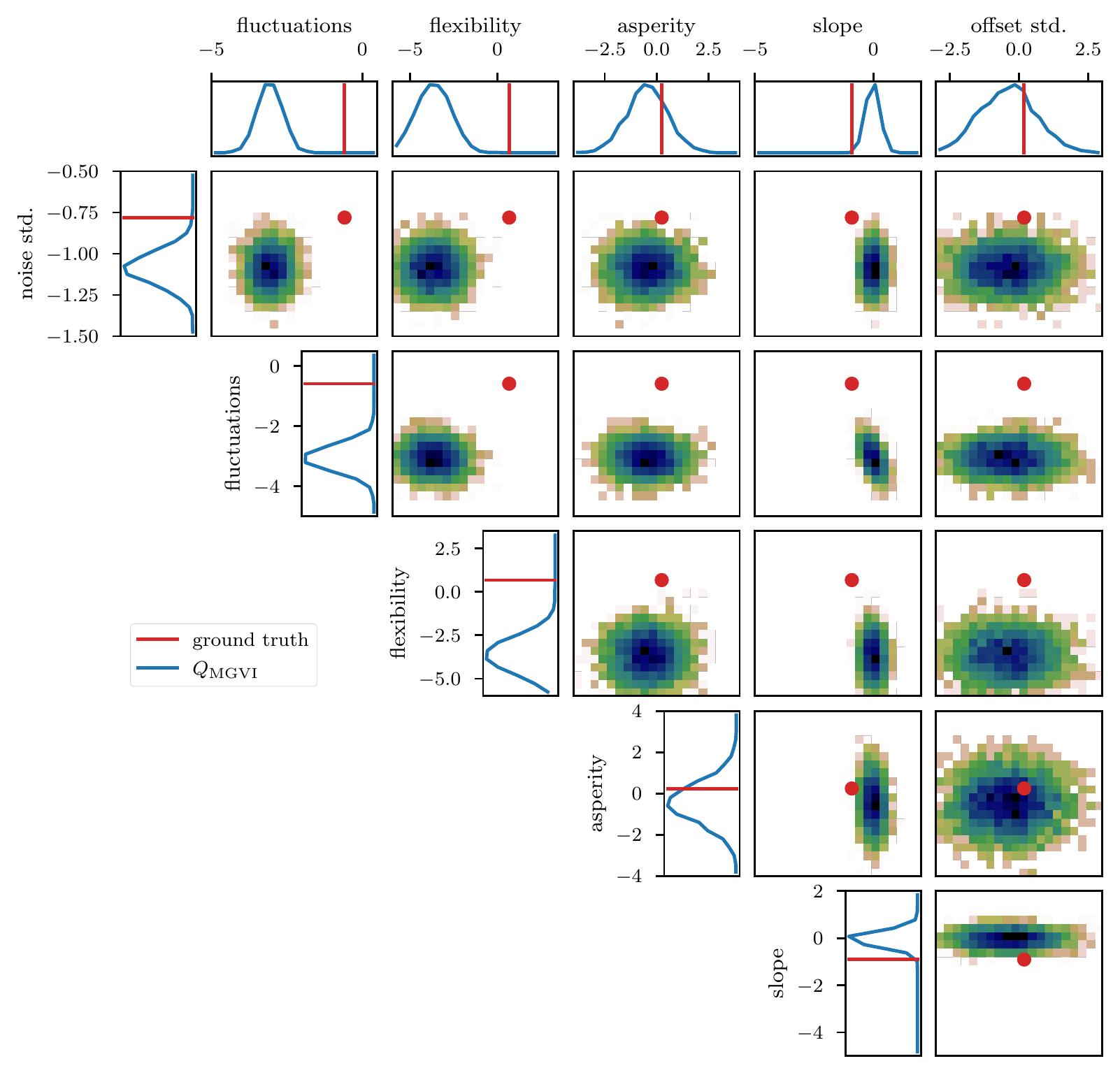}
	\centering
	\caption{Same setup as in figure \ref{fig:ne_panels_geo}, but for the approximation using the MGVI algorithm.}\label{fig:ne_panels_mgvi}
\end{figure*}

\begin{figure*}[htp]
	\centering
	\includegraphics[scale=1., angle=0]{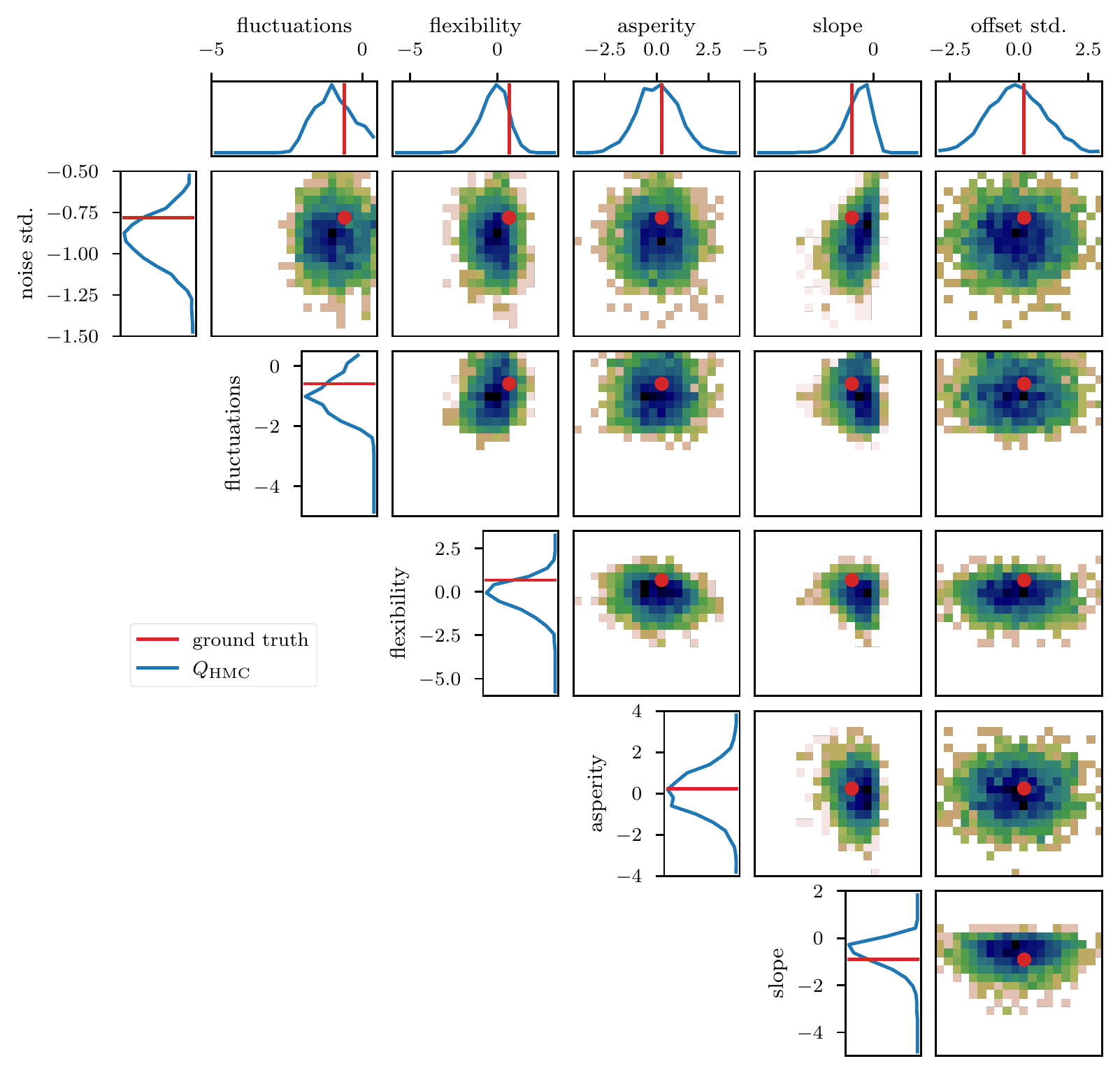}
	\centering
	\caption{Same setup as in figure \ref{fig:ne_panels_geo}, but for the approximation using HMC sampling.}\label{fig:ne_panels_hmc}
\end{figure*}

To further study the posterior distribution of the various scalar parameters that enter the power spectrum model (see Table \ref{table:cf_params}), as well as the noise standard deviation $\sigma_n$, we depict the reconstructed marginal posterior distributions for all pairs of inferred scalar parameters. Figures \ref{fig:ne_panels_geo}, \ref{fig:ne_panels_mgvi}, and \ref{fig:ne_panels_hmc} show the posterior distributions recovered using geoVI, MGVI, and HMC, respectively. All parameters are displayed in their corresponding standard coordinate system, i.E.\ they all follow a zero-mean unit variance normal distribution prior to the measurement. From an inference perspective, some of these parameters are very challenging to reconstruct, as their coupling to the observed data is highly non-linear and influenced by many other parameters of the model. In turn, their values are highly influential to the statistical properties of more interpretable variables such as the observed signal $e^s$ and its spectrum $P_s$. We see that despite these challenges the geoVI posterior appears to give reasonable results, that are largely in agreement with the ground truth, within uncertainties. Thus the algorithm appears to be able to pick up parts of the non-linear structure of the posterior, which is validated when compared to the MGVI algorithm, as for these parameters the MGVI reconstruction (figure \ref{fig:ne_panels_mgvi}) does not yield reliable results any more. This is to be expected in case of significant non-linearity as MGVI is the first order approximation of geoVI. In comparison to HMC (figure \ref{fig:ne_panels_hmc}), however, we find that there remain some differences in the recovered posterior distributions. The HMC results regarding the ``fluctuations'' and ``noise std.'' parameters are more centered on the ground truth and in particular the posterior distribution of the ``slope'' parameter is significantly different and more constrained, compared to the geoVI results. These differences indicate that there remain some limitations to the recovered geoVI results in the regime of highly non-linear parameters of the model which we may associate to the theoretical limitations discussed in section \ref{sec:properties_error}.

\begin{figure*}[htp]
	\centering
	\includegraphics[scale=1., angle=0]{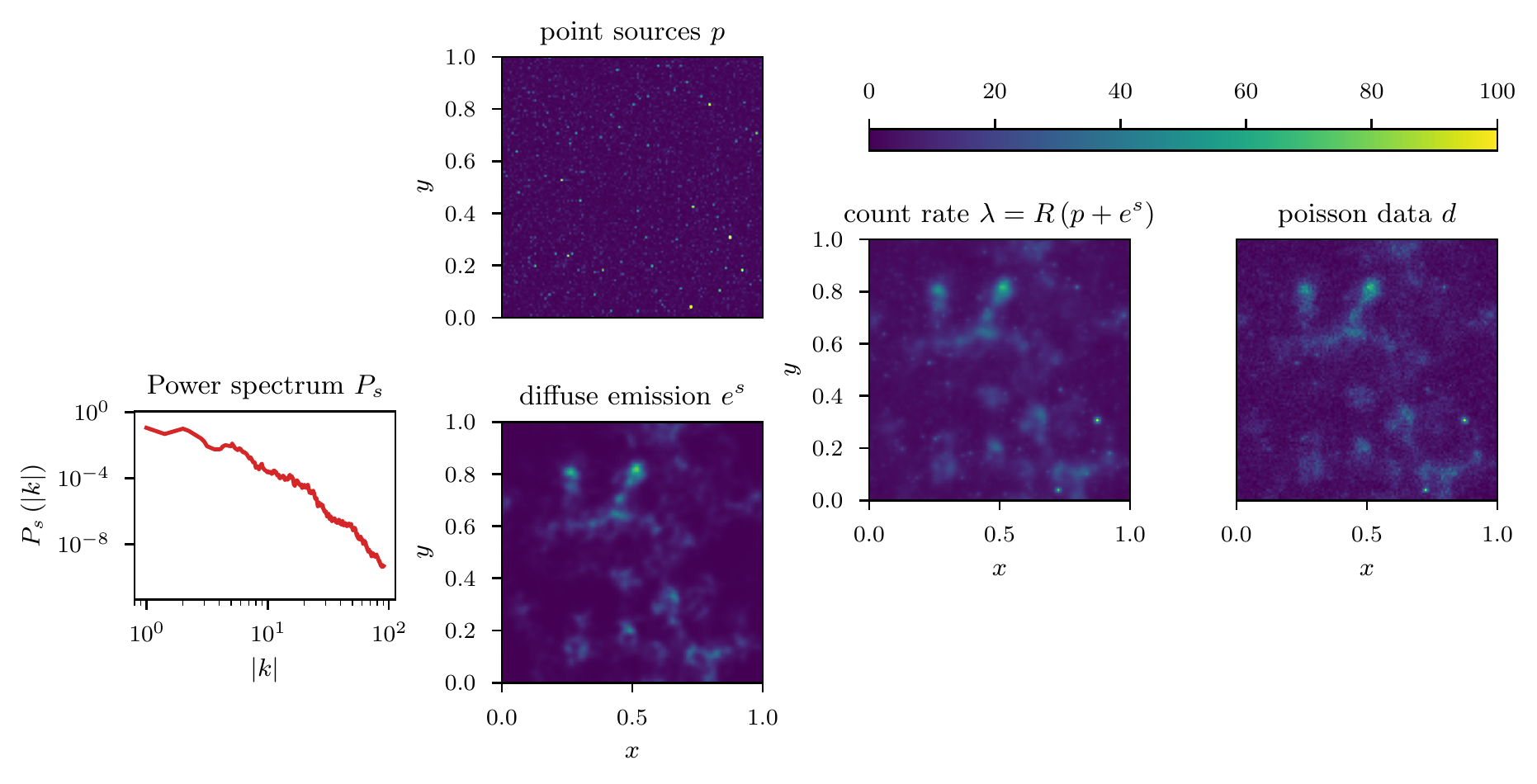}
	\centering
	\caption{Graphical setup of the separation problem discussed in section \ref{sec:poisson}. Random realization of the power spectrum $P_s$ (left) which is used to generate the log-signal $s$, which, after exponentiation, models the diffuse emission on the sky $e^s$. The point sources $p$ (top panel in the middle), which are a realization of the position-independent inverse-gamma process, get combined with the diffuse emission and the result is convolved with a spherical symmetric point spread function $R$ to yield the per-pixel count rate $\lambda$ which is ultimately used as the rate in a Poisson distribution used to realize the count data $d$.}\label{fig:poisson_setup}
\end{figure*}

\subsection{Separation of diffuse emission from point sources}\label{sec:poisson}
In a second inference problem we consider the imaging task of separating diffuse, spatially extended and correlated emission $e^s$ from, bright, but uncorrelated point sources $p$ in an image. This problem is often encountered within certain astrophysical imaging problems \cite{bertin1996sextractor} where the goal is to recover the emission of spatially extended structures such as gas or galactic dust. This emission usually gets superimposed by the bright emission of stars (point sources) in the image plane, and only their joint emission can be observed. In this example we assume that the emission is observed through a detection device that convolves the incoming emission with a spherical symmetric point spread function $R$ and ultimately measures photon counts on a pixelated grid. Specifically we may define a Poisson process with count rate

\begin{equation}
	\lambda = R(p + e^s) \quad \text{with} \quad P(d|\lambda) = \prod_i \frac{\left(\lambda_i\right)^{d_i} e^{-\lambda_i} }{\left(k_i\right)!} \ ,
\end{equation}
where $i$ labels the pixels of the detector. We assume the diffuse emission to follow a statistically homogeneous and isotropic log-normal distribution with unknown prior power spectrum. Thus $s$ is again distributed according to the prior process previously given in section \ref{sec:gauprocess}. The point sources follow an inverse-gamma distribution at every point $(x,y)$ of the image plane, given as
\begin{equation}
	P(p_{xy}) = \frac{q^\alpha}{\Gamma(\alpha)}  \left(p_{xy}\right)^{-\alpha -1} \exp\left(-\frac{q}{p_{xy}}\right) \ ,
\end{equation}
where in the particular example we used $(\alpha, q) = (2,3)$. A visualization of the problem setup with the various stages of the observation process is given in figure \ref{fig:poisson_setup}. 

\begin{figure*}[htp]
	\centering
	\includegraphics[scale=1., angle=0]{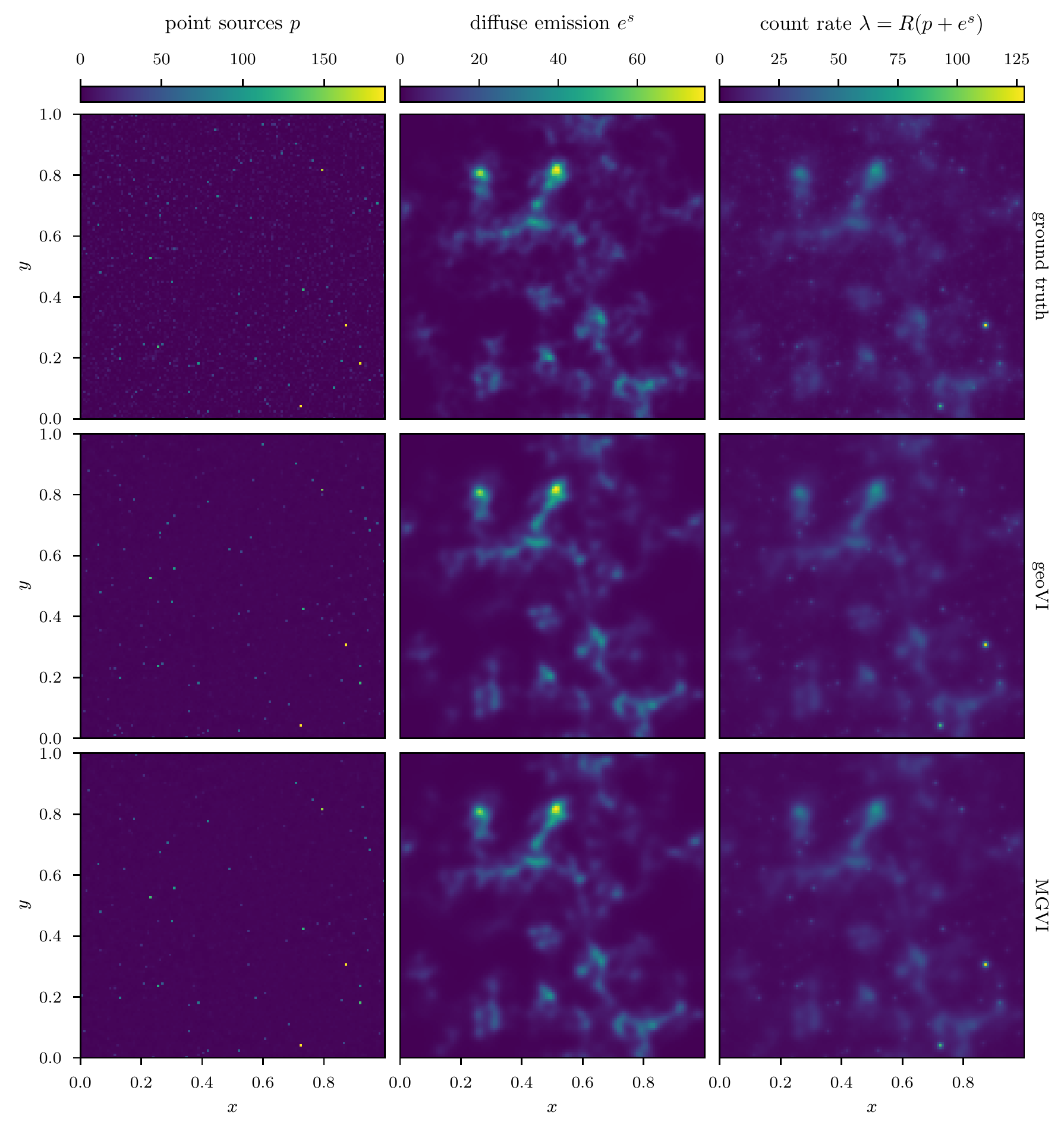}
	\centering
	\caption{Comparison of the ground truth (top row) to the geoVI (middle row) and the MGVI (bottom row) algorithms. The middle and bottom rows show the posterior means for (from left to right) the point sources $p$, the diffuse emission $e^s$, and the count rate $\lambda$.}\label{fig:poisson_signal}
\end{figure*}

\begin{figure*}[htp]
	\centering
	\includegraphics[scale=1., angle=0]{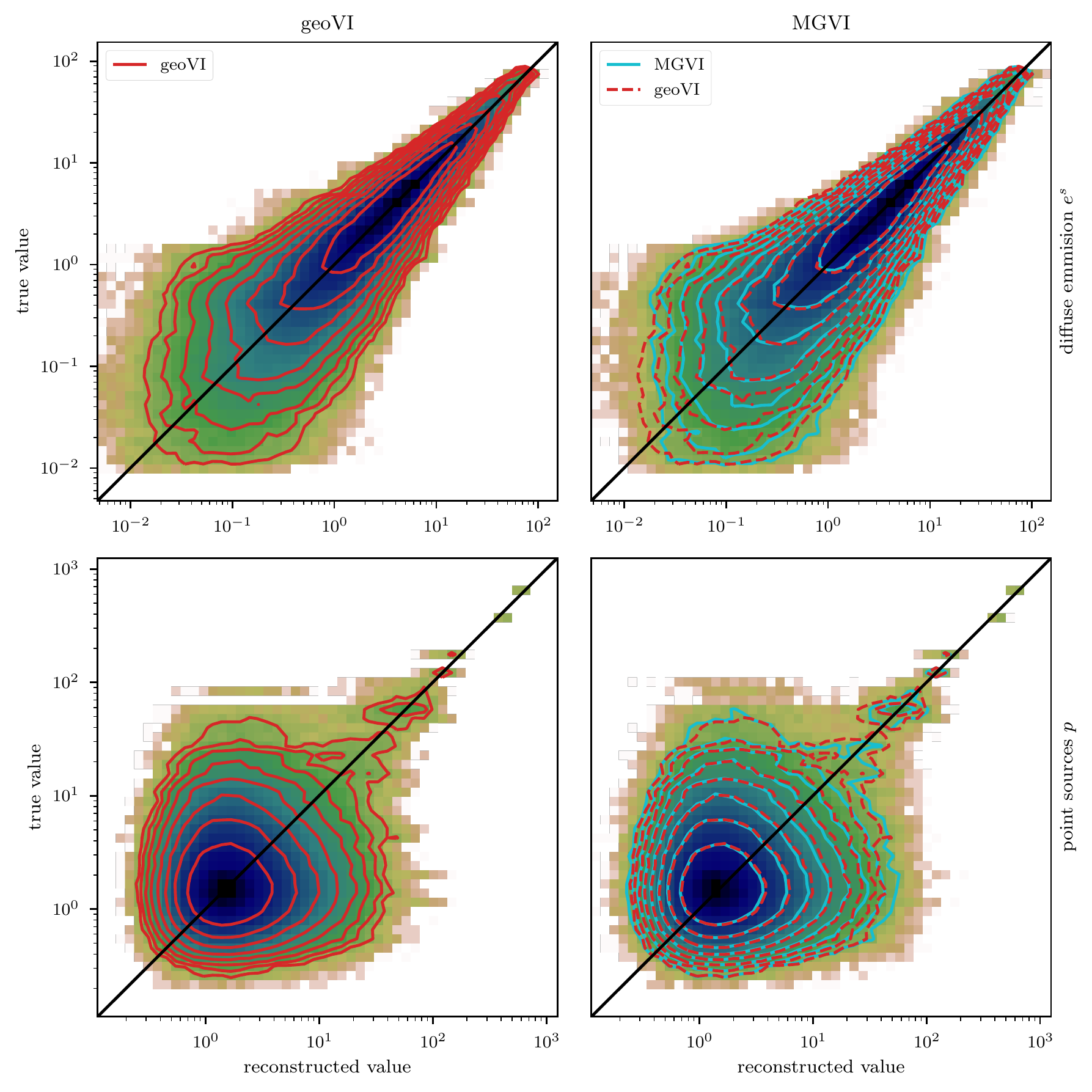}
	\centering
	\caption{Comparison of the per-pixel flux between the ground truth (y-axis) and the reconstruction (x-axis) for the diffuse emission $e^s$ (top row), and the point sources $p$ (bottom row). The left column shows the geoVI result where the density of pixels is color-coded ranging from blue, where the density is highest, to green towards lower densities. The red lines indicate contours of equal density. The right column displays the same for the MGVI reconstruction, with the corresponding density contours now displayed in light blue. The red dashed contours are the density contours of the geoVI case, shown for comparison.}\label{fig:poisson_density}
\end{figure*}

\begin{figure*}[htp]
	\centering
	\includegraphics[scale=1., angle=0]{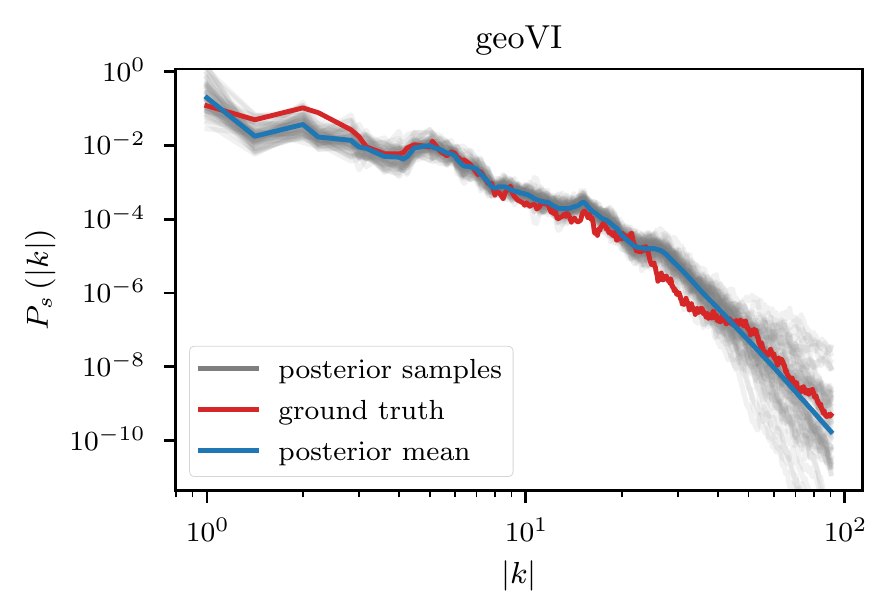}
	\includegraphics[scale=1., angle=0]{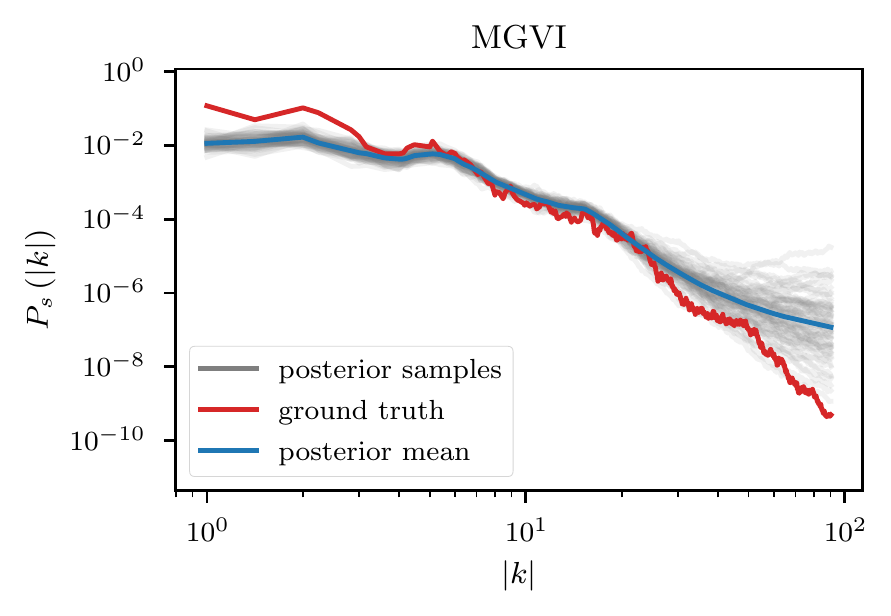}
	\centering
	\caption{Power spectrum $P_s$ of the logarithm of the diffuse emission $s$. The red line is the ground truth, the blue line the posterior mean, and the gray lines a subset of posterior samples for the geoVI (left) and MGVI (right) approximations.}\label{fig:poisson_pspec}
\end{figure*}

We employ the geoVI and MGVI algorithms to infer all, a priori standard distributed, degrees of freedom of the model and recover the power spectrum $P_s$ for the diffuse emission together with its hyper parameters, the realized emission $e^s$ and the point sources $p$, from the Poisson count data $d$. The reconstructed two dimensional images of $p$ and $e^s$ are displayed in figure \ref{fig:poisson_signal} together with the recovered count rate $\lambda$, and compared to their respective ground truth. We find that in this example there is barely a visible difference between the reconstructed diffuse emission of MGVI and geoVI. Both reconstructions are in good agreement with the ground truth. For the point sources, we find that the brightest sources are well recovered by both algorithms, while geoVI manages to infer a few more of the fainter sources as opposed to MGVI. Nevertheless, for both algorithms, the posterior mean does not recover very faint sources present in the true source distribution. This can also be seen in figure \ref{fig:poisson_density}, where we depict the per-pixel flux values for all locations in the image against their reconstructed values, for both, the diffuse emission and the point sources. We find that the MGVI and the geoVI are, on average, in very good agreement. It is noteworthy that the deviations between the true and the reconstructed flux values increases towards smaller values, which is to be expected due to the larger impact of the Poisson noise. For the spatially independent point sources, there appears to be a transition regime around a flux of $\approx 50$, below which point sources become barely detectable. All in all, for the diffuse emission as well as the point sources, both reconstruction methods apparently yield similar results, consistent with the ground truth.
In addition, in figure \ref{fig:poisson_pspec}, we depict the inferred power spectra. We find that the overall shape is reconstructed well by both algorithms, but smaller, more detailed features can only be recovered using geoVI. In addition we find that the statistical properties of the spectrum, as indicated by e.g.\ the roughness of the spectrum as a function of the Fourier modes $|k|$, are well reconstructed by geoVI and in agreement with the true spectrum, whereas the MGVI reconstruction, including the posterior samples, appear to be systematically too smooth compared to the ground truth. As discussed in the previous example in more detail, the parameters that enter the model to determine these properties of the spectrum are highly non-linearly coupled and influenced by the observed data and therefore the linear approximation as used in MGVI becomes, at some point, invalid.

\section{Further properties and challenges}
Aside from the apparent capacity to approximate non-linear and high-dimensional posterior distributions, there are some further properties that can be derived from geoVI and the associated coordinate transformation. In the following, we show how to obtain a lower bound to the evidence using the geoVI results. Furthermore, we outline a way to utilize the coordinate transformation in the context of Hamilton Monte-Carlo (HMC) sampling. Finally, some limitations remain to the approximation capacity of geoVI in its current form, which are discussed in section \ref{sec:pathological}

\subsection{Evidence lower bound (ELBO)}
With the results of the variational approximation at hand, we can provide an Evidence lower bound (ELBO).
To this end consider the Hamiltonian $\mathcal{H}(\xi|d)$ of the posterior which takes the form

\begin{equation}\label{eq:ham_post}
\mathcal{H}(\xi|d) = \mathcal{H}(\xi, d) - \mathcal{H}(d) = \mathcal{H}(d|\xi) + \frac{1}{2} \xi^T\xi + \frac{1}{2} \log\left(\left|2 \pi \mathds{1}\right|\right) - \mathcal{H}(d) \ ,
\end{equation}
and the Hamiltonian of the approximation $\mathcal{H}_Q$ as a function of $r$, given as

\begin{equation}\label{eq:ham_approx}
\mathcal{H}_Q(r|\bar{\xi}) = \frac{1}{2} g(\bar{\xi}+r;\bar{\xi})^T g(\bar{\xi}+r;\bar{\xi}) + \frac{1}{2} \log\left(\left|2 \pi \mathds{1}\right|\right) - \frac{1}{2} \log\left(\left|\tilde{\mathcal{M}}(\bar{\xi}+r)\right|\right) \ .
\end{equation}
Using these Hamiltonians, we may write the variational approximation as

\begin{align}
	\mathrm{KL}(Q;P) &= \left<\mathcal{H}(\xi = \bar{\xi}+r|d)\right>_{Q(r|\bar{\xi})} - \left<\mathcal{H}_Q(r|\bar{\xi})\right>_{Q(r|\bar{\xi})} \notag\\&= \left<\mathcal{H}(\xi = \bar{\xi}+r,d)\right>_{Q(r|\bar{\xi})} - \mathcal{H}(d) - \left<\mathcal{H}_Q(r|\bar{\xi})\right>_{Q(r|\bar{\xi})} \ .
\end{align}
As $\mathcal{H}(d) = - \log(P(d))$, we can derive a lower bound for the logarithmic evidence $P(d)$ using the KL as

\begin{equation}
\log(P(d)) \geq \log(P(d)) - \mathrm{KL}(Q;P) \ ,
\end{equation}
where the lower bound becomes maximal if the KL becomes minimal. Thus we may use our final expansion point $\bar{\xi}$ obtained from minimizing the KL together with the Hamiltonians (equations \eqref{eq:ham_post} and \eqref{eq:ham_approx}) to arrive at

\begin{align}
	&\log(P(d)) - \mathrm{KL}(Q;P) = \notag\\&= \frac{1}{2} \mathrm{tr}(\mathds{1}) - \left<\mathcal{H}(d|\xi = \bar{\xi}+r) + \frac{1}{2} \left(\bar{\xi}+r\right)^T\left(\bar{\xi}+r\right) + \frac{1}{2} \log\left(\left|\tilde{\mathcal{M}}(\bar{\xi}+r)\right|\right) \right>_{Q(r|\bar{\xi})} \notag\\ &\approx \frac{1}{2} \mathrm{tr}(\mathds{1}) - \frac{1}{N} \sum_{i=1}^N \left(\mathcal{H}(d|\xi = \bar{\xi}+r^*_i) + \frac{1}{2} \left(\bar{\xi}+r^*_i\right)^T\left(\bar{\xi}+r^*_i \right) + \frac{1}{2} \log\left(\left|\tilde{\mathcal{M}}(\bar{\xi}+r^*_i)\right|\right)\right) \ ,
\end{align}
where $\left\lbrace r^*_i\right\rbrace_{i \in \lbrace 1, ... , N \rbrace}$ are a set of samples drawn from the approximation $Q(r|\bar{\xi})$. Under the assumption that the log determinant of $\tilde{\mathcal{M}}$ is approximately constant throughout the typical set reached by $Q$, we may replace its sample average with the value obtained at $\bar{\xi}$ to arrive at

\begin{align}\label{eq:elbo_simplified}
	\log(P(d)) - \mathrm{KL}(Q;P) &\approx \frac{1}{2} \mathrm{tr}(\mathds{1}) - \frac{1}{2} \log\left(\left|\bar{\mathcal{M}}\right|\right) \notag\\&- \frac{1}{N} \sum_{i=1}^N \left(\mathcal{H}(d|\xi = \bar{\xi}+r^*_i) + \frac{1}{2} \left(\bar{\xi}+r^*_i\right)^T\left(\bar{\xi}+r^*_i \right) \right) \ ,
\end{align}
where we also replaced the metric of the expansion $\tilde{\mathcal{M}}$ with the metric of the posterior $\mathcal{M}$ as they are identical when evaluated at the expansion point $\bar{\xi}$ (see equation \eqref{eq:metric_consistency}). The assumption that the log determinant does not vary significantly within the typical set is also a requirement for the approximation $Q$ to be a close match for the posterior $P$ and in turn it is a necessary condition for the ELBO to be a tight lower bound to the evidence as only in this case the KL may become small. Therefore equation \eqref{eq:elbo_simplified} is a justified simplification in case the entire variational approximation itself is justified. Nevertheless it may be useful to compute the log determinant also for the posterior samples if feasible, as it provides a valuable consistency check for the method itself.

\subsection{RMHMC with metric approximation}\label{sec:hmc}
As initially discussed in the introduction, aside from Variational inference methods there exist Markov chain Monte-Carlo (MCMC) methods that utilize the geometry of posterior to increase sampling efficiency. A recently introduced Hybrid Monte-Carlo (HMC) method called Riemannian manifold HMC (or RMHMC) utilizes the same posterior metric as discussed in this work in order to define a Riemannian manifold on which the HMC integration is performed. As one of the key results presented here yields an approximate isometry for this manifold, we like to study the impact of the proposed coordinate transformation on RMHMC. To do so, recall that in HMC the random variable $\xi \in \mathds{R}^M$, which is distributed according to a posterior distribution $P(\xi)$, is accompanied by another random variable $p \in \mathds{R}^M$, called momentum, and their joint distribution $P(\xi,p)$ is factorized by means of the posterior $P(\xi)$, and the conditional distribution $P(p|\xi)$. The main idea of HMC is to regard the joint Hamiltonian $\mathcal{H}(\xi,p) = - \log(P(\xi,p))$ as an artificial Hamiltonian system that can be used to construct a new posterior sample from a previous one by following trajectories of the Hamiltonian dynamics. In particular suppose that we are given some random realization $\xi^0$ of $P(\xi)$, we may use the conditional distribution $P(p|\xi^0)$ to generate a random realization $p^0$. Given a pair $(\xi^0,p^0)$, HMC solves the dynamical system associated with the Hamiltonian $\mathcal{H}(\xi,p)$ for some integration time $t^*$, to obtain a new pair $(\xi^*,p^*)$. As Hamiltonian dynamics is both energy and volume preserving by construction, one can show that if $(\xi^0,p^0)$ is a random realization of $P(\xi,p)$, then also $(\xi^*,p^*)$ is. This procedure may be repeated until a desired number of posterior samples is collected.
In practice, the performance of an HMC implementation for a specific distribution $P(\xi)$ strongly depends on the choice of conditional distribution $P(p|\xi)$. To simplify the Hamiltonian trajectories and enable a fast traversion of the posterior, RMHMC has been proposed which utilizes a position dependent metric for the conditional distribution of the momentum which takes the form
\begin{equation}
P(p|\xi) = \mathcal{N}(p; 0, \mathcal{M}(\xi)) \ ,
\end{equation}
where $\mathcal{M}(\xi)$ denotes the metric associated with the posterior $P(\xi)$ as introduced in section \ref{sec:geopost} in equation \eqref{eq:postmetric}. The associated Hamiltonian takes the form

\begin{equation}\label{eq:hamilton_rmhmc}
\mathcal{H}(p, \xi) = \frac{1}{2} p^T \mathcal{M}(\xi)^{-1} p + \frac{1}{2} \log\left(\left|\mathcal{M}(\xi)\right|\right) + \mathcal{H}(\xi) \ .
\end{equation}
In direct analogy of the discussion in section \ref{sec:geopost}, the motivation of utilizing the metric is that the resulting Hamiltonian system can be understood as being defined on the Riemannian manifold associated with $\mathcal{M}$. Therefore the geometric complexity is absorbed into the shape of the manifold, and the trajectories become particularly simple. In practice, however, numerical integration of the system related to equation \eqref{eq:hamilton_rmhmc} is challenging, as in general $\mathcal{H}(p, \xi)$ is non-separable. Here, our coordinate transformation may come in handy, as a Hamiltonian using the approximated metric $\tilde{\mathcal{M}}$ (equation \eqref{eq:approxmet}) instead of $\mathcal{M}$ becomes separable. Specifically replacing $\mathcal{M}$ in equation \eqref{eq:hamilton_rmhmc} yields

\begin{align}\label{eq:hamilton_rmhmc_approx}
	\mathcal{H}(p, \xi) &= \frac{1}{2} p^T \tilde{\mathcal{M}}(\xi;\bar{\xi})^{-1} p + \frac{1}{2} \log\left(\left|\tilde{\mathcal{M}}(\xi;\bar{\xi})\right|\right) + \mathcal{H}(\xi) \notag\\&= \frac{1}{2} p^T \left(\left(\frac{\partial g(\xi; \bar{\xi})}{\partial \xi}\right)^T \frac{\partial g(\xi; \bar{\xi})}{\partial \xi}\right)^{-1} p + \tilde{\mathcal{H}}(\xi; \bar{\xi}) \ .
\end{align}
This modified system allows for a canonical transformation of the form

\begin{equation}
\begin{pmatrix}
y \\ v
\end{pmatrix} \leftarrow \begin{pmatrix}
g(\xi;\bar{\xi}) \\ \left(\frac{\partial g(\xi;\bar{\xi})}{\partial \xi}\right)^T p
\end{pmatrix} \ ,
\end{equation}
in which the Hamiltonian (equation \ref{eq:hamilton_rmhmc_approx}) takes the form

\begin{equation}\label{eq:hamtrafo}
\mathcal{H}(v, y) = \frac{1}{2} v^T v + \left.\tilde{\mathcal{H}}(\xi; \bar{\xi})\right|_{\xi = g^{-1}(y; \bar{\xi})} \equiv T(v) + V(y) \ ,
\end{equation}
and therefore $\mathcal{H}$ is separable in the momenta $v$ and the position $y$. This separability is an interesting property as it has the potential to simplify the integration step used within RMHMC. However, in its current form, we find that there are multiple issues with this approach that prevent an efficient implementation in practice. For one, the transformation $g$ depends on an expansion point $\bar{\xi}$, which becomes a hyper-parameter of the method that has to be determined (possibly in the warm-up phase). In addition, unlike the direct approach discussed in section \ref{sec:direct approx}, we cannot circumvent the inversion of $g$, which is only implicitly available in general, as it has to be computed for every integration step related to $\mathcal{H}(v, y)$ (equation \eqref{eq:hamtrafo}). Therefore, numerical integration of the system may be simpler, but evaluation of $\mathcal{H}(v, y)$ becomes more expensive. Finally, the approximation of the metric may become invalid as we move far away from the expansion point $\bar{\xi}$, and therefore the applicability compared to an RMHMC implementation using the full metric $\mathcal{M}$ is limited. Nevertheless we find the existence of a separable approximation to the Hamiltonian system very interesting, and think that the (or a similar) transformation $g$ and its associated coordinate system $(y,v)$ might be of relevance in the future development of RMHMC algorithms.

\begin{figure*}[htp]
	\includegraphics[scale=1., angle=0]{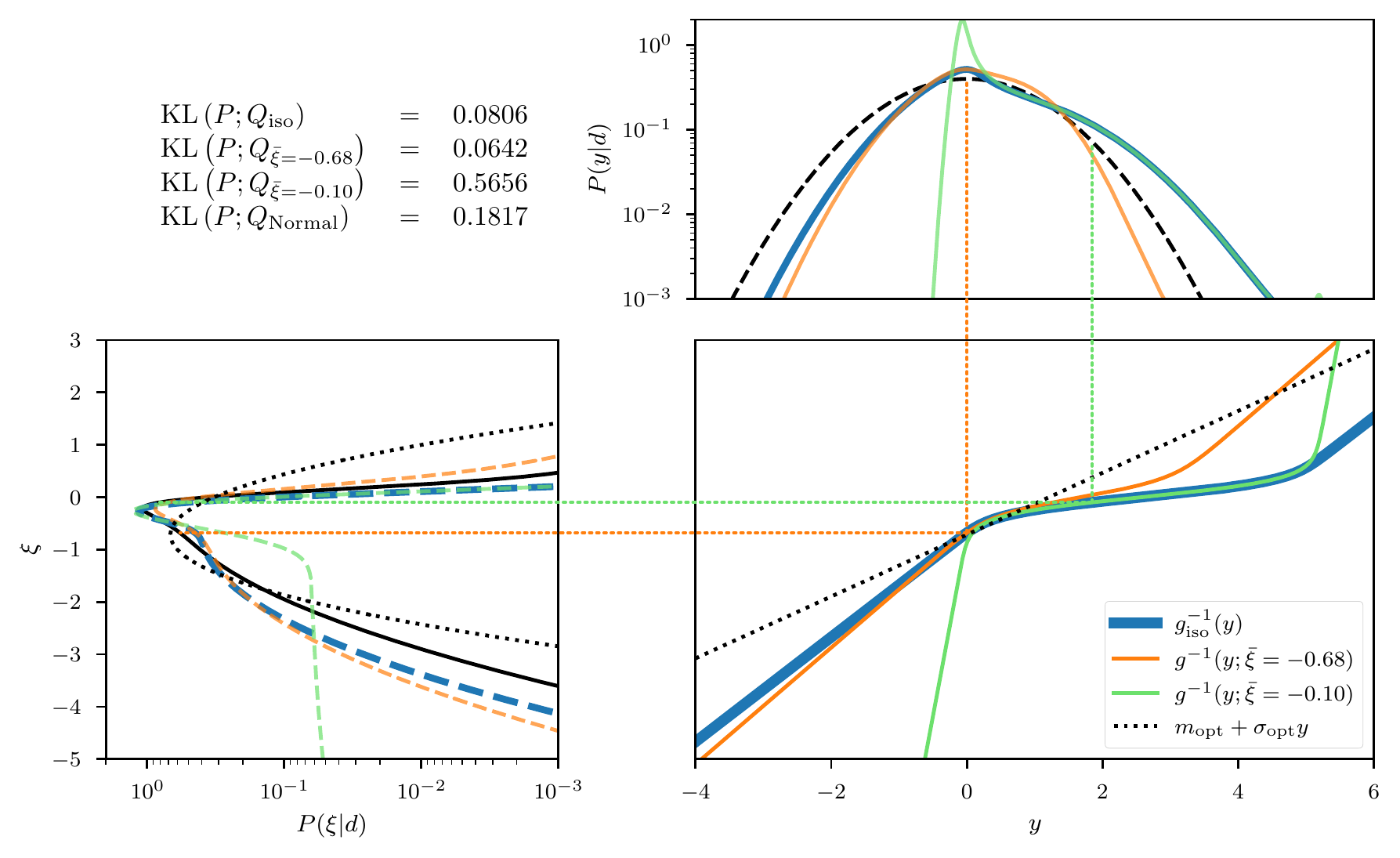}
	\centering
	\caption{Same setup as in figure \ref{fig:lognormal_1d}, but for the sigmoid-normal distributed case. In addition to the exact isometry $g_{\mathrm{iso}}$, the approximation using the optimal expansion point $\bar{\xi} = -0.68$ and a pathological heavy-tail example using $\bar{\xi} = -0.1$ is displayed.}\label{fig:pathological_sigmoid}
\end{figure*}

\subsection{Pathological cases}\label{sec:pathological}
As discussed in section \ref{sec:coord_trafo}, one property that can violate our assumptions are non-monotonic changes in the metric. To this end, consider a sigmoid-normal distributed random variable, and a measurement subject to additive, independent noise of the form
\begin{equation}
P(d|\xi) = \mathcal{N}(d; \sigma\left(\sigma_p \xi\right), \sigma_n^2) \quad \text{with}
\quad P(\xi) = \mathcal{N}(\xi; 0, 1) \ ,
\end{equation}
where $\sigma(\bullet)$ denotes the sigmoid function. The resulting posterior, its associated coordinate transformation, as well as its geoVI approximation, is displayed in figure \ref{fig:pathological_sigmoid} for a case with $(\sigma_p, \sigma_n, d) = (3, 0.2, 0.2)$. We find that similar to the one-dimensional log-normal example of section \ref{sec:simple_demos}, the approximation quality depends on the chosen expansion point. However, the changes in approximation quality are much more drastic as in the log-normal example. In particular, due to the sigmoid non-linearity, there exists a turning point in the coordinate transformation $g$, and if we choose an expansion point close to this point, we see that the approximation to the transformation strongly deviates from the optimal transformation as we move away from this point. As a result, in this case the approximation to the posterior (left panel of figure \ref{fig:pathological_sigmoid}) obtains a heavy tail that is neither present in the true posterior nor the approximation using the optimal transformation. Nevertheless there may very well also exist a case where such a heavy tail is present in the optimal approximation to the transformation. Even in the depicted case, where the tail is only present for sub-optimal choices of the expansion point, an optimization algorithm might have to traverse this sub-optimal region to reach the optimum. Thus the heavy tail can lead to extreme samples for some intermediate approximation, and therefore the geoVI algorithm could become unstable.

\begin{figure*}[ht]
	\includegraphics[scale=1., angle=0]{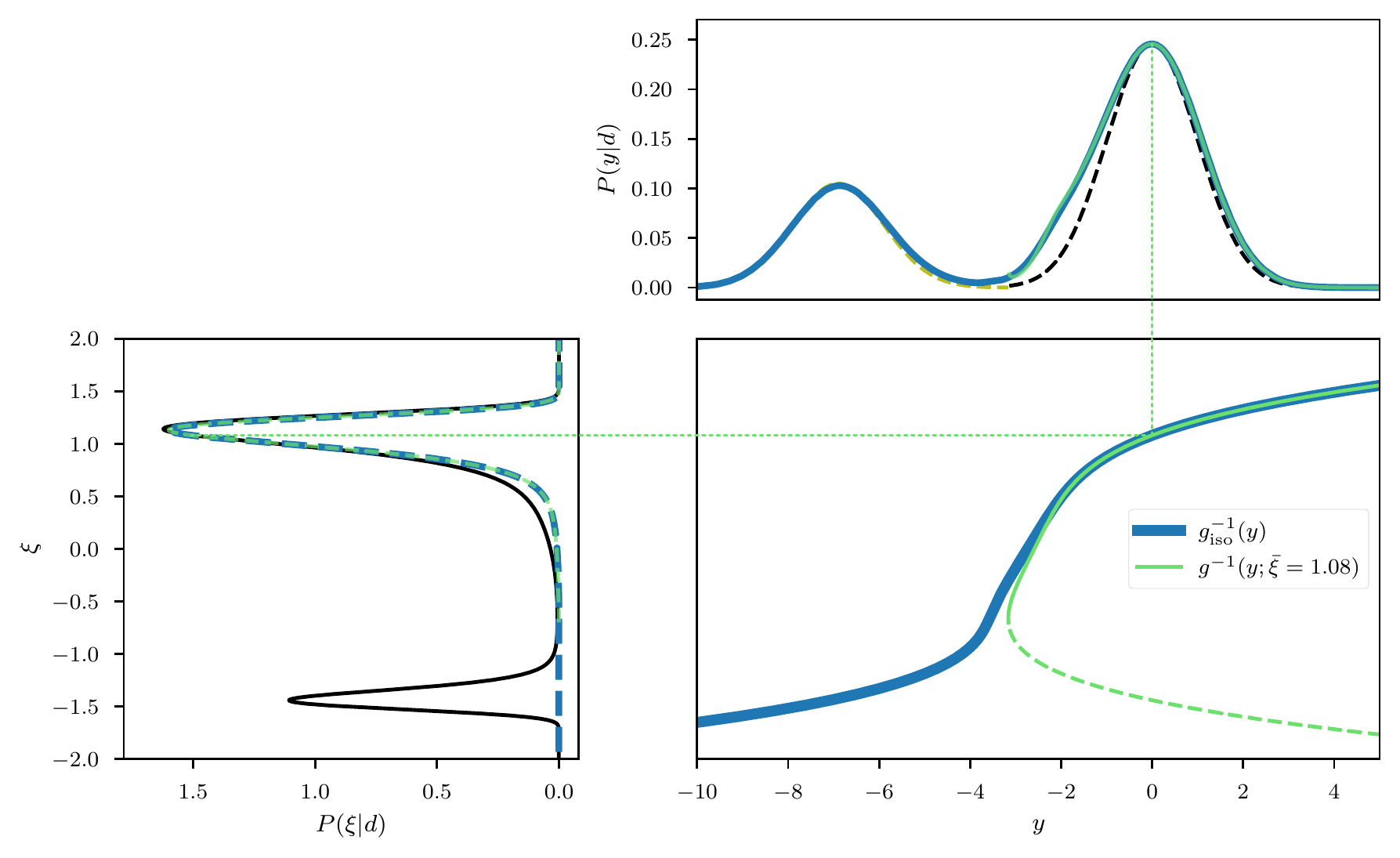}
	\centering
	\caption{Second pathological example, given as a bi-modal posterior distribution. The setup is similar to figures \ref{fig:lognormal_1d} and \ref{fig:pathological_sigmoid}, where in this example only the (locally) optimal expansion point $\bar{\xi}=1.08$ is used.}\label{fig:pathological_multimode}
\end{figure*}

In a second example we consider a bi-modal posterior distribution, generated from a Gaussian measurement of a polynomial. Specifically we consider a likelihood of the form

\begin{equation}
P(d|\xi) = \mathcal{N}\left(d; \xi^4 + \xi, 1\right) \ ,
\end{equation}
with $\xi$ being a priori standard distributed. As can be seen in figure \ref{fig:pathological_multimode}, this scenario leads to a bi-modal posterior distribution with two well separated, asymmetric modes. We find that the geometrically optimal transformation $g_{\mathrm{iso}}$ also leads to a bi-modal distribution in the transformed coordinates, however the local asymmetry and curvature of each mode has approximately been removed. Thus while an approximation of the posterior by means of a single unit Gaussian distribution is apparently not possible, each mode may be approximated individually, at least in case the modes are well separated.
If we consider the approximation of the coordinate transformation used within geoVI, and choose as an expansion point the optimal point associated with one of the two modes, we get that for the chosen mode the approximation remains valid and the transformation is close to the optimal transformation. However, if we move away from the mode towards the other mode, the approximation quickly deviates from $g_{\mathrm{iso}}$ and eventually becomes non-invertible. Therefore only the approximation of one of the modes is possible. Here, care must be taken, as in practical applications the inversion of $g$ is performed numerically and one has to ensure that the inversion does not end up on the second branch of the transformation.

This summarizes the two main issues that may render a geoVI approximation of a posterior distribution invalid. The challenges and issues related to multi modality appear to be quite fundamental, as in its current form, the geoVI method falls into the category of methods that utilize local information of the posterior which all suffer from the inability to deal with more than a single mode. The problems related to turning points are more specific for geoVI, and its implications need to be further studied in order to generalize its range of applicability in the future. One promising finding is that this issue appears to be solely related to the local approximation of the transformation with a ``bad'' expansion point, as the geometrically optimal transformation $g_{\mathrm{iso}}$ apparently does not show such behavior. Therefore an extension of the current approximation technique using e.g.\ multiple expansion points, or identifying and excluding these ``bad'' expansion points during optimization, may provide a solution to this problem. At the current stage of the development, however, it is unclear how to incorporate such ideas into the algorithm without loss of the functional form of $g$ that allows for the numerically efficient implementation at hand.

\section{Summary and Outlook}
In this work we introduced a coordinate transformation for posterior distributions that yields a coordinate system in which the distributions take a geometrically simple form. In particular we construct a metric as the sum of the Fisher metric of the likelihood and the identity matrix for a standard prior distribution, and construct the transformation that relates this metric to the Euclidean metric. Using this transformed coordinate system, we introduce geometric Variational Inference (geoVI), where we perform a variational approximation of the transformed posterior distribution with a normal distribution with unit covariance. As the coordinate transformation is only approximately available and utilizes an expansion point around which it is most accurate, the VI task reduces to finding the optimal expansion point such that the variational KL between the true posterior and the approximation becomes minimal. There exists a numerically efficient realization that enables high-dimensional applications of geoVI because even though the transformation is non-volume preserving, geoVI avoids a computation of the related log-determinant of the Jacobian of the transformation at any point. The expansion point used to generate intermediate samples is only passively updated. Furthermore, the application of the constructed coordinate transformation is similar to the cost of computing the gradient of the posterior Hamiltonian. In addition, to generate random realizations, computing the appearing matrix square root of the metric can be entirely avoided, and the inverse transformation is achieved implicitly by second order numerical inversion.

Despite being an approximation method, we find that geoVI is successfully applicable in non-linear, but uni-modal settings, which we demonstrated with multiple examples. We see that non-linear features of the posterior distribution can accurately be captured by the coordinate transformation in low-dimensional examples. This property may translate into high dimensions, as it increases the overall reconstruction quality there when compared to its linearized version MGVI. Nevertheless we also find remaining pathological cases in which further development is necessary to achieve a good approximation quality.

In addition to posterior approximation, geoVI results can be used in order to provide an evidence lower bound (ELBO) which is used for model comparison. Finally we demonstrate the overlap to another posterior sampling technique based on Hamilton Monte-Carlo (HMC), that utilizes the same metric used in geoVI, called Riemannian manifold HMC.

All in all, the geoVI algorithm, and more generally the constructed approximative coordinate transformation, are a fast and accurate way to approximate non-linear and high-dimensional posterior distributions.

\vspace{6pt} 



\authorcontributions{Conceptualization, Philipp Frank and Reimar Leike; methodology, Philipp Frank and Reimar Leike; software, Philipp Frank; validation, Reimar Leike and Torsten En{\ss}lin; writing---original draft preparation, Philipp Frank; writing---review and editing, Reimar Leike and Torsten En{\ss}lin; visualization, Philipp Frank; supervision, Torsten En{\ss}lin.; project administration, Torsten En{\ss}lin. All authors have read and agreed to the published version of the manuscript.}

\funding{This research received no external funding}

\acknowledgments{The authors would like to thank Philipp Arras for his detailed feedback on the manuscript, Sebastian Hutschenreuther for his hands on feedback to the early versions of geoVI, Jakob Knollm{\"u}ller for the development of the MGVI algorithm, and Martin Reinecke for his contributions to {\tt NIFTy}.}

\conflictsofinterest{The authors declare no conflict of interest.}


\abbreviations{Abbreviations}{The following abbreviations are used in this manuscript:\\

\noindent 
\begin{tabular}{@{}ll}
VI & Variational Inference\\
MCMC & Markov-Chain Monte-Carlo\\
geoVI & geometric Variational Inference\\
VB & Variational Bayes'\\
RMHMC & Riemannian manifold Hamilton Monte-Carlo\\
HMC & Hamilton (Hybrid) Monte-Carlo\\
MGVI & Metric Gaussian Variational Inference\\
{\tt NIFTy} & Numerical Information Field Theory\\
KL & Kullback-Leibler divergence\\
MVP & matrix vector product\\
MAP & Maximum a posterior\\
DFT & Discrete Fourier transform\\
FFT & Fast Fourier transform\\
ELBO & Evidence lower bound
\end{tabular}}

\appendixtitles{yes} 
\appendixstart
\appendix
\section{Likelihood transformations}\label{ap:isolhs}
In order to construct the coordinate transformation $x(\xi)$ introduced in section \ref{sec:coord_trafo}, we require that the Fisher metric of the likelihood $\mathcal{M}_{d|\xi}$ may be written as the pullback of the Euclidean metric. Recall that the likelihood expressed in coordinates $\xi$ is obtained from the likelihood $P(d|s')$ with $s' = f'(\xi)$ (see equation \eqref{eq:likelihood}). Therefore we may express $\mathcal{M}_{d|\xi}$ as

\begin{equation}
\mathcal{M}_{d|\xi}(\xi) = \left(\frac{\partial s'}{\partial \xi}\right)^T \mathcal{M}_{d|s'} \frac{\partial s'}{\partial \xi} \ .
\end{equation}
Thus the task reduces to construct a transformation $x(s')$ that recovers $\mathcal{M}_{d|s'}$ from the Euclidean metric if we set the full transformation to be $x(\xi) \equiv x(s'=f'(\xi))$.
Specifically we require for $x(s')$

\begin{equation}\label{eq:metlh}
	\mathcal{M}_{d|s'} \overset{!}{=} \left(\frac{\partial x}{\partial s'}\right)^T \frac{\partial x}{\partial s'} \ .
\end{equation}
Below, in table \ref{table:isolhs}, we give a summary of multiple commonly used likelihoods, their respective Fisher metric, and the associated transformation $x(s')$.

\begin{table}[htp]
\begin{tabular}{lccc}
	Name & $\mathcal{H}(d|s')$ & Metric $\mathcal{M}$ & Trafo. $x(s')$ \\
	Normal & $\frac{1}{2} (d-s')^T N^{-1} (d-s') + \mathrm{cst.}$ & $N^{-1}$ & $\sqrt{N^{-1}} s'$ \\
	Poisson & $1^T s' - d^T \log(s') + \mathrm{cst.}$&$\nicefrac{1}{s'}$ & $\frac{1}{2} \sqrt{s'}$ \\
	Inv. Gamma & $(\alpha+1)^T \log(s') + \beta^T \left(\frac{1}{s'}\right) + \mathrm{cst.}$ &$\frac{\alpha +1}{s'^2}$ & $\sqrt{\alpha+1} \log\left(s'\right)$ \\
	Student-T & $\frac{\theta + 1}{2} \log\left(1 + \frac{s'^2}{\theta}\right) + \mathrm{cst.}$&$\frac{\theta+1}{\theta+3}$ & $\sqrt{\frac{\theta+1}{\theta+3}} s'$ \\
	Bernoulli & $-d^T \log(s') - (1-d)^T \log(1-s') + \mathrm{cst.}$&$\frac{1}{s'(1-s')}$ & $-2 \tan^{-1}\left(\sqrt{s'}\right)$\\
\end{tabular}
\caption{List of common likelihood distributions with their respective Hamiltonian $\mathcal{H}(d|s')$, their Fisher Metric $\mathcal{M}(d|s')$, and the associated coordinate transformation $x(s')$ satisfying equation \eqref{eq:metlh}.}\label{table:isolhs}
\end{table}

For some likelihoods, however, such a decomposition is not accessible in a simple form. One example that is being used in this work is a normal distribution with unknown mean $m$ and variance $v$. The Hamiltonian of a one dimensional example takes the form

\begin{equation}
	\mathcal{H}(d|m,v) = \frac{1}{2} \frac{(d-m)^2}{v} + \frac{1}{2} \log(v) + \mathrm{cst.} \ ,
\end{equation}
and the corresponding fisher metric for $s' = (m,v)$ is

\begin{equation}
	\mathcal{M}_{d|s'} = \begin{pmatrix}
	\frac{1}{v} & 0 \\ 
	0 & \frac{1}{2 v^2} 
	\end{pmatrix} \ .
\end{equation}
While there is no simple decomposition by means of the Jacobian of some function $x$, there is an approximation available for which $x$ takes the form

\begin{equation}
	x(s') = \begin{pmatrix}
	\frac{d-m}{\sqrt{v}} \\
	\frac{1}{2} \log(v)
	\end{pmatrix} \quad \text{with} \quad \frac{\partial x}{\partial s'} = \begin{pmatrix}
	-\frac{1}{\sqrt{v}} & - \frac{d-m}{2 v^{\nicefrac{3}{2}}} \\
	0 & \frac{1}{2 v}
	\end{pmatrix} \ .
\end{equation}
We can compute the approximation to the metric and find

\begin{equation}
	\left(\frac{\partial x}{\partial s'}\right)^T \frac{\partial x}{\partial s'} = \begin{pmatrix}
	\frac{1}{v} & \frac{d-m}{2 v^2} \\
	\frac{d-m}{2 v^2} & \frac{1}{4 v^2} + \frac{(d-m)^2}{4 v^2}
	\end{pmatrix} \ .
\end{equation}
Note that as opposed to the Fisher metric, this approximation depends on the observed data $d$. In fact we can recover the Fisher metric from this approximation by taking the expectation value w.r.t.\ the likelihood. Specifically

\begin{equation}
	\left< \left(\frac{\partial x}{\partial s'}\right)^T \frac{\partial x}{\partial s'} \right>_{\mathcal{N}(d; m,v)} = \begin{pmatrix}
	\frac{1}{v} & 0 \\
	0 & \frac{1}{2 v^2}
	\end{pmatrix} = \mathcal{M}_{d|s'} \ ,
\end{equation}
and therefore it may be regarded as a local approximation using the observed data. All examples of this work that use a normal distribution where in addition to the mean also the variance is inferred, use this approximation.

\subsection{Multiple likelihoods}
In general, we may encounter measurement situations where multiple likelihoods are involved, e.g.\ if we aim to constrain $s'$ with multiple data-sets simultaneously. Specifically consider a set of $D$ data-sets $\left\lbrace d_i \right\rbrace_{i \in \lbrace 1, ... , D \rbrace}$, and an associated mutually independent set of likelihoods, such that the joint likelihood takes the form

\begin{equation}
	P(d_1, ..., d_D | s') = \prod_{i=1}^{D} P(d_i|s') \ ,
\end{equation}
we get that the corresponding Fisher metric takes the form

\begin{equation}
	\mathcal{M}_{d_1,...,d_D|s'}(s') = \sum_{i=1}^D \mathcal{M}_{d_i|s'} \ .
\end{equation}
If we assume that we have, for every individual metric $\mathcal{M}_{d_i|s'}$, an associated transformation $x_i(s')$ available that satisfies equation \eqref{eq:metlh}, we see that we can stack them together to form a combined transformation

\begin{equation}
	x(s') \equiv \left(x_1(s'), ..., x_D(s')\right)^T \ ,
\end{equation}
that automatically satisfies \eqref{eq:metlh} for the joint metric $\mathcal{M}_{d_1,...,d_D|s'}$.

\section{Correlated Field model}\label{ap:cf_model}
Here, we give a brief description of the generative model for power spectra and resulting Gaussian processes used in section \ref{sec:applications}. For a detailed and extended derivation please refer to \cite{arras2020variable}.

A random realization $s \in \mathcal{L}(\Lambda)$ of a statistically homogeneous and isotropic Gaussian process $P(s)$, defined over an $L$ dimensional domain $\Lambda = \left[0,1\right]^L$, subject to periodic boundary conditions along each dimension, may be represented as a Fourier series via

\begin{equation}\label{eq:fourierseries}
	s_x = \left(\mathcal{F}^\dagger A \xi \right)_x \equiv \sum_{k} e^{- 2 \pi i k x} A(|k|) \ \xi_k \quad \text{with} \quad \xi_k \sim \mathcal{N}(\xi; 0,1) \ \forall k \ ,
\end{equation}
where $k = \left(k_1, ..., k_L\right) \in \mathcal{Z}^L$ is a multi-index labeling the individual Fourier components, and $|k|$ denotes its Euclidean norm. In order to discretize $s$ on a computer, we may truncate this Fourier series, i.E.\ by replacing the infinite index $k$ with a finite index that truncates at some maximal $k_{\mathrm{max}}$. The operator $\mathcal{F}$ denotes the Fourier transformation and $\mathcal{F}^\dagger$ its corresponding back-transformation (or their discrete versions in case of truncation). The so-called amplitude spectrum $A$ may be identified with the square root of the power-spectrum $P_s$ of the process (specifically $P_s$ being the eigen-spectrum of the linear operator associated with the covariance of the prior probability $P(s)$). Therefore we proceed to construct a model for $A$ rather then $P_s$ as it is more convenient for a generative model. The non-parametric prior model for $A$ is largely built on the assumption that power spectra (and therefore also amplitude spectra) do not vary arbitrarily for similar $|k|$, which in turn allows us to assume that the values of $A$ are, to some degree, correlated. A prominent example of a physically plausible spectrum is a power-law $P_s = |k|^\alpha$ and therefore it turns out to be more convenient to represent $A$ on a log-log-scale, specifically

\begin{equation}
\tau_l \equiv \left.\log\left(A(|k|)\right)\right|_{|k| = e^l} \ ,
\end{equation}
since power-laws become straight lines on these scales. As $k$ is a regularly spaced index, the new index $l = \log(|k|)$ is an irregularly spaced index starting from the smallest non-zero mode labeled as $l_0$ (the origin with $|k|=0$ is treated separately). To exploit correlations in the prior of $\tau_l$, we define a random process $\tilde{\tau}(l)$ over a continuous domain $O = [l_0, \infty)$ ($O = [l_0,l_{\mathrm{max}} = \log(|k_{\mathrm{max}}|)]$ in the truncated case), and evaluate this process on the irregularly spaced locations on which $\tau_l$ is defined. The prior process used for $\tilde{\tau}$ is a Gauss-Markov process given in terms of a linear stochastic differential equation of the form

\begin{equation}
\frac{\partial}{\partial l} \begin{pmatrix} \tilde{\tau}(l) \\ y(l) \end{pmatrix} + 
\begin{pmatrix}
0 & -1 \\ 
0 & 0
\end{pmatrix} \begin{pmatrix} \tilde{\tau}(l) \\ y(l) \end{pmatrix}
= \sigma \begin{pmatrix} \epsilon \ \eta_l \\ 
\xi_l
\end{pmatrix} \ , \quad \text{with} \quad \nicefrac{\eta_l}{\xi_l} \sim \mathcal{N}(\nicefrac{\eta_l}{\xi_l}; 0, 1) \ \forall l \in O \ .
\end{equation}
A Markov process can easily be realized on an irregular grid utilizing its transition probability which in this case takes the form

\begin{equation}
	\left.P\left(\begin{pmatrix} \tau_l \\ y_l \end{pmatrix}\right| \begin{pmatrix} \tau_{l_0} \\ y_{l_0} \end{pmatrix}\right) = \mathcal{N}\left(\begin{pmatrix} \tau_l \\ y_l \end{pmatrix}; \begin{pmatrix} 1 & \Delta_l \\ 0 & 1 \end{pmatrix} \begin{pmatrix} \tau_{l_0} \\ y_{l_0} \end{pmatrix}, \sigma^2 \begin{pmatrix}
	\nicefrac{\Delta_l^3}{3} + \epsilon^2 \Delta_l & \nicefrac{\Delta_l^2}{2} \\
	\nicefrac{\Delta_l^2}{2} & \Delta_l
	\end{pmatrix}\right) \
\end{equation}
with $\Delta_l = l-l_0$. We notice that in absence of stochastic deviations (e.g.\ if $\sigma=0$), the solution is a straight line with slope $y_{l_0}$ that determines the exponent of the power-law, and therefore becomes a variable of the model on which we place a Gaussian prior with a negative prior mean (to a priori favor falling power laws). The offset $\tau_{l_0}$ becomes, after exponentiation, an overall scaling factor that sets the variance of the stochastic process $s$. Thus $\tau_{l_0}$ (specifically its exponential) is also a variable of the model which we refer to as ``fluctuations'' in Table \ref{table:cf_params}. Similarly, the zero-mode (i.E.\ $A(|k|=0)$), which is not included in $\tau$, is set to be a log-normal distributed random variable which we refer to as ``offset std.''.
Finally $\sigma$ (named flexibility) and $\epsilon$ (named asperity) become both log-normal distributed variables (again see table \ref{table:cf_params}) that determine the variance and shape of the deviations of $\tau$ from a straight line (i.E.\ the deviations of $A$ from a power-law).

\end{paracol}
\reftitle{References}


\externalbibliography{yes}
\bibliography{main.bib}

\begin{thebibliography}{999}

\bibitem[Geyer(1992)]{geyer1992practical}
Geyer, C.J.
\newblock Practical markov chain monte carlo.
\newblock {\em Statistical science} {\bf 1992}, pp. 473--483.

\bibitem[Brooks \em{et~al.}(2011)Brooks, Gelman, Jones, and
  Meng]{brooks2011handbook}
Brooks, S.; Gelman, A.; Jones, G.; Meng, X.L.
\newblock {\em Handbook of markov chain monte carlo}; CRC press,  2011.

\bibitem[Marjoram \em{et~al.}(2003)Marjoram, Molitor, Plagnol, and
  Tavar{\'e}]{marjoram2003markov}
Marjoram, P.; Molitor, J.; Plagnol, V.; Tavar{\'e}, S.
\newblock Markov chain Monte Carlo without likelihoods.
\newblock {\em Proceedings of the National Academy of Sciences} {\bf 2003},
  {\em 100},~15324--15328.

\bibitem[Blei \em{et~al.}(2017)Blei, Kucukelbir, and
  McAuliffe]{blei2017variational}
Blei, D.M.; Kucukelbir, A.; McAuliffe, J.D.
\newblock Variational inference: A review for statisticians.
\newblock {\em Journal of the American statistical Association} {\bf 2017},
  {\em 112},~859--877.

\bibitem[Hoffman \em{et~al.}(2013)Hoffman, Blei, Wang, and
  Paisley]{hoffman2013stochastic}
Hoffman, M.D.; Blei, D.M.; Wang, C.; Paisley, J.
\newblock Stochastic variational inference.
\newblock {\em Journal of Machine Learning Research} {\bf 2013}, {\em 14}.

\bibitem[Rezende and Mohamed(2015)]{rezende2015variational}
Rezende, D.; Mohamed, S.
\newblock Variational inference with normalizing flows.
\newblock  International Conference on Machine Learning. PMLR,  2015, pp.
  1530--1538.

\bibitem[Kucukelbir \em{et~al.}(2017)Kucukelbir, Tran, Ranganath, Gelman, and
  Blei]{kucukelbir2017automatic}
Kucukelbir, A.; Tran, D.; Ranganath, R.; Gelman, A.; Blei, D.M.
\newblock Automatic differentiation variational inference.
\newblock {\em The Journal of Machine Learning Research} {\bf 2017}, {\em
  18},~430--474.

\bibitem[Kingma and Welling(2013)]{kingma2013auto}
Kingma, D.P.; Welling, M.
\newblock Auto-encoding variational bayes.
\newblock {\em arXiv preprint arXiv:1312.6114} {\bf 2013}.

\bibitem[Fox and Roberts(2012)]{fox2012tutorial}
Fox, C.W.; Roberts, S.J.
\newblock A tutorial on variational Bayesian inference.
\newblock {\em Artificial intelligence review} {\bf 2012}, {\em 38},~85--95.

\bibitem[{\v{S}}m{\'\i}dl and Quinn(2006)]{vsmidl2006variational}
{\v{S}}m{\'\i}dl, V.; Quinn, A.
\newblock {\em The variational Bayes method in signal processing}; Springer
  Science \& Business Media,  2006.

\bibitem[Girolami and Calderhead(2011)]{rmhmc}
Girolami, M.; Calderhead, B.
\newblock Riemann manifold Langevin and Hamiltonian Monte Carlo methods.
\newblock {\em Journal of the Royal Statistical Society: Series B (Statistical
  Methodology)} {\bf 2011}, {\em 73},~123--214,
  \href{http://xxx.lanl.gov/abs/https://rss.onlinelibrary.wiley.com/doi/pdf/10.1111/j.1467-9868.2010.00765.x}{{\normalfont
  [https://rss.onlinelibrary.wiley.com/doi/pdf/10.1111/j.1467-9868.2010.00765.x]}}.
\newblock
  doi:{\changeurlcolor{black}\href{https://doi.org/https://doi.org/10.1111/j.1467-9868.2010.00765.x}{\detokenize{https://doi.org/10.1111/j.1467-9868.2010.00765.x}}}.

\bibitem[Duane \em{et~al.}(1987)Duane, Kennedy, Pendleton, and
  Roweth]{duane1987hybrid}
Duane, S.; Kennedy, A.D.; Pendleton, B.J.; Roweth, D.
\newblock Hybrid monte carlo.
\newblock {\em Physics letters B} {\bf 1987}, {\em 195},~216--222.

\bibitem[Betancourt(2017)]{betancourt2017conceptual}
Betancourt, M.
\newblock A conceptual introduction to Hamiltonian Monte Carlo.
\newblock {\em arXiv preprint arXiv:1701.02434} {\bf 2017}.

\bibitem[Saha \em{et~al.}(2020)Saha, Bharath, and
  Kurtek]{doi:10.1080/01621459.2019.1585253}
Saha, A.; Bharath, K.; Kurtek, S.
\newblock A Geometric Variational Approach to Bayesian Inference.
\newblock {\em Journal of the American Statistical Association} {\bf 2020},
  {\em 115},~822--835,
  \href{http://xxx.lanl.gov/abs/https://doi.org/10.1080/01621459.2019.1585253}{{\normalfont
  [https://doi.org/10.1080/01621459.2019.1585253]}}.
\newblock PMID: 33041402,
  doi:{\changeurlcolor{black}\href{https://doi.org/10.1080/01621459.2019.1585253}{\detokenize{10.1080/01621459.2019.1585253}}}.

\bibitem[Knollm{\"u}ller and En{\ss}lin(2019)]{knollmuller2019metric}
Knollm{\"u}ller, J.; En{\ss}lin, T.A.
\newblock Metric gaussian variational inference.
\newblock {\em arXiv preprint arXiv:1901.11033} {\bf 2019}.

\bibitem[Arras \em{et~al.}(2019)Arras, Baltac, Ensslin, Frank, Hutschenreuter,
  Knollmueller, Leike, Newrzella, Platz, Reinecke, et~al.]{arras2019nifty5}
Arras, P.; Baltac, M.; Ensslin, T.A.; Frank, P.; Hutschenreuter, S.;
  Knollmueller, J.; Leike, R.; Newrzella, M.N.; Platz, L.; Reinecke, M.;
  others.
\newblock Nifty5: Numerical information field theory v5.
\newblock {\em Astrophysics Source Code Library} {\bf 2019}, pp. ascl--1903.

\bibitem[Bogachev \em{et~al.}(2005)Bogachev, Kolesnikov, and
  Medvedev]{bogachev2005triangular}
Bogachev, V.I.; Kolesnikov, A.V.; Medvedev, K.V.
\newblock Triangular transformations of measures.
\newblock {\em Sbornik: Mathematics} {\bf 2005}, {\em 196},~309.

\bibitem[Fisher(1925)]{fisher1925theory}
Fisher, R.A.
\newblock Theory of statistical estimation.
\newblock  Mathematical Proceedings of the Cambridge Philosophical Society.
  Cambridge University Press,  1925, Vol.~22, pp. 700--725.

\bibitem[Rao(1992)]{rao1992information}
Rao, C.R.
\newblock Information and the accuracy attainable in the estimation of
  statistical parameters. In {\em Breakthroughs in statistics}; Springer,
  1992; pp. 235--247.

\bibitem[Amari and Nagaoka(2000)]{amari2000methods}
Amari, S.; Nagaoka, H.
\newblock {\em Methods of Information Geometry}; Translations of mathematical
  monographs, American Mathematical Society,  2000.

\bibitem[Cencov(2000)]{cencov2000statistical}
Cencov, N.N.
\newblock {\em Statistical decision rules and optimal inference}; Number~53,
  American Mathematical Soc.,  2000.

\bibitem[Betancourt(2013)]{betancourt2013general}
Betancourt, M.
\newblock A general metric for Riemannian manifold Hamiltonian Monte Carlo.
\newblock  International Conference on Geometric Science of Information.
  Springer,  2013, pp. 327--334.

\bibitem[Nocedal and Wright(2006)]{NoceWrig06}
Nocedal, J.; Wright, S.J.
\newblock {\em Numerical Optimization}, second ed.; Springer: New York, NY,
  USA,  2006; p. 168.

\bibitem[Hestenes \em{et~al.}(1952)Hestenes, Stiefel,
  et~al.]{hestenes1952methods}
Hestenes, M.R.; Stiefel, E.; others.
\newblock {\em Methods of conjugate gradients for solving linear systems};
  Vol.~49, NBS Washington, DC,  1952.

\bibitem[Hutchinson(1989)]{hutchinson1989stochastic}
Hutchinson, M.F.
\newblock A stochastic estimator of the trace of the influence matrix for
  Laplacian smoothing splines.
\newblock {\em Communications in Statistics-Simulation and Computation} {\bf
  1989}, {\em 18},~1059--1076.

\bibitem[Han \em{et~al.}(2015)Han, Malioutov, and Shin]{han2015large}
Han, I.; Malioutov, D.; Shin, J.
\newblock Large-scale log-determinant computation through stochastic Chebyshev
  expansions.
\newblock  International Conference on Machine Learning. PMLR,  2015, pp.
  908--917.

\bibitem[Ubaru \em{et~al.}(2017)Ubaru, Chen, and Saad]{ubaru2017fast}
Ubaru, S.; Chen, J.; Saad, Y.
\newblock Fast estimation of tr(f(a)) via stochastic lanczos quadrature.
\newblock {\em SIAM Journal on Matrix Analysis and Applications} {\bf 2017},
  {\em 38},~1075--1099.

\bibitem[Fitzsimons \em{et~al.}(2017)Fitzsimons, Granziol, Cutajar, Osborne,
  Filippone, and Roberts]{fitzsimons2017entropic}
Fitzsimons, J.; Granziol, D.; Cutajar, K.; Osborne, M.; Filippone, M.; Roberts,
  S.
\newblock Entropic trace estimates for log determinants.
\newblock  Joint European Conference on Machine Learning and Knowledge
  Discovery in Databases. Springer,  2017, pp. 323--338.

\bibitem[Hutschenreuter \em{et~al.}(2021)Hutschenreuter, Anderson, Betti,
  Bower, Brown, Br{\"u}ggen, Carretti, Clarke, Clegg, Costa,
  et~al.]{hutschenreuter2021galactic}
Hutschenreuter, S.; Anderson, C.S.; Betti, S.; Bower, G.C.; Brown, J.A.;
  Br{\"u}ggen, M.; Carretti, E.; Clarke, T.; Clegg, A.; Costa, A.; others.
\newblock The Galactic Faraday rotation sky 2020.
\newblock {\em arXiv preprint arXiv:2102.01709} {\bf 2021}.

\bibitem[Welling \em{et~al.}(2021)Welling, Frank, En{\ss}lin, and
  Nelles]{welling2021reconstructing}
Welling, C.; Frank, P.; En{\ss}lin, T.; Nelles, A.
\newblock Reconstructing non-repeating radio pulses with Information Field
  Theory.
\newblock {\em Journal of Cosmology and Astroparticle Physics} {\bf 2021}, {\em
  2021},~071.

\bibitem[Arras \em{et~al.}(2021)Arras, Bester, Perley, Leike, Smirnov,
  Westermann, and En{\ss}lin]{arras2021comparison}
Arras, P.; Bester, H.L.; Perley, R.A.; Leike, R.; Smirnov, O.; Westermann, R.;
  En{\ss}lin, T.A.
\newblock Comparison of classical and Bayesian imaging in radio
  interferometry-Cygnus A with CLEAN and resolve.
\newblock {\em Astronomy \& Astrophysics} {\bf 2021}, {\em 646},~A84.

\bibitem[Arras \em{et~al.}(2020)Arras, Frank, Haim, Knollm{\"u}ller, Leike,
  Reinecke, and En{\ss}lin]{arras2020variable}
Arras, P.; Frank, P.; Haim, P.; Knollm{\"u}ller, J.; Leike, R.; Reinecke, M.;
  En{\ss}lin, T.
\newblock The variable shadow of M87.
\newblock {\em arXiv preprint arXiv:2002.05218} {\bf 2020}.

\bibitem[{Leike, R. H.} \em{et~al.}(2020){Leike, R. H.}, {Glatzle, M.}, and
  {En\ss{}lin, T. A.}]{reimardust}
{Leike, R. H.}.; {Glatzle, M.}.; {En\ss{}lin, T. A.}.
\newblock Resolving nearby dust clouds.
\newblock {\em A\&A} {\bf 2020}, {\em 639},~A138.
\newblock
  doi:{\changeurlcolor{black}\href{https://doi.org/10.1051/0004-6361/202038169}{\detokenize{10.1051/0004-6361/202038169}}}.

\bibitem[Arras \em{et~al.}(2019)Arras, Frank, Leike, Westermann, and
  En{\ss}lin]{arras2019unified}
Arras, P.; Frank, P.; Leike, R.; Westermann, R.; En{\ss}lin, T.A.
\newblock Unified radio interferometric calibration and imaging with joint
  uncertainty quantification.
\newblock {\em Astronomy \& Astrophysics} {\bf 2019}, {\em 627},~A134.

\bibitem[{Hutschenreuter, Sebastian} and {En\ss{}lin, Torsten
  A.}(2020)]{hutschfaraday}
{Hutschenreuter, Sebastian}.; {En\ss{}lin, Torsten A.}.
\newblock The Galactic Faraday depth sky revisited.
\newblock {\em A\&A} {\bf 2020}, {\em 633},~A150.
\newblock
  doi:{\changeurlcolor{black}\href{https://doi.org/10.1051/0004-6361/201935479}{\detokenize{10.1051/0004-6361/201935479}}}.

\bibitem[Wiener(1950)]{wiener_extrapolation_1950}
Wiener, N.
\newblock {\em Extrapolation, interpolation, and smoothing of stationary time
  series, with engineering applications.}; Number ix, 163 p. in Stationary time
  series, Technology Press of the Massachusetts Institute {ofTechnology},
  1950.

\bibitem[Matti~Lassas(2009)]{1930-8337_2009_1_87}
Matti~Lassas, Eero~Saksman, S.S.
\newblock Discretization-invariant Bayesian inversion and Besov space priors.
\newblock {\em Inverse Problems \& Imaging} {\bf 2009}, {\em 3},~87.
\newblock
  doi:{\changeurlcolor{black}\href{https://doi.org/10.3934/ipi.2009.3.87}{\detokenize{10.3934/ipi.2009.3.87}}}.

\bibitem[Frank \em{et~al.}(2021)Frank, Leike, and Enßlin]{fdi_adp}
Frank, P.; Leike, R.; Enßlin, T.A.
\newblock Field Dynamics Inference for Local and Causal Interactions.
\newblock {\em Annalen der Physik} {\bf 2021}, {\em 533},~2000486,
  \href{http://xxx.lanl.gov/abs/https://onlinelibrary.wiley.com/doi/pdf/10.1002/andp.202000486}{{\normalfont
  [https://onlinelibrary.wiley.com/doi/pdf/10.1002/andp.202000486]}}.
\newblock
  doi:{\changeurlcolor{black}\href{https://doi.org/https://doi.org/10.1002/andp.202000486}{\detokenize{https://doi.org/10.1002/andp.202000486}}}.

\bibitem[Bertin and Arnouts(1996)]{bertin1996sextractor}
Bertin, E.; Arnouts, S.
\newblock SExtractor: Software for source extraction.
\newblock {\em Astronomy and astrophysics supplement series} {\bf 1996}, {\em
  117},~393--404.

\end{thebibliography}

\end{document}